\documentclass[longbibliography,aps,prx,amsmath,amssymb,superscriptaddress,twocolumn,10pt]{revtex4-2}

\usepackage{preamble_revtex}

\makeatletter
\def\@date{}      % clears the date internally
\def\print@date{}  % disables printing the date line
\makeatother

\usepackage{newtxtext}
\usepackage{newtxmath}

\begin{document}

\title{Understanding the complexity of frequency and phase angle fluctuations in power grids}

\author{Alessandro Lonardi}
\affiliation{Centre for Complex Systems, School of Mathematical Sciences, Queen Mary University of London, Mile End Road, London E1 4NS, United Kingdom}
\author{Jacques M. Maritz}
\affiliation{Department of Engineering Sciences, University of the Free State, Bloemfontein 9301, South Africa}
\author{Leonardo Rydin Gorjão}
\affiliation{Department of Environmental Sciences, Faculty of Science, Open University
of The Netherlands, Heerlen, 6419AT, The Netherlands}
\affiliation{Faculty of Science and Technology, Norwegian University of Life Sciences,
Ås, 1432, Norway}
\author{Christian Beck}
\affiliation{Centre for Complex Systems, School of Mathematical Sciences, Queen Mary University of London, Mile End Road, London E1 4NS, United Kingdom}

\begin{abstract}
    Power grids must modernize to meet climate goals while maintaining reliable and stable operating conditions. Yet progress is hindered by a limited understanding of the stochastic processes underlying grid frequency and phase-angle fluctuations, which are induced by the growing penetration of renewable generation, consumer demand fluctuations, and market trading.
    This issue is particularly acute in Africa, where grids often face weak investment.
    Here, we present results from a newly collected, large-scale, high-resolution dataset of grid frequency and phase angles for the United Kingdom and South Africa, comprising close to one billion data points.
    Using superstatistical modeling, we treat market-driven power fluctuations as a slowly varying parameter driving grid dynamics and incorporate nonlinear frequency control.
    As a result, we derive an analytical model that reproduces multimodal frequency distributions previously obtained from numerical simulations, as well as heavy-tailed fluctuations and double-exponential frequency autocorrelation decays, all in excellent agreement with experimental measurements.
    Beyond frequency, we also address the so far largely overlooked problem of characterizing spatial phase-angle fluctuations.
    By comparing our predictions with measurement data, we demonstrate that a low-dimensional effective grid model accurately fits South African data despite the grid's complexity. We also highlight significant differences between the grids of South Africa and the United Kingdom.
    Our results clarify how energy markets and control policies shape grid dynamics across countries with contrasting infrastructure maturity.
\end{abstract}

\maketitle

\section{Introduction}

The 2024 United Nations Climate Change Conference pledged to modernize power grids globally to meet the Paris Agreement temperature goal~\cite{cop29pledge}. Such a commitment faces significant hurdles in many countries, particularly in Africa~\cite{wef2025scaling}, where electricity transmission often lacks sustained investment and the possibility of clean energy adoption is limited~\cite{iea2024world}. Designing future-proof power grids requires collecting high-precision, large-scale data and developing realistic physical models. Research in these directions progressed~\cite{weissbach2009high,rohden2012self,motter2013spontaneous,manik2014supply,mele2016impact,anvari2016short,schafer2018isolating,schafer2018dynamically,schaefer2018non,deng2019frequency,haehne2019propagation,vorobev2019deadbands,gorjao2020open,anvari2020stochastic,gorjao2020data,gorjao2021spatio,delgiudice2021effects,witthaut2022collective,anvari2022data,han2022complexity,wen2023non,kraljic2023towards,oberhofer2023non,kruse2023physics,openacess2025grid,drewnick2025analyzing,boettcher2026impact}, but much of it has focused on countries with well-developed infrastructures, prompting a geographical bias. Nevertheless, understanding mechanisms that govern power grid dynamics is important for systems that have received comparatively little study.

Typical drivers that influence grid dynamics include energy market fluctuations ~\cite{wood2013power} and control mechanisms that stabilize the grid frequency to prevent failures~\cite{machowski2020power,kundur2007power}. Regional markets balance power supply and demand by trading electricity at discrete intervals~\cite{mayer2018electricity}. Transaction volume fluctuates due to factors like the volatility of electricity prices~\cite{han2022complexity} and residential consumption patterns~\cite{anvari2022data}, which, in turn, continuously affect grid dynamics. As a result, frequency deviations from the nominal value can exhibit large fluctuations~\cite{weissbach2009high,schaefer2018non,schafer2018isolating,haehne2019propagation,gorjao2020data,anvari2020stochastic,kraljic2023towards}. Heavy-tailed frequency distributions can be reproduced by incorporating power transactions analytically~\cite{gorjao2020data}, through non-Gaussian noise~\cite{schaefer2018non,schafer2018isolating,gorjao2021spatio,anvari2016short,kashima2015modeling}, via numerical simulations~\cite{weissbach2009high,haehne2019propagation} and neural networks~\cite{kruse2023physics}, or by data extraction~\cite{anvari2020stochastic}. However, these models assume linear frequency control, whereby generators inject power proportional to frequency deviations to damp perturbations. This is a good first approximation, but it is unrealistic for large grids~\cite{kundur2007power,machowski2020power} where linear control yields unimodal frequency distributions that differ from typically observed multimodal distributions~\cite{mele2016impact,deng2019frequency,delgiudice2021effects,wen2023non,kraljic2023towards}. Recent studies employ nonlinear control~\cite{vorobev2019deadbands,kraljic2023towards,oberhofer2023non} but rely on simulations and preclude direct data fitting.

Grid dynamics is also characterized by phase angles describing the relative rotation of generators. During regular operations, they are phase-locked, namely, generators operate with nearly constant phase-angle differences to prevent disruptions~\cite{kundur2007power,machowski2020power}. Phase-locked states are widely studied with dynamical systems methods~\cite{witthaut2022collective} to investigate grid stability~\cite{manik2014supply,manik2017cycle}, spontaneous synchronization~\cite{rohden2012self,motter2013spontaneous}, and cascades~\cite{schafer2018dynamically}, among other phenomena. Contrarily, their stochastic properties are seldom explored, and mainly in the context of synchronization of synthetic systems~\cite{schafer2017escape,hindes2019network,schmietendorf2017impact,tumash2018effect}. An empirically validated description of phase-angle fluctuations in real-world grids is still missing. 

\begin{figure}[t]
\centering
\includegraphics[width=1.0\linewidth]{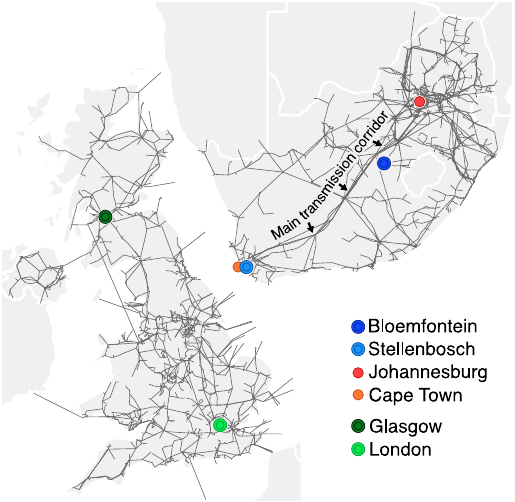}
\caption{Power grids in the UK and SA. Grid nodes include power plants and substations, while links represent transmission lines connecting them. Node and link data were retrieved separately from Open Infrastructure Map (\href{https://openinframap.org}{https://openinframap.org}) and matched by aggregating link endpoints and nodes within \qty{3.33}{km}. After merging, loops and duplicate lines were removed. The final networks consist of a single connected component with 2059 nodes and 2614 links for SA, and 2115 nodes and 2467 links for the UK. Cities where PMUs were installed are marked, along with Johannesburg and Cape Town, which are connected by a main transmission corridor running diagonally across SA and relevant to phase-angle fluctuations.}
\label{fig: grid networks}
\end{figure}

To fill these gaps, we collected and analyzed approximately one billion frequency and phase‑angle measurements at a 0.1-s resolution. Our dataset far exceeds prevailing coarser, frequency-only datasets resolved at seconds \cite{gorjao2020open,schaefer2018non,schafer2018isolating,anvari2020stochastic,gorjao2020data,delgiudice2021effects,wen2023non,kraljic2023towards,kruse2023physics,drewnick2025analyzing,openacess2025grid}. We concentrate on four example cities: Stellenbosch and Bloemfontein in South Africa (SA), and London and Glasgow in the United Kingdom (UK) (\Cref{fig: grid networks}). 
The British grid is an ideal testbed for theory, owing to extensive research~\cite{schaefer2018non,gorjao2020data,gorjao2020open,anvari2020stochastic,manik2014supply,vorobev2019deadbands,rohden2012self,kraljic2023towards,delgiudice2021effects,motter2013spontaneous,schafer2018dynamically,kruse2023physics}, tightly regulated control~\cite{nesonote,ofgemnote}, and its centrality in the European energy market~\cite{epexukmarket}.
The South African grid is the largest in Africa and is synchronous with eleven other countries in the Southern African market~\cite{sappmarket}. Notwithstanding its significance, the grid has faced disruptions for the past two decades~\cite{conradie2000symphony,csir2025utility} and heavily depends on fossil fuels~\cite{iea2024southafrica}, putting a strain on economic growth~\cite{janse2023reflections}, healthcare~\cite{manoki2023power}, and population well-being~\cite{greyling2019gross}. The influence of grid features like lenient control~\cite{nersacode} on frequency and phase-angle fluctuations remains insufficiently quantified, as research has primarily focused on load shedding~\cite{gorjao2023stochastic} and private microgrids~\cite{maritz2024data}.

Here, we study typical properties of the grid dynamics in SA and the UK with a new kind of superstatistical method~\cite{beck2001dynamical,beck2003superstatistics,beck2005from}. Generally, a superstatistics is a mixture of multiple statistics that describes a nonequilibrium system in a stationary state with a slowly fluctuating intensive parameter. Previously, superstatistics for power grids have considered fluctuating control coefficients \cite{schaefer2018non}. By contrast, building on recent studies~\cite{anvari2022data,kruse2023physics}, we treat the market‑driven power imbalance between grid generation and load as the fluctuating superstatistical parameter that drives the slow evolution of the grid dynamics. We assume that this parameter evolves on timescales much longer than those of control, so that over short intervals the grid relaxes to local equilibrium states around the nominal frequency. The average of the local equilibrium states, weighted by slow power fluctuations, describes the frequency distributions. For phase angles, we employ a reduced grid model, where phase-angle differences are described by a stochastic damped-driven motion in a tilted washboard potential~\cite{risken1989fokker,schafer2017escape,manik2014supply}.

When confronted with measured data, our models explain and quantitatively reproduce key empirical features of frequency distributions, including peaks and dips induced by nonlinear control in both SA and the UK, as well as heavy tails. We also observe a double-exponential decay of the frequency autocorrelation in SA. For phase angles, we find empirical support for a reduced-grid model in SA that accurately fits the measured phase-angle difference distribution despite the grid's underlying complexity. In the UK, no such reduction to a low-dimensional model is observed. Overall, our results illustrate and quantify the subtle differences that can arise across different power grids.

\section*{Results}

For our investigation, fluctuating power-grid frequencies and phase angles are the observables of interest. They are measured using newly installed Phasor Measurement Units (PMUs), which capture electrical waveforms at their installation sites and report frequency and phase-angle measurements synchronized to a standard GPS time reference. The measurements are processed to remove spurious data and yield six fine-grained time series (two for frequencies in London and Stellenbosch, four for phase angles in each city) with a 0.1-s resolution over six months, from April to October 2025 (see {\crefmethods} for details).

For convenience, in all figures and calculations, we convert the frequency $f(t)$ in $\unit{Hz}$ to the angular velocity $\omega(t) := 2\pi(f(t)-f_\mathrm{R})$ in $\unit{rad/s}$, where $f_\mathrm{R} = \qty{50}{Hz}$ is the nominal frequency in the UK and SA. Phase angles $\theta(t)$ are in radians.

%%%%%%%%%%%%%%% FIGURE 1 %%%%%%%%%%%%%%%
%%%%%%%%%%%%%%%%%%%%%%%%%%%%%%%%%%%%%%%%

\subsection*{Frequency}

\begin{figure*}[t]
\centering
\includegraphics[width=1.0\linewidth]{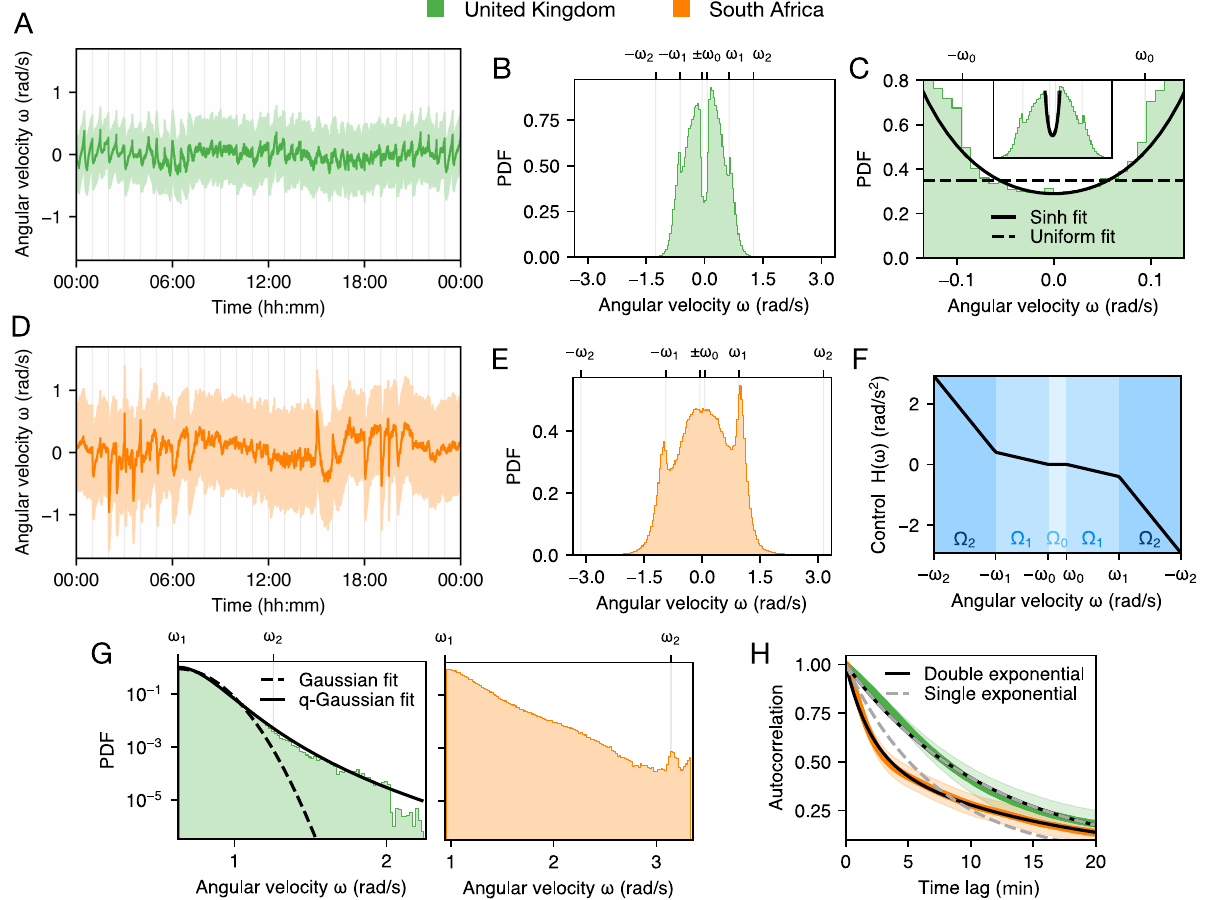}
\caption{
Frequency measurements in SA and the UK. We conventionally use data from Stellenbosch and London to represent their countries, as, during regular operations, they are effectively identical to those in Bloemfontein and Glasgow. MLE fits in panels C and G are done with one million points sampled uniformly from the frequency signal in the deadband and tails.
{(A, D)} Average frequency time series aggregated over \qty{24}{h}. Lighter areas are standard deviations.  
{(B, E)} Frequency histograms. Vertical lines indicate the control regions of panel F. The peaks at $\pm \omega_1$ are asymmetric due to finite-size effects. The histogram's robustness is validated by splitting the data into monthly segments in \emph{SI Appendix}, Note S1, Fig. S4, S5.  
{(C)} MLE fit in the deadband. Data is fitted in $\Omega_0$ (see panel F) and plotted over a larger interval. The dashed line is the uniform fit that neglects the slow power contribution \cite{schaefer2018non}.  
{(F)} Control function. In the deadband $\Omega_0$ there is no control. In $\Omega_1$ and $\Omega_2$ control is linear with coefficients $\gamma_1 < \gamma_2$. The $y$-axis is arbitrarily scaled.  
{(G)} Tails. The left and right tails are aggregated. In the UK, the tail at $|\omega| > \omega_1$ is fitted with a Gaussian and a $q$-Gaussian. See \emph{SI Appendix}, Note S7, S11, Fig. S9.  
{(H)} Autocorrelation at short time lags $\Delta t \leq \qty{20}{min}$ with lighter areas being standard deviations. We fit a single- and a double-decay ansatz. See \emph{SI Appendix}, Note S8.
}
\label{fig: panel frequency}
\end{figure*}

To observe typical daily variations of frequency dynamics, we aggregate the measurements into \qty{24}{h} intervals. The resulting profiles are shown in \Cref{fig: panel frequency}{A, D}, where the frequencies fluctuate around the nominal value. In the UK, fluctuations are tightly clustered, whereas in SA they are much broader, with average standard deviations  $\sigma_{\text{SA}} \approx 1.6 \, \sigma_{\text{UK}}$. This difference reflects the distinct frequency control mechanisms of the two grids. The UK's control linearly dampens fluctuations below \qty{0.2}{Hz}, with an additional linear layer between \qty{0.1}{Hz} and \qty{0.2}{Hz}~\cite{nesonote,ofgemnote}. In SA, contracted generators linearly dampen fluctuations above \qty{0.15}{Hz}, but no control is mandated below this threshold. Non-contracted generators only dampen fluctuations above \qty{0.5}{Hz}~\cite{nersacode}. In both grids, control has finite precision: if fluctuations are small, for instance, below \qty{0.01}{Hz}, it stays idle~\cite{rebours2007survey}. The region around the utility frequency without control is known as the grid frequency deadband.

\Cref{fig: panel frequency}{A, D} also display intermittent large fluctuations due to market transactions \cite{schaefer2018non,schafer2018isolating,gorjao2020data}. These occur every \qty{30}{min} in the UK and \qty{1}{h} in SA, consistent with energy markets' intraday calendars \cite{sappmarket,epexukmarket}. Such fluctuations dampen in the UK between 8-11 a.m. and 2-8 p.m., and in SA between 8 a.m.-2 p.m. and 9 p.m.-1 a.m., when heterogeneous consumer activity rises.
Details on control and markets in the UK and SA are in \emph{SI Appendix}, Notes S2, S3, Fig. S6, S7.

Control, market, and consumer demand produce the multimodal, heavy-tailed distributions in \Cref{fig: panel frequency}{B, E}. 
Accurately describing their features is our first main modeling challenge.

We model frequency dynamics with the aggregated swing equation \cite{ulbig2014impact},
\begin{equation}
    \label{eq: swing equation}
    \frac{\mathrm{d} \omega}{\mathrm{d}t} = H(\omega) + P(t) + \epsilon \xi \,,
\end{equation}
which expresses energy conservation in the grid (see {\crefmethods} for its derivation and full description). 
In \Cref{eq: swing equation}, $H(\omega)$ represents frequency control, which we model as a piecewise linear function as shown in \Cref{fig: panel frequency}F. We split the imbalance between generated and consumed power into two components: $P(t)$, the slowly varying term central to our superstatistical model, and the noise $\epsilon \xi$, incorporating the fast fluctuations not resolved in $P(t)$. The noise also accounts for other fast fluctuating sources, such as short-term volatility generation or rapidly changing consumer demand \cite{anvari2016short,milan2013turbulent}.

%%%%%%%%%%%%%%%%%%%%%%%%%%%%%%%%%%%%%%%%%%%%%%%%
%%%%%%%%%%%%%%%%%%%%%%%%%%%%%%%%%%%%%%%%%%%%%%%%

In detail, we assume that the evolution of $P(t)$ is slow relative to the control damping rates, which describe how quickly the grid responds to a frequency fluctuation. To get quantitative estimates, we extract the damping rates with Kernel Regression (KR) \cite{lamoroux2009kernel,gorjao2019kramersmoyal} and find $\tau = \qty{4.06 \pm 0.03}{min}$ in the UK and $\tau = \qty{9.07 \pm 0.04}{min}$ in SA (see {\crefmethods}). Power data, namely measured values for the difference $P(t)$ between power production and demand, are often scarce \cite[references therein]{anvari2022data}, and in our case, unavailable. However, large-scale power adjustments are generally organized on market time scales ($\qty{30}{min}$ in the UK and $\qty{1}{h}$ in SA), which set a horizon for slow changes in the system imbalance \cite{kruse2023physics}. More importantly, \Cref{eq: swing equation} allows us to choose the separation of time scales conventionally, absorbing all fluctuations of the same order as $\tau$ into noise and control and retaining only slower variations in $P(t)$. The inferred grid coefficients should therefore be understood as effective parameters that depend on the chosen time-scale separation. Taken together, this evidence provides quantitative support for the use of superstatistics.
We assume Gaussian white noise $\xi \sim \mathcal{N}(0,1)$ with amplitude $\epsilon$, which gives an analytically tractable superstatistical model with a multimodal frequency distribution, fitting those in \Cref{fig: panel frequency}{B, E}. A discussion on alternative noise choices is given in \emph{SI Appendix}, Note S5.

Employing superstatistics, we treat $P(t) = P$ as effectively constant over short time intervals (of the order of $\tau$) and model the frequency distribution as an ensemble average of microstates $\omega$-given-$P$:
\begin{equation}
    \label{eq: frequency distribution}
    p(\omega) \propto \int f(\omega \, | \, P) \, \varphi(P) \, \mathrm{d}P \,.
\end{equation}
The distribution $f(\omega \, | \, P)$ describes the short-term equilibrated grid microstate $\omega$-given-$P$, before the slow power change causes it to drift. Integrating against $\varphi(P)$ accounts for the distribution of slow power fluctuations. We calculate $f( \omega \, | \, P)$ in closed-form by solving a Fokker--Planck equation obtained from \Cref{eq: swing equation} (see {\crefmethods}). Its expression, in conjunction with \Cref{eq: frequency distribution}, yields rich quantitative insights and captures all features of the frequency histograms.

The frequency deadband in the UK exhibits a pronounced suppression. Using moment-generating functions, we show that a dip emerges in the deadband for any non-degenerate power distribution with $\mathbb{E}[P] = 0$ and $\mathbb{E}[P^2] > 0$. The first condition follows from energy conservation, while the second is true if $P(t)$ is not trivially $P(t) \equiv 0$. Beyond this first formal argument, we refine the estimation of deadband dips. We assume constant power, $\varphi(P) = c$, within a small interval of width $\delta$ around $P = 0$. This yields the closed-form expression $p(\omega) \propto (c \epsilon^2 / \omega) \, \sinh(\delta \omega / \epsilon^2)$ for the deadband distribution (see {\crefmethods}). In \Cref{fig: panel frequency}{C}, we fit the distribution to the UK data via Maximum-Likelihood Estimation (MLE) and obtain excellent agreement between theory and observations. SA data exhibit a less pronounced dip, which, although similar in interpretation, precludes MLE. See \emph{SI Appendix}, Note S11 for algorithmic details.

Two peaks arise at the boundary between control regions, namely at $\pm\qty{0.1}{Hz}$ in the UK and $\pm \qty{0.15}{Hz}$ in SA, with substantially larger magnitudes in SA. Their emergence can also be explained by piecewise linear control. Formally, $f(\omega \, | \, P)$ takes the form of a Gaussian branch in each linearly damped region. At the boundary, branches from different regions merge under the integral in \Cref{eq: frequency distribution}, leading to an accumulation of probability mass that manifests as local maxima, with varying damping strengths influencing peak heights. We validate this argument by extracting damping coefficients in the linearly damped control regions through fitting the frequency autocorrelation in \Cref{fig: panel frequency}{H}. Specifically, we derive a closed-form ansatz for the autocorrelation as a weighted sum of two exponentials $C(\Delta t) = A \exp({-\gamma_1 \Delta t}) + (1-A) \exp({-\gamma_2 \Delta t})$, where $\gamma_1$ and $\gamma_2$ are the damping coefficients for the inner and outer regions, while $0 \leq A \leq 1$ is a weight parameter (see {\crefmethods}). We interpolate the autocorrelation using nonlinear least squares. The fit supports our argument: in the UK, the autocorrelation is well-described by a single exponential, indicating similar damping strengths, hence attenuated peaks. The best fit for SA requires two distinct coefficients, $\gamma_1 = (\num{8.6 \pm 0.2}) {\cdot} 10^{-4}\,\unit{rad/s}$ and $\gamma_2 = (\num{1.15 \pm 0.02}){\cdot} 10^{-4}\,\unit{rad/s}$. Double-exponential decays, indicative of nonlinear control, have been observed in grids worldwide but had not yet been quantified \cite{schaefer2018non,gorjao2020open,wen2023non}.

Finally, we examine the tails of the frequency distributions. In \Cref{fig: panel frequency}{G}, the UK data exhibit heavy tails that are accurately captured by a $q$-Gaussian MLE fit with power law heavy tails \cite{touchette2005asymptotics}.
In our superstatistical model, heavy tails are inherited by $\varphi(P)$. We show with saddlepoint approximation \cite{butler2007saddlepoint} that power law tails arise in $p(\omega)$ when $\varphi(P)$ follows a power law distribution and the Gaussian profile of  $f(\omega \,|\, P)$ is sufficiently peaked (see {\crefmethods}). $q$-Gaussian distributions also arise from superstatistics applied to power grids where the frequency control is modeled by an effective friction following a chi-squared profile \cite{schaefer2018non}. Our model offers a complementary explanation for heavy tails where control remains constant, and slow power fluctuations mix the system's microstates. SA data exhibit a small peak in the tails, indicative of a transition between control regions at $\qty{0.5}{Hz}$. The data beyond this peak are insufficient to fit the tail distributions in a reliable manner.

\subsection*{Phase angles}

\begin{figure}[t]
\centering
\includegraphics[scale=0.9]{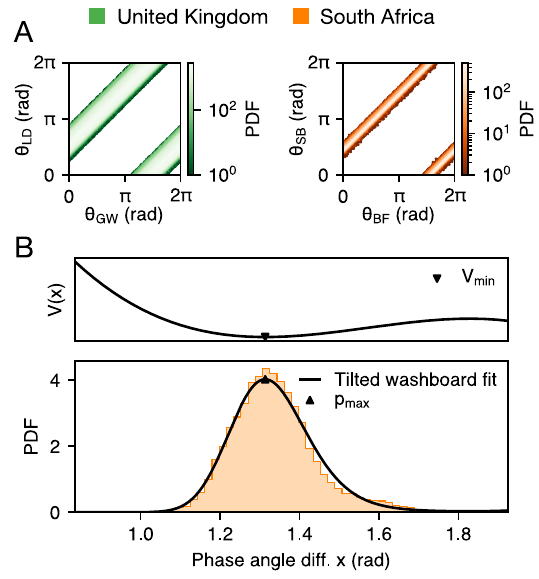}
\caption{
Phase angle measurements in SA and the UK. 
(A) Phase-locked hexbin distributions. A companion figure showing data processing is in \emph{SI Appendix}, Fig. S1. 
(B) MLE fit of the phase-angle difference histogram in SA. Both the potential $V(x)$ and the corresponding distribution $p(x)$ obtained from the MLE are drawn. The fit was performed using one million points sampled uniformly from $x(t)$.
}
\label{fig: panel phase angle}
\end{figure}

Our second main challenge is to characterize the statistics of phase-angle fluctuations.
To this end, we evaluate phase-angle differences $x(t) = \theta_{1}(t) - \theta_{2}(t)$ between $\theta_{1}(t)$ and $\theta_{2}(t)$ measured in Stellenbosch and Bloemfontein, and in London and Glasgow. During regular operations, the grid is phase-locked. Namely, $x(t)$ stays approximately constant over time as shown in \Cref{fig: panel phase angle}{A}, where phase-locked states appear as diagonal bands with unit slope. Contrary to the frequencies, phase-difference bands in the UK are broader than in SA, highlighting a structural difference between the two grids. Large frequency fluctuations in SA due to lenient control are coherent across the grid and shift all phase angles around their locked values. Instead, phase-difference fluctuations are location-specific. We ascribe their greater magnitude in the UK to the high penetration of renewable energy sources. This interpretation is corroborated by the fact that renewables account for approximately 40\% and 8\% of power generation in the UK and SA, respectively \cite{iea2024southafrica, iea2024unitedkingdom}.

%%%%%%%%%%%%%%% FIGURE 1 %%%%%%%%%%%%%%%
%%%%%%%%%%%%%%%%%%%%%%%%%%%%%%%%%%%%%%%%

We use a reduced grid model to describe the dynamics of $x(t)$. For tractability, we neglect superstatistics and assume constant power generation, consider only the effective interaction between two generators $\theta_{1}(t)$ and $\theta_{2}(t)$, and take linear control for simplicity. Under these assumptions, we can recast energy conservation as a stochastic damped-driven harmonic oscillator in a tilted washboard potential $V(x)$ \cite{schafer2017escape,manik2014supply} (see \crefmethods):
\begin{align}
    \label{eq: tilted washboard equation}
    \frac{\mathrm{d}^2 x}{\mathrm{d}t^2} + \gamma \frac{\mathrm{d}x}{\mathrm{d}t} + \frac{\mathrm{d}V}{\mathrm{d}x} = \epsilon \xi \, \\
    \label{eq: tilted washboard potential}
    V(x) := - \delta x - 2 \kappa \cos x \,.
\end{align}
In \Cref{eq: tilted washboard equation}, $\gamma$ denotes the constant control coefficient, and $\epsilon \xi$ is Gaussian white noise as in \Cref{eq: swing equation}. We reuse $\epsilon$ for simplicity, but generally, the two noise amplitudes in the two equations differ. In \Cref{eq: tilted washboard potential}, $\delta$ quantifies the difference of mechanical power between generators, and $\kappa$ is a transmission coupling controlling power exchange.

\Cref{eq: tilted washboard equation} admits the stationary distribution $p(x) \propto \exp(- \gamma V(x) / \epsilon^2)$ \cite{risken1989fokker}, which describes the probability of finding the system near a stable potential minimum located at $x_{\text{min}} = \arcsin(\delta / 2\kappa)$ (see {\crefmethods}).
We fit $p(x)$ to the SA data using MLE and find close agreement between model and observations. This outcome is surprisingly good, given that the grid is a highly complex system, yet it is well-captured by the simple model in \Crefrange{eq: tilted washboard equation}{eq: tilted washboard potential}. A plausible explanation is the strong dominance of the two regions around Johannesburg and Cape Town, which account for most of the grid activity and are linked by a transmission corridor running diagonally across the country, as shown in \Cref{fig: grid networks}. The reduced model does not fit the UK data, as the ansatz $p(x)$ appears to be incompatible with the regional characteristics of the UK; \emph{SI Appendix}, Note S10, Fig. S3. We expect phase-angle differences in the UK to require more refined modeling assumptions.

\section*{Discussion}

In this work, we analyzed high-resolution frequency and phase-angle time series from the British and South African power grids using methods rooted in statistical physics.

The frequency data exhibit multimodal, heavy-tailed distributions (\Cref{fig: panel frequency}{B, E}). We show that their features are accurately described by a superstatistical model in which slow fluctuations in the grid power imbalance drive the system's long-term evolution. Within this framework, the stationary distribution of the aggregated swing equation (\Cref{eq: swing equation}) becomes an average of short-time equilibrated microstates weighted by power-imbalance fluctuations (\Cref{eq: frequency distribution}).
The salient advantage of such a formulation is that it enables the integration of nonlinear frequency control, allowing us to derive and fit an analytical expression in the deadband dip (\Cref{fig: panel frequency}{C}), explain control-induced multimodality (\Cref{fig: panel frequency}{B, E}), describe and fit heavy tails well represented by a $q$-Gaussian distribution (\Cref{fig: panel frequency}{G}), and account for the double-exponential decay of the frequency autocorrelation (\Cref{fig: panel frequency}{H}). Earlier models employed a different version of superstatistics with fluctuating control coefficients \cite{schaefer2018non} or were based on simpler constructive assumptions for the power-imbalance profile \cite{gorjao2020data} combined with linear control, thereby capturing only partially the histogram features. Nonlinear control has been modeled mainly with numerical schemes so far~\cite{vorobev2019deadbands,kraljic2023towards,oberhofer2023non}. By contrast, our new results presented here advance a data-driven analytical model in close quantitative agreement with empirical observations.

The stochastic properties of phase-angle dynamics have, to date, received comparatively little attention in the literature. Although they have been investigated with synthetic experiments \cite{schafer2017escape,hindes2019network,schmietendorf2017impact}, a large-scale empirical validation of theoretical predictions has been lacking. We show that an effective two-generator grid model (\Crefrange{eq: tilted washboard equation}{eq: tilted washboard potential}) accurately describes the phase-angle difference distribution between Stellenbosch and Bloemfontein. The level of agreement observed in \Cref{fig: panel phase angle}{B} is remarkable given the structural and dynamical complexity of the real-world grid (\Cref{fig: grid networks}). As a plausible explanation for this closer-than-expected coincidence, we argue that the South African grid is dominated by two generation-load centers in Johannesburg and Cape Town, effectively reducing an otherwise high-dimensional system to a low-dimensional one.

Our findings are supported by a newly obtained set of data points consisting of about one billion measurements sampled at 0.1-s resolution, demonstrating that large-scale, high-resolution measurements are essential for a detailed understanding of the statistical signatures of grid dynamics.

Overall, we have demonstrated how country-specific factors such as control regulations, energy-market organization and scheduling, energy mix, and infrastructure maturity shape the statistical properties of grid observables. We also hypothesize that the higher fraction of renewables in the United Kingdom, together with structural grid differences, additionally contribute to the differences observed in the stochastic properties of both frequency and phase angles.

Regarding implementation details and further refinements of the superstatistical model developed in this paper, it would be of interest to directly extract the statistics of the slow power imbalance $P(t)$ from measured frequency data and then integrate \Cref{eq: frequency distribution}. Doing so could also enable fitting the long-time power-law decay observed in the frequency autocorrelation using a superstatistics-motivated ansatz that would explicitly account for frequency jumps caused by market transactions. At present, these detailed features related to jump processes are not covered by our model.

\section*{Materials and Methods}

\phantomsection
\label{sec: materials and methods}

\subsection*{Data Collection and Processing}
We accessed the PMUs data using the GridRadar API (\href{https://gridradar.net/en}{https://gridradar.net}). The raw time series span six months and comprise approximately one billion data points. We processed them to remove spurious measurements returned by the PMUs following these steps:
(1) Rare ($< 1\%$) \texttt{NaN} entries were present in the data and were accounted for using \texttt{NaN}-robust operations.
(2)
Numerical instabilities occurred after a few hourly intervals with missing data, when the PMUs returned no value. To remove them, we discarded a 10-hour window after each empty interval, which was chosen at convenience to cover all spurious measurements.
(3)
We trimmed phase-angle artifacts lying outside the bands in \Cref{fig: panel phase angle}{A} by filtering out points outside the range $\Delta_1 \leq x \leq \Delta_2$. We chose $\Delta_1, \Delta_2$ separately for each band such that less than $0.02\%$ of the data is removed.  
(4)
PMUs introduced artifacts due to GPS clock corrections \cite{almas2018vulnerability}. These appeared as stepwise jumps in the phase-angle differences $x(t)$. We removed them by extracting the lower envelope of the signal, which preserves its stochastic profile while excluding artifacts. All results for phase-angle differences are given in terms of lower envelopes.

For details on (3)-(4), see \emph{SI Appendix}, Note S1, Figs. S1, S3.

\subsection*{Aggregated Swing Equation}

The dynamics of a grid with $i = 1, \dots, N$ generators obeys in good approximation the swing equations \cite{machowski2020power,kundur2007power}, which express energy conservation in terms of the generators' phase angles $\theta_i(t) \in [0,2\pi)$ and angular velocities $\omega_i(t) := 2\pi (f_i(t) - f_{\mathrm{R}})$ \cite[for a derivation]{manik2014supply}. Since during regular operation $\omega_i(t) = \omega(t)$ for all $i$, the grid dynamics can be recast into the aggregated \Cref{eq: swing equation}, controlling the angular velocity $\omega(t)$, independent of $i$. We outline the essential steps of its derivation below. Details are in \emph{SI Appendix}, Note S4.

For each generator $i$, the swing equations read
\begin{alignat}{2}
    \label{eq: swing equation 1}
    \frac{\mathrm{d} \theta_i}{\mathrm{d} t} &= \omega_i\\
    \label{eq: swing equation 2}
    M_i \frac{\mathrm{d} \omega_i}{ \mathrm{d} t} &= H_i(\omega_i) + P_i(t )+\sum_{j = 1}^{N} K_{ij} \sin (\theta_j -  \theta_i) +\epsilon_i \xi_i \,.
\end{alignat}
In \Cref{eq: swing equation 2}, the $M_i$ denote the generator inertia. The function $H_i(\omega_i)$ represents frequency control, which we explicitly define below for the aggregated equation. The term $P_i(t)$ is the slow net power output of the generator, capturing its supply-demand imbalance with consumers. We assume a superstatistical separation of time scales: each $P_i(t)$ evolves slowly compared to the control rates. Fast fluctuations are incorporated into the i.i.d. Gaussian white noise $\xi_i$ with amplitude $\epsilon_i$. The non-negative coupling coefficients $K_{ij}$ regulate the power exchanged between generators. If $i$ and $j$ are not connected, then $K_{ij} = 0$.

The grid regulates power production to balance the aggregated slow power $P(t)$ in \Cref{eq: swing equation}, ensuring equilibrium on average \cite{gorjao2020data,schaefer2018non}:
\begin{equation}
\label{eq: equilibrium}
P(t) := \sum_{i=1}^N \frac{P_i(t)}{\sum_{i=1}^N M_i} \quad \langle P \rangle := \lim_{T  \to \infty} \int_0^T P(t) \,\mathrm{d}t = 0 \,.
\end{equation}

By rescaling and summing \Cref{eq: swing equation 2} over $i$, the sine contributions vanish. Hence, we recover \Cref{eq: swing equation}. In \Cref{eq: swing equation}, we also swap $H_i(\omega_i)$ with $H(\omega)$ by assuming homogeneous control for all generators.
We model control as $H(\omega) := - \sign{\omega}h(|\omega|)$, with
\begin{align}
    \label{eq: control definition}
    h(x) &:= \begin{cases}
    0 \quad &x \in \Omega_0\\
    \gamma_{1} \, (x - \omega_0) \quad &x \in \Omega_1 \\
     \gamma_{2} \, (x - \omega_1) + \gamma_{1} \, (\omega_1 - \omega_0) \quad &x \in \Omega_2 \,,
    \end{cases}
\end{align}
where $\Omega_0$, $\Omega_1$, and $\Omega_2$  denote the deadband, the inner control region, and the outer control region of \Cref{fig: panel frequency}{F}, respectively.

\subsection*{Superstatistics Validation: Kernel regression (KR)}

We estimate the control damping coefficients and noise amplitudes from the data and use them to obtain the control damping rates. The estimation is performed using KR \cite{lamoroux2009kernel,gorjao2019kramersmoyal}, as explained below. Extensive details are given in \emph{SI Appendix}, Note S9, Fig. S11.

First, we detrend frequencies to remove slow drifts, like those caused by market transactions. Specifically, we focus on $\omega_{\text{det}}(t) := \omega(t) - \omega_{\text{filter}}(t)$, where $\omega_{\text{filter}}(t)$ is the frequency signal filtered with a Gaussian kernel with $\sigma = \qty{60}{s}$ \cite{drewnick2025analyzing,gorjao2019kramersmoyal}.

For validation purposes only, we approximate $H(\omega)$ by a linear function with a single damping coefficient $\gamma$. Estimating region-specific damping reliably with KR is hard, as it requires continuous signals, whereas the data fluctuate across control regions. Here, $\gamma$ can be interpreted as describing the aggregate contribution of the control function across the regions in \Cref{eq: control definition}.

Then, we extract the finite-difference drift estimator
\begin{equation}
    \label{eq: drift estimator}
    {D}_1(\omega) := \frac{1}{\Delta t_\mathrm{PMU}} \left\langle \omega_{\text{det}}(t + \Delta t_{\mathrm{PMU}}) - \omega_{\text{det}}(t) \, | \, \omega_{\text{det}}(t) = \omega \right\rangle \,.
\end{equation}
In \Cref{eq: drift estimator}, $
\left\langle \cdot \, | \, \omega_{\text{det}}(t) = \omega \right\rangle
$
denotes a kernel-weighted conditional average at $\omega_{\text{det}}(t) = \omega$ and $\Delta t_{\mathrm{PMU}} = \qty{0.1}{s}$. Following previous works \cite{gorjao2023stochastic}, we use a scaled Epanechnikov kernel $K_h(x) = 3 [1 - {(x/h)}^2 ]/4h$, $|x| < h$ \cite{gorjao2023stochastic}, with $h = \qty{0.1}{rad/s}$.

The estimator ${D_1}(\omega)$ is linear with a negative slope. We use linear least squares to extract its slope, which is the estimate of $\gamma$; the extracted value is consistent with previous work \cite{gorjao2020data,drewnick2025analyzing,schaefer2018non,kruse2023physics,oberhofer2023non}. The damping rates are calculated as $\tau = 1 / \gamma $, since $\gamma$ sets an exponential relaxation timescale on the frequency. We also obtain an estimate of the deadband escape time by extracting $\epsilon$ with KR and employing mean first-passage time theory \cite{redner2001guide}. All these estimates back up the use of superstatistics.

\subsection*{Frequency Distribution: Analytical Derivation} Below, we outline calculations; details are in \emph{SI Appendix}, Note S6, Fig. S8.

To compute $f(\omega \, | \, P)$, we solve the Fokker--Planck equation corresponding to the Langevin equation in \Cref{eq: swing equation}. Specifically, we integrate its ``stationary'' solution satisfying $\partial f / \partial t = 0$ in the short time intervals where $P$ can be kept constant, and enforce continuity at the control boundaries $\pm \omega_0, \pm \omega_1$. This yields
\begin{align}
    \label{eq: quasi stationary distribution}
   f(\omega \, | \, P ) = \frac{1}{Z} \times 
    \begin{cases}
    \displaystyle \exp \left( \frac{2 P }{\epsilon^2} \omega\right) \quad &\omega \in \Omega_0\\[0.5em]
    \displaystyle C_1 \exp \left( - \frac{\gamma_1}{\epsilon^2} \left( \omega - \Lambda_1 \right)^2  \right) &\omega \in \Omega_1 \\[0.5em]
    \displaystyle C_2  \exp \left( - \frac{\gamma_2}{\epsilon^2} \left( \omega - \Lambda_2 \right)^2  \right) &\omega \in \Omega_2 \, ,
    \end{cases}
\end{align}
where we defined
\begin{align}
    \label{eq: gaussian shift 1}
    \Lambda_1&:= \sign\omega \omega_{0} + {P}/{\gamma_1} \\
    \label{eq: gaussian shift 2}
    \Lambda_2 &:= \sign\omega \omega_{1} - ({\gamma_1}/{\gamma_2}) \, \sign\omega (\omega_{1} - \omega_{0}) + {P}/{\gamma_2} \,.
\end{align}
The scaling factors $C_1$ and $C_2$ ensure continuity. These factors, together with $\Lambda_1$ and $\Lambda_2$, are constant over the two branches $\omega < 0$, $\omega > 0$, and therefore depend only on the sign of $\omega$ and $P$. The global normalization $Z$ depends on $P$. Regrettably, $Z$ is not analytically integrable and precludes solving directly \Cref{eq: frequency distribution}.

\subsection*{Frequency deadband}

To study the frequency deadband, we focus on small, nonzero values of $P$, as these effectively capture $p(\omega)$ in $\Omega_0$: for large $P$, $Z$ suppresses the distribution in $\Omega_0$. In the small-$P$ regime, we approximate $Z$ by a constant thereby obtaining $p(\omega) \propto M_{\varphi} ({2\omega}/{\epsilon^{2}})$ from \Cref{eq: quasi stationary distribution} in $\Omega_0$, where $M_{\varphi}$ is the moment-generating function of $P$. Expanding it at $\omega=0$ gives
\begin{equation}
\label{eq: expain mgf}
\frac{\mathrm{d} p}{\mathrm{d} \omega} \bigg|_{\omega=0}
\propto \mathbb{E} [P] \quad
\frac{\mathrm{d} ^{2}p}{\mathrm{d} \omega^{2}}\bigg|_{\omega=0}
\propto \mathbb{E} [P^{2}] \,,
\end{equation}
showing us that any non-degenerate distribution $\varphi(P)$ with positive second moment produces a minimum at $\omega = 0$ that arises from averaging $f(\omega \,|\, P)$ in $\Omega_0$. The condition $\mathbb{E}_\varphi[P] = 0$ holds from energy conservation expressed in \Cref{eq: equilibrium} and assuming ergodicity.

Assuming further that $\varphi(P) = c$ for small $|P| \leq \delta/2$, we can integrate \Cref{eq: frequency distribution} to accurately fit the deadband in \Cref{fig: panel frequency}{C}. Details are in \emph{SI Appendix}, Note S7.

\subsection*{Peaks and Autocorrelation}

The frequency distribution peaks at $\pm \omega_1$ are captured by \Cref{eq: quasi stationary distribution}. Integrating \Cref{eq: frequency distribution} over $\Omega_1$ and $\Omega_2$ amounts to doing a convolution of the Gaussian branches of \Cref{eq: quasi stationary distribution} with $\varphi(P)$. If $\varphi(P)$ has sufficient support around $P = \gamma (\omega_1 - \omega_0)$ and does not strongly favor one branch over the other, the convolution roughly preserves the relative prominence of each Gaussian. Particularly, the branches in $\Omega_2$ produce taller peaks than those in $\Omega_1$, as $\gamma_2 > \gamma_1$; \emph{SI Appendix}, Note S7.

To verify that $\gamma_2 > \gamma_1$ experimentally, we study the frequency autocorrelation. KR does not allow us to extract region-specific damping coefficients. We circumvent this limitation by computing the second moment of $\omega$ conditioned on $P$, which, when integrated against $\varphi(P)$, yields the autocorrelation for short time lags. Even without knowing $\varphi(P)$, this approach shows that the autocorrelation is well described by a weighted sum of two exponential decays, with decay rates $\gamma_1$, $\gamma_2$; \emph{SI Appendix}, Note S8. The estimates are consistent with those found with KR.

At long time lags, the autocorrelation decays as a power law, consistent with previous findings \cite{kraljic2023towards} where this decay is modeled with fractional noise \cite{mandelbrot1968fractional}; \emph{SI Appendix}, Fig. S10. In our model, the slow evolution of $P(t)$ may induce long-term correlations in frequency, warranting further investigation.

\subsection*{Heavy Tails} 
$q$-Gaussian tails are compatible with our model. In particular, we show that power law tails, characteristic of $q$-Gaussian distributions, arise if $\varphi(P) \propto |P|^{-\beta}$. We approximate $H(\omega)$ by a linear function with a single damping coefficient $\gamma$. This is not restrictive, as we are now focusing on the tails of $p(\omega)$, where the non-linear profile of $H(\omega)$ is negligible. With this approximation, \Cref{eq: quasi stationary distribution} reduces to a Gaussian convolved with $\varphi(P)$. For sufficiently large $\omega$ compared to the width of the Gaussian, specifically, when $\omega \gg \epsilon / \sqrt{2\gamma}$, the convolution can be evaluated using saddlepoint approximation and gives
\begin{equation}
    \label{eq: saddle point approx result}
    p(\omega)  \propto \sqrt{\frac{\pi}{\gamma}} {\epsilon |\omega|^{-\beta}} \left[ 1 +  O\left(\frac{\epsilon^2}{2 \gamma \omega^2}\right){}\right] \simeq {\sqrt{\frac{\pi}{\gamma}}} \epsilon {|\omega|^{-\beta}} \,.
\end{equation}
For all mathematical details, see \emph{SI Appendix}, Note S7.

\subsection*{Tilted-Washboard Potential}
We outline the essential steps to derive \Crefrange{eq: tilted washboard equation}{eq: tilted washboard potential} from the swing equation in \Crefrange{eq: swing equation 1}{eq: swing equation 2}. Details are in \emph{SI Appendix}, Note S10, Fig. S12.

We substitute \Cref{eq: swing equation 1} into \Cref{eq: swing equation 2} and take $N=2$, which gives us two second-order stochastic differential equations controlling $\theta_1(t)$ and $\theta_2(t)$. We further assume $\gamma = D_1 / M_1 = D_2 / M_2$, $\kappa = K_{12} / M_1 = K_{21} / M_2$ \cite{schafer2017escape,manik2014supply}. Subtracting the swing equation for $\theta_1(t)$ from that for $\theta_2(t)$, we obtain \Cref{eq: tilted washboard equation}, where $\delta := P_1/M_1 - P_2/M_2$. The noise $\epsilon \xi$ is Gaussian if $\xi_1$, $\xi_2$ are i.i.d. Gaussian. Neglecting the noise, we classify the equilibria of \Cref{eq: tilted washboard equation} and find that, during regular operations, the reduced grid is locked at the stable point $x_{\text{min}} = \arcsin(\delta / 2 \kappa)$.

\Cref{eq: tilted washboard equation} yields a Smoluchowski (overdamped Fokker--Planck) equation, which we couple with a continuity equation for $p(x)$ and its probability current $J(x)$. These equations can be solved in a periodic potential, such as $V(x)$ in \Cref{eq: tilted washboard potential} \cite{risken1989fokker}. The SA data reveal that phase differences $x$ are tightly localized in the potential well at $x_{\text{min}}$ and are accurately described by $p(x)$ when $J(x) \equiv 0$, which we fit in \Cref{fig: panel phase angle}{B}.

\subsection*{Data, Materials, and Software Availability}
All numerical routines are open-sourced and tested using synthetic data at \href{https://github.com/aleable/power-grid-complexity}{https://github.com/aleable/power-grid-complexity}. Representative empirical datasets, including frequency measurements from London and phase-angle measurements from Stellenbosch and Bloemfontein of August 2025, are also publicly available at \href{https://zenodo.org/records/19397526}{https://zenodo.org/records/19397526}

\subsection*{Acknowledgements}{The authors acknowledge funding by the UKRI-STFC grant UKRI467 {\emph{``Stability of the South African power grid---a statistical physics-based approach''}}. L.R.G acknowledges the support of the Norwegian FME SOLAR grant, RCN 350244. A.L. acknowledges discussions with Günther Neuwirth, providing technical information on the PMUs. The authors thank Arash Rezaeinazhad and Deniz Ero\u{g}lu for processing and providing the data for \Cref{fig: grid networks}.}

\bibliography{bib}

@ARTICLE{almas2018vulnerability,
  author={Almas, M. S. and Vanfretti, L. and Singh, Ravi S. and Jonsdottir, Gudrun M.},
  journal={IEEE Transactions on Smart Grid}, 
  title={Vulnerability of Synchrophasor-Based WAMPAC Applications’ to Time Synchronization Spoofing}, 
  year={2018},
  volume={9},
  number={5},
  pages={4601-4612},
  doi={10.1109/TSG.2017.2665461}}

@Article{anvari2022data,
author={Anvari, Mehrnaz
and Proedrou, Elisavet
and Sch{\"a}fer, Benjamin
and Beck, Christian
and Kantz, Holger
and Timme, Marc},
title={Data-driven load profiles and the dynamics of residential electricity consumption},
journal={Nature Communications},
year={2022},
month={Aug},
day={06},
volume={13},
number={1},
pages={4593},
issn={2041-1723},
doi={10.1038/s41467-022-31942-9},
url={https://doi.org/10.1038/s41467-022-31942-9}
}

@article{anvari2016short,
doi = {10.1088/1367-2630/18/6/063027},
url = {https://dx.doi.org/10.1088/1367-2630/18/6/063027},
year = {2016},
month = {jun},
publisher = {IOP Publishing},
volume = {18},
number = {6},
pages = {063027},
author = {Anvari, M and Lohmann, G and Wächter, M and Milan, P and Lorenz, E and Heinemann, D and Tabar, M Reza Rahimi and Peinke, Joachim},
title = {Short term fluctuations of wind and solar power systems},
journal = {New Journal of Physics}
}

@article{anvari2020stochastic,
  title = {Stochastic properties of the frequency dynamics in real and synthetic power grids},
  author = {Anvari, Mehrnaz and Rydin Gorj\~ao, Leonardo and Timme, Marc and Witthaut, Dirk and Sch\"afer, Benjamin and Kantz, Holger},
  journal = {Phys. Rev. Res.},
  volume = {2},
  issue = {1},
  pages = {013339},
  numpages = {11},
  year = {2020},
  month = {Mar},
  publisher = {American Physical Society},
  doi = {10.1103/PhysRevResearch.2.013339},
  url = {https://link.aps.org/doi/10.1103/PhysRevResearch.2.013339}
}

@article{beck2001dynamical,
  title = {{Dynamical Foundations of Nonextensive Statistical Mechanics}},
  author = {Beck, Christian},
  journal = {Phys. Rev. Lett.},
  volume = {87},
  issue = {18},
  pages = {180601},
  numpages = {4},
  year = {2001},
  month = {Oct},
  publisher = {American Physical Society},
  doi = {10.1103/PhysRevLett.87.180601},
  url = {https://link.aps.org/doi/10.1103/PhysRevLett.87.180601}
}

@article{beck2005from,
  title = {From time series to superstatistics},
  author = {Beck, Christian and Cohen, Ezechiel G. D. and Swinney, Harry L.},
  journal = {Phys. Rev. E},
  volume = {72},
  issue = {5},
  pages = {056133},
  numpages = {8},
  year = {2005},
  month = {Nov},
  publisher = {American Physical Society},
  doi = {10.1103/PhysRevE.72.056133},
  url = {https://link.aps.org/doi/10.1103/PhysRevE.72.056133}
}

@article{beck2003superstatistics,
title = {Superstatistics},
journal = {Physica A: Statistical Mechanics and its Applications},
volume = {322},
pages = {267-275},
year = {2003},
issn = {0378-4371},
doi = {https://doi.org/10.1016/S0378-4371(03)00019-0},
url = {https://www.sciencedirect.com/science/article/pii/S0378437103000190},
author = {C. Beck and E. G. D. Cohen}
}

@book{butler2007saddlepoint,
  title={Saddlepoint approximations with applications},
  author={Butler, Ronald W},
  volume={22},
  year={2007},
  publisher={Cambridge University Press}
}

@misc{boettcher2026impact,
      title={{Impact of Market Reforms on Deterministic Frequency Deviations in the European Power Grid}}, 
      author={Philipp C. Böttcher and Carsten Hartmann and Andrea Benigni and Thiemo Pesch and Dirk Witthaut},
      year={2026},
      eprint={2602.09645},
      archivePrefix={arXiv},
      primaryClass={physics.soc-ph},
      url={https://arxiv.org/abs/2602.09645}, 
}

@book{conradie2000symphony,
  title     = {{A Symphony of Power: The Eskom Story}},
  author    = "Conradie, S R and Messerschmidt, L J M",
  publisher = "C. van Rensburg Publications",
  year      =  2000
}

@note{cop29pledge,
    title = {{COP29 Global Energy Storage and Grids Pledge}},
    author = {{United Nations Climate Change Conference}},
    url = {https://cop29.az/en/pages/cop29-global-energy-storage-and-grids-pledge},
    year = {2024},

}

@note{csir2025utility,
    title = {{Utility-scale power generation statistics in South Africa}},
    author = {{CSIR Energy Research Centre}},
    url = {https://www.csir.co.za/sites/default/files/2025-09/Utility%20Statistics%20Report_Jan%202025_Final.pdf},
    year = {2025},

}

@article{delgiudice2021effects,
title = {Effects of inertia, load damping and dead-bands on frequency histograms and frequency control of power systems},
journal = {International Journal of Electrical Power \& Energy Systems},
volume = {129},
pages = {106842},
year = {2021},
issn = {0142-0615},
doi = {https://doi.org/10.1016/j.ijepes.2021.106842},
url = {https://www.sciencedirect.com/science/article/pii/S014206152100082X},
author = {Davide {del Giudice} and Angelo Brambilla and Samuele Grillo and Federico Bizzarri}
}

@INPROCEEDINGS{deng2019frequency,
  author={Deng, Xianda and Li, Hongyu and Yu, Wenpeng and Weikang, Wang and Liu, Yilu},
  booktitle={2019 IEEE Power \& Energy Society General Meeting (PESGM)}, 
  title={{Frequency Observations and Statistic Analysis of Worldwide Main Power Grids Using FNET/GridEye}}, 
  year={2019},
  volume={},
  number={},
  pages={1-5},
  doi={10.1109/PESGM40551.2019.8973560}}

@article{drewnick2025analyzing,
    author = {Drewnick, Tim and Wen, Xinyi and Oberhofer, Ulrich and Rydin Gorjão, Leonardo and Beck, Christian and Hagenmeyer, Veit and Schäfer, Benjamin},
    title = {Analyzing deterministic and stochastic influences on the power grid frequency dynamics with explainable artificial intelligence},
    journal = {Chaos: An Interdisciplinary Journal of Nonlinear Science},
    volume = {35},
    number = {3},
    pages = {033153},
    year = {2025},
    month = {03},
    issn = {1054-1500},
    doi = {10.1063/5.0239371},
    url = {https://doi.org/10.1063/5.0239371},
}

@misc{epexnotes,
  title = {{Basics of the Power Market}},
  url = {https://www.epexspot.com/en/basicspowermarket},
  author = {{EPEX SPOT}},
  year = {2025},

}

@misc{epexukmarket,
  title = {{Annual Trading Results of 2024 - Power Trading on EPEX SPOT reaches all-time high}},
  url = {https://www.epexspot.com/sites/default/files/download_center_files/2025-01-28_EPEX%20SPOT_Annual%20Power%20Trading%20Results%202024_finaldraft_0.pdf},
  author = {{EPEX SPOT}},
  year = {2024},
}

@article{gorjao2019kramersmoyal,
doi = {10.21105/joss.01693}, url = {https://doi.org/10.21105/joss.01693}, year = {2019}, publisher = {The Open Journal}, volume = {4}, number = {44}, pages = {1693}, author = {Rydin Gorjão, Leonardo and Meirinhos, Francisco}, title = {{kramersmoyal: Kramers--Moyal coefficients for stochastic processes}}, 
journal = {Journal of Open Source Software}}

@ARTICLE{gorjao2020data,
  author={Rydin Gorjão, Leonardo and Anvari, Mehrnaz and Kantz, Holger and Beck, Christian and Witthaut, Dirk and Timme, Marc and Schäfer, Benjamin},
  journal={IEEE Access}, 
  title={{Data-Driven Model of the Power-Grid Frequency Dynamics}}, 
  year={2020},
  volume={8},
  number={},
  pages={43082-43097},
  doi={10.1109/ACCESS.2020.2967834}}

@article{gorjao2023stochastic,
doi = {10.1088/2632-072X/acb629},
url = {https://dx.doi.org/10.1088/2632-072X/acb629},
year = {2023},
month = {feb},
publisher = {IOP Publishing},
volume = {4},
number = {1},
pages = {015007},
author = {Rydin Gorjão, Leonardo and Maritz, Jacques},
title = {The stochastic nature of power-grid frequency in South Africa},
journal = {Journal of Physics: Complexity}
}

@article{gorjao2021spatio,
doi = {10.1088/1367-2630/ac08b3},
url = {https://dx.doi.org/10.1088/1367-2630/ac08b3},
year = {2021},
month = {jul},
publisher = {IOP Publishing},
volume = {23},
number = {7},
pages = {073016},
author = {Rydin Gorjão, Leonardo and Schäfer, Benjamin and Witthaut, Dirk and Beck, Christian},
title = {Spatio-temporal complexity of power-grid frequency fluctuations},
journal = {New Journal of Physics}
}

@misc{greyling2019gross,
  author       = {Greyling, T. and Rossouw, S. and AFSTEREO},
  title        = {Gross National Happiness.today Index},
  year         = {2019},
  howpublished = {\url{http://gnh.today}},
}

@article{haehne2019propagation,
  title = {Propagation of wind-power-induced fluctuations in power grids},
  author = {Haehne, Hauke and Schmietendorf, Katrin and Tamrakar, Samyak and Peinke, Joachim and Kettemann, Stefan},
  journal = {Phys. Rev. E},
  volume = {99},
  issue = {5},
  pages = {050301},
  numpages = {5},
  year = {2019},
  month = {May},
  publisher = {American Physical Society},
  doi = {10.1103/PhysRevE.99.050301},
  url = {https://link.aps.org/doi/10.1103/PhysRevE.99.050301}
}

@article{hindes2019network,
  title = {Network desynchronization by non-Gaussian fluctuations},
  author = {Hindes, Jason and Jacquod, Philippe and Schwartz, Ira B.},
  journal = {Phys. Rev. E},
  volume = {100},
  issue = {5},
  pages = {052314},
  numpages = {11},
  year = {2019},
  month = {Nov},
  publisher = {American Physical Society},
  doi = {10.1103/PhysRevE.100.052314},
  url = {https://link.aps.org/doi/10.1103/PhysRevE.100.052314}
}

@article{han2022complexity,
  title = {{Complexity and Persistence of Price Time Series of the European Electricity Spot Market}},
  author = {Han, Chengyuan and Hilger, Hannes and Mix, Eva and B{\"o}ttcher, Philipp C. and Reyers, Mark and Beck, Christian and Witthaut, Dirk and Gorj{\~a}o, Leonardo Rydin},
  journal = {PRX Energy},
  volume = {1},
  issue = {1},
  pages = {013002},
  numpages = {17},
  year = {2022},
  month = {Apr},
  publisher = {American Physical Society},
  doi = {10.1103/PRXEnergy.1.013002},
  url = {https://link.aps.org/doi/10.1103/PRXEnergy.1.013002}
}

@article{janse2023reflections,
  title={{Reflections on load-shedding and potential GDP}},
  author={Janse van Rensburg, T and Morema, K},
  journal={{South African Reserve Bank Occasional Bulletin of Economic Notes OBEN /23/01}},
  volume={2301},
  year={2023}
}

@INPROCEEDINGS{kashima2015modeling,
  author={Kashima, Kenji and Aoyama, Hiroki and Ohta, Yoshito},
  booktitle={2015 54th IEEE Conference on Decision and Control (CDC)}, 
  title={Modeling and linearization of systems under heavy-tailed stochastic noise with application to renewable energy assessment}, 
  year={2015},
  volume={},
  number={},
  pages={1852-1857},
  doi={10.1109/CDC.2015.7402480}}

@article{kundur2007power,
  title={Power system stability},
  author={Kundur, Prabha},
  journal={Power system stability and control},
  volume={10},
  number={1},
  pages={7--1},
  year={2007}
}

@ARTICLE{kraljic2023towards,
  author={Kraljic, David},
  journal={IEEE Transactions on Power Systems}, 
  title={{Towards Realistic Statistical Models of the Grid Frequency}}, 
  year={2023},
  volume={38},
  number={1},
  pages={256-266},
  doi={10.1109/TPWRS.2022.3163336}
}

@article{kruse2023physics,
  title = {{Physics-Informed Machine Learning for Power Grid Frequency Modeling}},
  author = {Kruse, Johannes and Cramer, Eike and Sch\"afer, Benjamin and Witthaut, Dirk},
  journal = {PRX Energy},
  volume = {2},
  issue = {4},
  pages = {043003},
  numpages = {20},
  year = {2023},
  month = {Oct},
  publisher = {American Physical Society},
  doi = {10.1103/PRXEnergy.2.043003},
  url = {https://link.aps.org/doi/10.1103/PRXEnergy.2.043003}
}

@article{lamoroux2009kernel,
title = {Kernel-based regression of drift and diffusion coefficients of stochastic processes},
journal = {Physics Letters A},
volume = {373},
number = {39},
pages = {3507-3512},
year = {2009},
issn = {0375-9601},
doi = {https://doi.org/10.1016/j.physleta.2009.07.073},
url = {https://www.sciencedirect.com/science/article/pii/S037596010900944X},
author = {David Lamouroux and Klaus Lehnertz}
}

@Article{manik2014supply,
author={Manik, Debsankha
and Witthaut, Dirk
and Sch{\"a}fer, Benjamin
and Matthiae, Moritz
and Sorge, Andreas
and Rohden, Martin
and Katifori, Eleni
and Timme, Marc},
title={Supply networks: Instabilities without overload},
journal={The European Physical Journal Special Topics},
year={2014},
month={Oct},
day={01},
volume={223},
number={12},
pages={2527-2547},
issn={1951-6401},
doi={10.1140/epjst/e2014-02274-y},
url={https://doi.org/10.1140/epjst/e2014-02274-y}
}

@book{machowski2020power,
  title={Power system dynamics: stability and control},
  author={Machowski, Jan and Lubosny, Zbigniew and Bialek, Janusz W and Bumby, James R},
  year={2020},
  publisher={John Wiley \& Sons}
}

@article{manik2017cycle,
    author = {Manik, Debsankha and Timme, Marc and Witthaut, Dirk},
    title = {Cycle flows and multistability in oscillatory networks},
    journal = {Chaos: An Interdisciplinary Journal of Nonlinear Science},
    volume = {27},
    number = {8},
    pages = {083123},
    year = {2017},
    month = {08},
    issn = {1054-1500},
    doi = {10.1063/1.4994177},
    url = {https://doi.org/10.1063/1.4994177},
}

@article{manoki2023power,
title = {{Power cuts and South Africa's health care}},
journal = {The Lancet},
volume = {401},
number = {10386},
pages = {1413},
year = {2023},
issn = {0140-6736},
doi = {https://doi.org/10.1016/S0140-6736(23)00850-4},
url = {https://www.sciencedirect.com/science/article/pii/S0140673623008504},
author = {Munyaradzi Makoni}
}

@article{mandelbrot1968fractional,
  title={Fractional Brownian motions, fractional noises and applications},
  author={Mandelbrot, Benoit B and Van Ness, John W},
  journal={SIAM review},
  volume={10},
  number={4},
  pages={422--437},
  year={1968},
  publisher={SIAM}
}

@article{maritz2024data,
  title        = {{Data-Driven Modeling of Frequency Dynamics Observed in Operating Microgrids: A South African University Campus Case Study}},
  author       = {Maritz, Jacques and Rydin Gorj{\~a}o, Leonardo and Bester, P Armand and Esterhuysen, Nicolaas and Erasmus, Stefaans and Riekert, Stephanus and Immelman, Reuben and Geldenhuys, Tiaan and Viljoen, Alexandra and Bodenstein, Charl},
  journal      = {IEEE Access},
  volume       = {12},
  pages        = {14466--14473},
  year         = {2024},
  publisher    = {IEEE},
  doi          = {10.1109/ACCESS.2024.3357945}
}

@article{mayer2018electricity,
title = {Electricity markets around the world},
journal = {Journal of Commodity Markets},
volume = {9},
pages = {77-100},
year = {2018},
issn = {2405-8513},
doi = {https://doi.org/10.1016/j.jcomm.2018.02.001},
url = {https://www.sciencedirect.com/science/article/pii/S2405851318300059},
author = {Klaus Mayer and Stefan Trück},
}

@article{milan2013turbulent,
  title = {{Turbulent Character of Wind Energy}},
  author = {Milan, Patrick and W\"achter, Matthias and Peinke, Joachim},
  journal = {Phys. Rev. Lett.},
  volume = {110},
  issue = {13},
  pages = {138701},
  numpages = {5},
  year = {2013},
  month = {Mar},
  publisher = {American Physical Society},
  doi = {10.1103/PhysRevLett.110.138701},
  url = {https://link.aps.org/doi/10.1103/PhysRevLett.110.138701}
}

@INPROCEEDINGS{mele2016impact,
  author = {Mele, Francesca Madia and Ortega, {\'A}lvaro and Z{\'a}rate-Mi{\~n}ano, Rafael and Milano, Federico},
  booktitle = {2016 Power Systems Computation Conference (PSCC)},
  title = {Impact of variability, uncertainty and frequency regulation on power system frequency distribution},
  year = {2016},
  pages = {1-8},
  doi = {10.1109/PSCC.2016.7540970}
}

@Article{motter2013spontaneous,
author={Motter, Adilson E.
and Myers, Seth A.
and Anghel, Marian
and Nishikawa, Takashi},
title={Spontaneous synchrony in power-grid networks},
journal={Nature Physics},
year={2013},
month={Mar},
day={01},
volume={9},
number={3},
pages={191-197},
issn={1745-2481},
doi={10.1038/nphys2535},
url={https://doi.org/10.1038/nphys2535}
}

@misc{nesonote,
  title = {{Dynamic Services (DC/DM/DR)}},
  url = {https://www.neso.energy/industry-information/balancing-services/frequency-response-services/dynamic-services-dcdmdr#Document-library},
  author = {{NESO: National Energy System Operator}},
  year = {2025},

}

@misc{nersacode,
  title = {{The South African Grid Code: 2022}},
  url = {https://www.nersa.org.za/files/files/2022/02/SAGC-Network-Version-10.1_January-2022.pdf},
  author = {{NERSA: National Energy Regulator of South Africa}},
  year = {2022},

}

@INPROCEEDINGS{oberhofer2023non,
  author={Oberhofer, Ulrich and Gorjao, Leonardo Rydin and Yalcin, G. Cigdem and Kamps, Oliver and Hagenmeyer, Veit and Schafer, Benjamin},
  booktitle={2023 IEEE Belgrade PowerTech}, 
  title={Non-linear, bivariate stochastic modelling of power-grid frequency applied to islands}, 
  year={2023},
  volume={},
  number={},
  pages={1-1},
  doi={10.1109/PowerTech55446.2023.10202986}
}

@misc{openacess2025grid,
  author       = {Rydin Gorj{\~a}o, Leonardo and Jumar, Richard and Maass, Heiko and Hagenmeyer, Veit and Yalcin, G. Cigdem and Kruse, Johannes and Timme, Marc and Beck, Christian and Witthaut, Dirk and Sch{\"a}fer, Benjamin},
  title        = {Open Access Power-Grid Frequency Database},
  year         = {2025},
  url = {https://doi.org/10.17616/R31NJMT0},
  publisher    = {re3data.org - Registry of Research Data Repositories},
  doi          = {10.17616/R31NJMT0}
}

@misc{ofgemnote,
  title = {{Decision on Dynamic Regulation, Dynamic Moderation, and Dynamic
Containment in relation to an update to the Terms and Conditions related to
Balancing}},
  url = {https://www.ofgem.gov.uk/sites/default/files/2023-02/Decision%20on%20Dynamic%20Regulation%2C%20Dynamic%20Moderation%20and%20Dynamic%20Containment%20in%20relation%20to%20an%20update%20to%20the%20Terms%20and%20Conditions%20related%20to%20Balancing.pdf},
  author = {{Ofgem}},
  year = {2023},

}

@article{rebours2007survey,
  author={Rebours, Yann G. and Kirschen, Daniel S. and Trotignon, Marc and Rossignol, Sbastien},
  journal={IEEE Transactions on Power Systems}, 
  title={{A Survey of Frequency and Voltage Control Ancillary Services—Part I: Technical Features}}, 
  year={2007},
  volume={22},
  number={1},
  pages={350-357},
  doi={10.1109/TPWRS.2006.888963}
}

@book{redner2001guide,
  title={A guide to first-passage processes},
  author={Redner, Sidney},
  year={2001},
  publisher={Cambridge University Press}
}

@incollection{risken1989fokker,
  title={Fokker-planck equation},
  author={Risken, Hannes},
  booktitle={The Fokker-Planck equation: methods of solution and applications},
  pages={63--95},
  year={1989},
  publisher={Springer}
}

@article{rohden2012self,
  title = {Self-Organized Synchronization in Decentralized Power Grids},
  author = {Rohden, Martin and Sorge, Andreas and Timme, Marc and Witthaut, Dirk},
  journal = {Phys. Rev. Lett.},
  volume = {109},
  issue = {6},
  pages = {064101},
  numpages = {5},
  year = {2012},
  month = {Aug},
  publisher = {American Physical Society},
  doi = {10.1103/PhysRevLett.109.064101},
  url = {https://link.aps.org/doi/10.1103/PhysRevLett.109.064101}
}

@Article{gorjao2020open,
author={Rydin Gorj{\~a}o, Leonardo and Jumar, Richard and Maass, Heiko and Hagenmeyer, Veit and Yalcin, G. Cigdem and Kruse, Johannes and Timme, Marc and Beck, Christian and Witthaut, Dirk and Sch{\"a}fer, Benjamin}, title={{Open database analysis of scaling and spatio-temporal properties of power grid frequencies}}, journal={Nature Communications}, year={2020}, month={Dec}, day={11}, volume={11}, number={1}, pages={6362}, issn={2041-1723}, doi={10.1038/s41467-020-19732-7}, url={https://doi.org/10.1038/s41467-020-19732-7} }

@book{samoradnitsky2017stable,
  title={{Stable non-Gaussian random processes: stochastic models with infinite variance}},
  author={Samoradnitsky, Gennady},
  year={2017},
  publisher={Routledge}
}

@misc{sappmarket,
  title = {{SAPP Statistics 2020}},
  url = {https://www.sapp.co.zw/sites/default/files/Statistics%202019-20.pdf},
  author = {{SAPP: Southern African Power Pool}},
  year = {2020},
}

@Article{schafer2018dynamically,
author={Sch{\"a}fer, Benjamin
and Witthaut, Dirk
and Timme, Marc
and Latora, Vito},
title={Dynamically induced cascading failures in power grids},
journal={Nature Communications},
year={2018},
month={May},
day={17},
volume={9},
number={1},
pages={1975},
issn={2041-1723},
doi={10.1038/s41467-018-04287-5},
url={https://doi.org/10.1038/s41467-018-04287-5}
}

@Article{schaefer2018non,
author={Sch{\"a}fer, Benjamin
and Beck, Christian
and Aihara, Kazuyuki
and Witthaut, Dirk
and Timme, Marc},
title={{Non-Gaussian power grid frequency fluctuations characterized by L{\'e}vy-stable laws and superstatistics}},
journal={Nature Energy},
year={2018},
month={Feb},
day={01},
volume={3},
number={2},
pages={119-126},
issn={2058-7546},
doi={10.1038/s41560-017-0058-z},
url={https://doi.org/10.1038/s41560-017-0058-z}
}

@article{schafer2017escape,
  title = {Escape routes, weak links, and desynchronization in fluctuation-driven networks},
  author = {Sch{\"a}fer, Benjamin and Matthiae, Moritz and Zhang, Xiaozhu and Rohden, Martin and Timme, Marc and Witthaut, Dirk},
  journal = {Phys. Rev. E},
  volume = {95},
  issue = {6},
  pages = {060203},
  numpages = {5},
  year = {2017},
  month = {Jun},
  publisher = {American Physical Society},
  doi = {10.1103/PhysRevE.95.060203},
  url = {https://link.aps.org/doi/10.1103/PhysRevE.95.060203}
}

@INPROCEEDINGS{schafer2018isolating,
  author={Sch{\"a}fer, Benjamin and Timme, Marc and Witthaut, Dirk},
  booktitle={2018 IEEE PES Innovative Smart Grid Technologies Conference Europe (ISGT-Europe)}, 
  title={{Isolating the Impact of Trading on Grid Frequency Fluctuations}}, 
  year={2018},
  volume={},
  number={},
  pages={1-5},
  doi={10.1109/ISGTEurope.2018.8571793}}

@Article{schmietendorf2017impact,
author={Schmietendorf, Katrin
and Peinke, Joachim
and Kamps, Oliver},
title={The impact of turbulent renewable energy production on power grid stability and quality},
journal={The European Physical Journal B},
year={2017},
month={Nov},
day={15},
volume={90},
number={11},
pages={222},
issn={1434-6036},
doi={10.1140/epjb/e2017-80352-8},
url={https://doi.org/10.1140/epjb/e2017-80352-8}
}

@article{touchette2005asymptotics,
  title = {Asymptotics of superstatistics},
  author = {Touchette, Hugo and Beck, Christian},
  journal = {Phys. Rev. E},
  volume = {71},
  issue = {1},
  pages = {016131},
  numpages = {6},
  year = {2005},
  month = {Jan},
  publisher = {American Physical Society},
  doi = {10.1103/PhysRevE.71.016131},
  url = {https://link.aps.org/doi/10.1103/PhysRevE.71.016131}
}

@article{tsallis1988possible,
author={Tsallis, Constantino},
title={{Possible generalization of Boltzmann-Gibbs statistics}},
journal={Journal of Statistical Physics},
year={1988},
month={Jul},
day={01},
volume={52},
number={1},
pages={479-487},
issn={1572-9613},
doi={10.1007/BF01016429},
url={https://doi.org/10.1007/BF01016429}
}

@article{tumash2018effect,
doi = {10.1209/0295-5075/123/20001},
url = {https://doi.org/10.1209/0295-5075/123/20001},
year = {2018},
month = {aug},
publisher = {EDP Sciences, IOP Publishing and Società Italiana di Fisica},
volume = {123},
number = {2},
pages = {20001},
author = {Tumash, L. and Olmi, S. and Schöll, E.},
title = {Effect of disorder and noise in shaping the dynamics of power grids},
journal = {Europhysics Letters}
}

@article{ulbig2014impact,
  title={Impact of low rotational inertia on power system stability and operation},
  author={Ulbig, Andreas and Borsche, Theodor S and Andersson, G{\"o}ran},
  journal={IFAC Proceedings Volumes},
  volume={47},
  number={3},
  pages={7290--7297},
  year={2014},
  publisher={Elsevier}
}

@ARTICLE{vorobev2019deadbands,
  author={Vorobev, Petr and Greenwood, David M. and Bell, John H. and Bialek, Janusz W. and Taylor, Philip C. and Turitsyn, Konstantin},
  journal={IEEE Transactions on Power Systems}, 
  title={{Deadbands, Droop, and Inertia Impact on Power System Frequency Distribution}}, 
  year={2019},
  volume={34},
  number={4},
  pages={3098-3108},
  doi={10.1109/TPWRS.2019.2895547}}

@INPROCEEDINGS{weissbach2009high,
  author={Weissbach, T. and Welfonder, E.},
  booktitle={2009 IEEE/PES Power Systems Conference and Exposition}, 
  title={{High frequency deviations within the European Power System: Origins and proposals for improvement}}, 
  year={2009},
  volume={},
  number={},
  pages={1-6},
  doi={10.1109/PSCE.2009.4840180}}

@book{wood2013power,
  title={Power generation, operation, and control},
  author={Wood, Allen J and Wollenberg, Bruce F and Shebl{\'e}, Gerald B},
  year={2013},
  publisher={John wiley \& sons}
}

@misc{wef2025scaling,
    title = {{Why scaling clean energy in the Global South is a three-legged balancing act}},
    author = {{World Economic Forum}},
    URL = {https://www.weforum.org/stories/2025/01/clean-energy-renewables-global-south/},
    year = {2025}
}

@Article{wen2023non,
author={Wen, Xinyi
and Anvari, Mehrnaz
and Rydin Gorj{\~a}o, Leonardo
and Yalcin, G. Cigdem
and Hagenmeyer, Veit
and Sch{\"a}fer, Benjamin},
title={Nonstandard power grid frequency statistics across continents},
journal={Scientific Reports},
year={2025},
month={Nov},
day={04},
volume={15},
number={1},
pages={38470},
issn={2045-2322},
doi={10.1038/s41598-025-25334-4},
url={https://doi.org/10.1038/s41598-025-25334-4}
}

@article{witthaut2022collective,
  title = {Collective nonlinear dynamics and self-organization in decentralized power grids},
  author = {Witthaut, Dirk and Hellmann, Frank and Kurths, J\"urgen and Kettemann, Stefan and Meyer-Ortmanns, Hildegard and Timme, Marc},
  journal = {Rev. Mod. Phys.},
  volume = {94},
  issue = {1},
  pages = {015005},
  numpages = {52},
  year = {2022},
  month = {Feb},
  publisher = {American Physical Society},
  doi = {10.1103/RevModPhys.94.015005},
  url = {https://link.aps.org/doi/10.1103/RevModPhys.94.015005}
}

@misc{iea2024world,
    title = {{World Energy Outlook 2024}},
    author = {{International Energy Agency}},
    URL = {https://iea.blob.core.windows.net/assets/c036b390-ba9c-4132-870b-ffb455148b63/WorldEnergyOutlook2024.pdf},
    year         = {2024}
}

@misc{iea2024southafrica,
    title = {{South Africa Energy Mix}},
    author = {{International Energy Agency}},
    URL = {https://www.iea.org/countries/south-africa/energy-mix},
    year         = {2024}
}

@misc{iea2024unitedkingdom,
    title = {{United Kindom Energy Mix}},
    author = {{International Energy Agency}},
    URL = {https://www.iea.org/countries/united-kingdom},
    year         = {2024}
}

% !TEX root = main.tex

%%%%%%%%%% Merge with supplemental materials %%%%%%%%%%
\onecolumngrid
\mbox{}
\clearpage
\newpage

%%%%%%%%%% Merge with supplemental materials %%%%%%%%%%
%%%%%%%%%% Prefix a "S" to all equations, figures, tables and reset the counter %%%%%%%%%%
\setcounter{equation}{0}
\setcounter{figure}{0}
\setcounter{section}{0}
\setcounter{table}{0}
\makeatletter
\renewcommand{\theequation}{S\arabic{equation}}
\renewcommand{\thefigure}{S\arabic{figure}}
\renewcommand{\thetable}{S\arabic{table}}
\renewcommand{\thesection}{S\arabic{section}}

\begin{center}
    {\large{\textbf{Supporting Information:\\Understanding the complexity of frequency and phase angle fluctuations in power grids}}}
\end{center}

\section{Data}

\subsection{Data collection}

Phasor Measurement Units (PMUs) are installed in Stellenbosch and Bloemfontein, South Africa, and in London and Glasgow, the United Kingdom. Each PMU measures electrical signals from the power grid and extracts frequencies $f_i(t)$ (equivalently, angular velocities $\omega_i(t) = 2 \pi (f_i(t) - f_{\mathrm{R}})$, with $f_{\mathrm{R}} = \qty{50}{Hz}$) and phase angles $\theta_i(t)$. We denote with the subscript $i$ the PMU installation sites, i.e., $i =$ Stellenbosch (SB), Bloemfontein (BF), London (LD), and Glasgow (GW). Since, during regular operations, $\omega_i(t) = \omega(t)$ for all $i$ in a grid, namely, only negligible fluctuations are observed in the frequency signal, we use frequency measurements from London and Stellenbosch to represent their respective countries. Phase angle differences are computed as $x(t) = \theta_1(t) - \theta_2(t)$, with $\theta_1(t) = \theta_{\text{LD}}(t)$ and $\theta_2(t) = \theta_{\text{GW}}(t)$ in the United Kingdom, and $\theta_1(t) = \theta_{\text{SB}}(t)$ and $\theta_2(t) = \theta_{\text{BF}}(t)$ in South Africa.

In practice, each PMU gives us discrete time frequency and phase-angle time series with a uniform resolution of $\Delta t_{\mathrm{PMU}} = \qty{0.1}{s}$. To retrieve these time series, it employs a Schmitt trigger that converts the grid voltage signal into a square wave, thereby mitigating noise. A frequency counter counts the number of oscillations $N_0$ of a high-precision reference frequency $f_0$ occurring between two successive square-wave edges at times $t$. The reference frequency $f_0$ is synchronized with an atomic clock via GPS. The frequencies measurements are then calculated as $f_i(t) = f_0 / N_0$, while the phase angles $\theta_i(t)$ are computed from how much earlier or later the latest voltage zero crossing occurs compared to the ideal grid period.

We query the PMUs via an API provided by Gridradar (\href{https://gridradar.net/en}{https://gridradar.net}) at 1-h intervals. For each hourly interval, we retrieve $36$ thousand points since $\Delta t_{\mathrm{PMU}} = \qty{0.1}{s}$. Our data range from the 28th of April 2025 to the 28th of October 2025, yielding a total of $\num{948 672 000} \approx \num{950}$ million data points to work with.

Extracting fine-grained, long-term measurements gives us an advantage over the literature, which typically reports frequencies measured every $\qty{1}{s}$ or more  \cite{gorjao2020open,schaefer2018non,schafer2018isolating,anvari2020stochastic,gorjao2020data,delgiudice2021effects,wen2023non,kraljic2023towards,kruse2023physics,drewnick2025analyzing,openacess2025grid}. A coarser sampling interval might produce unimodal frequency distributions (for instance, see Schäfer et al. \cite{schaefer2018non}) that fail to capture the features needed to fit and support our superstatistical model with nonlinear control.

\subsection{Data processing}

Phase angles yield ``hard-to-spot'' outliers. That is, spurious data that we would not have been able to detect by examining the $\theta_1(t)$, $\theta_2(t)$ in isolation, but only by analyzing $\theta_1(t)$, $\theta_2(t)$ against each other and their difference $x(t)$.

A first kind of ``hard-to-spot'' outlier is shown in \Cref{apx-fig: phase angle manual trimming}, where a small percentage of the total measurements lies outside the phase-locked bands. We filter out these spurious data by manually trimming all values outside $\Delta_1 \leq x \leq \Delta_2$. For the upper bands:
\begin{alignat*}{2}
  \text{United Kingdom:} \quad \Delta_1 &= \qty{0.58}{rad} \qquad
  \text{South Africa:}
    \quad && \Delta_1 = \qty{0.85}{rad} \\
     \Delta_2 &= \qty{2.9}{rad} 
     && \Delta_2  = \qty{1.9}{rad} \,.
\end{alignat*}
For the lower bands:
\begin{alignat*}{2}
  \text{United Kingdom:} \quad \Delta_1 &= -\qty{5.7}{rad} \qquad
  \text{South Africa:}
    \quad && \Delta_1 = -\qty{5.45}{rad} \\
     \Delta_2 &= -\qty{3.4}{rad} 
     && \Delta_2  = -\qty{4.3}{rad} \,.
\end{alignat*}
The total percentage of data points removed by manual filtering is less than $0.2\%$. This proportion is sufficiently small that it is not expected to affect our results. As a sanity check, we fit the upper diagonal bands of \Cref{apx-fig: phase angle manual trimming} with linear least squares. Since the angles are phase-locked, we expect the fit slope to be close to $m = 1$. We find:
\begin{equation*}
  \text{United Kingdom:} \quad m \approx \num{0.969} \qquad
  \text{South Africa:}
    \quad m \approx \num{0.997} \,.
\end{equation*}

A second kind of ``hard-to-spot'' outlier is shown in \Cref{apx-fig: phase processing sa}. In \Cref{apx-fig: phase processing sa}A, we show a snapshot of the signal $x(t)$, which is zoomed over its first (approximately) minute and a half in \Cref{apx-fig: phase processing sa}B. The signal exhibits stepwise deviations from its profile. We attribute these deviations to PMU clock corrections \cite{almas2018vulnerability}. Such corrections are typically introduced when phase angles cross a specific value, causing the artifacts to appear as discrete steps in $x(t)$. It is difficult to filter out these corrections directly from $x(t)$, as the time series are inherently noisy, making it challenging to systematically distinguish PMU corrections from genuine grid fluctuations. Nevertheless, to exclude artifacts, we use a simple yet effective solution and take the lower envelope of the signal. The envelope is computed using a mask with a width of $\qty{10}{s}$ and a stride of $\qty{5}{s}$. As shown in \Cref{apx-fig: phase processing sa}{A, B}, the lower envelope effectively removes the artifacts while preserving the signal profile. In \Cref{apx-fig: phase processing sa}C, we show the phase-angle-difference distribution and observe that the lower envelope preserves the overall shape of the histogram.

For consistency, we perform the same analysis with data from the United Kingdom. Results are shown in \Cref{apx-fig: frequency validation uk}. Here, the signal is cleaner, with no step-like artifacts. The lower-envelope histogram deviates only slightly from the original data.

We use the phase angle's lower envelope to fit the tilted washboard potential in the main text.

\begin{figure*}[htpb]
\centering
\includegraphics[width=0.85\textwidth]{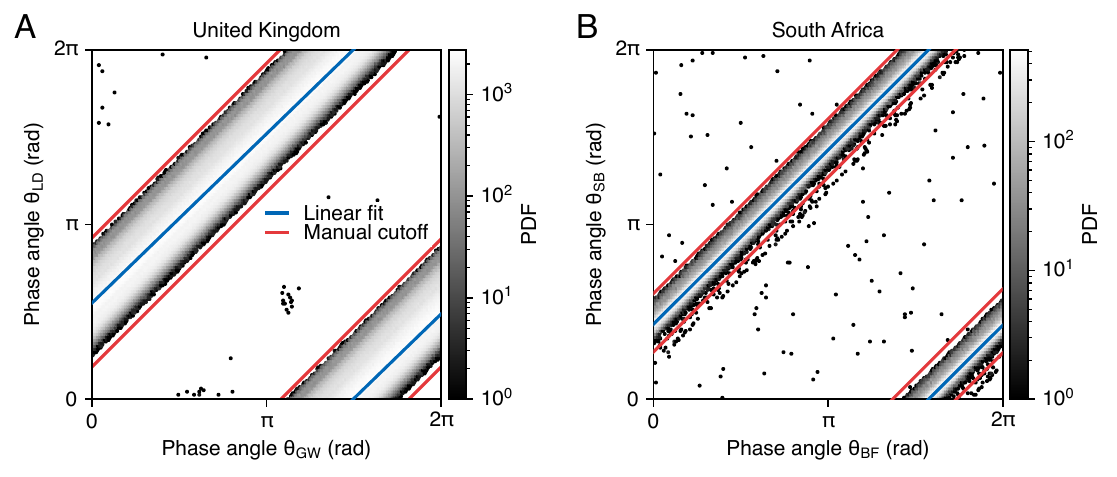}
\caption{Trimming the phase-locked bands. Phase angles are phase-locked, as evidenced by the hexbin plots that form diagonal bands. Outliers outside $\Delta_1 \leq x \leq \Delta_2$ are those points lying beyond the red diagonal lines. The upper bands are fitted. To visually verify the goodness of this fit, we plot the upper-band fit on the lower bands by opportunity shifting the line by $2\pi$. (A) Data for the United Kingdom. To improve the visualization, we subsample the phase time series, taking one point every 10. (B) Data for South Africa. To improve the visualization, we subsample the phase time series, taking one point every 200.}
\label{apx-fig: phase angle manual trimming}
\end{figure*}

\begin{figure*}[htpb]
\centering
\includegraphics[width=0.57\textwidth]{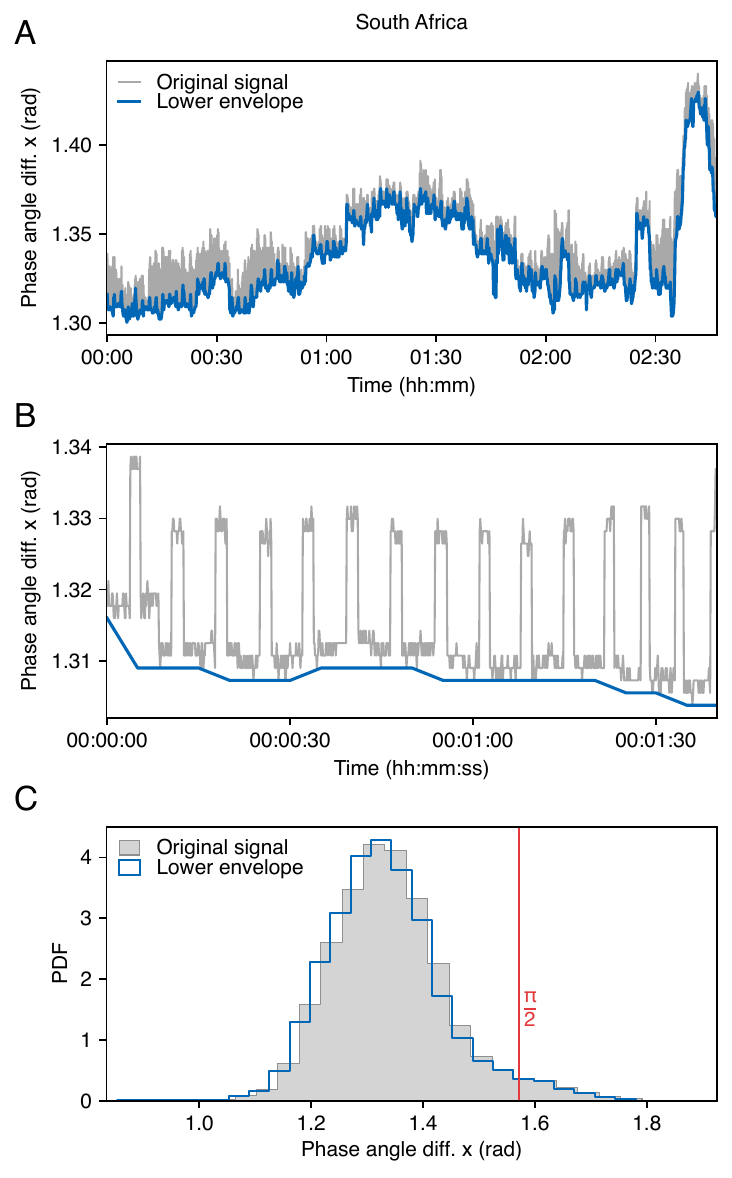}
\caption{Processing the phase-angle difference for South Africa. (A) A snapshot of the original signal $x(t)$ is shown together with its lower envelope. (B) Panel A is zoomed in to approximately the first $\qty{1.5}{min}$. (C) The distribution of the original signal $x(t)$ and its lower envelope is plotted. The lower envelope is used to fit the tilted washboard potential described in the main text. The distribution mode lies well below $\pi / 2$, which is the boundary of the stability region allowed by the tilted washboard potential (see \Cref{apx-fig: tilted washboard panel}B and \Cref{apx-sec: Analytical Derivations of Phase Angle Difference Distribution}). Therefore, MLE yields a good fit.}
\label{apx-fig: phase processing sa}
\end{figure*}

\begin{figure*}[htpb]
\centering
\includegraphics[width=0.57\textwidth]{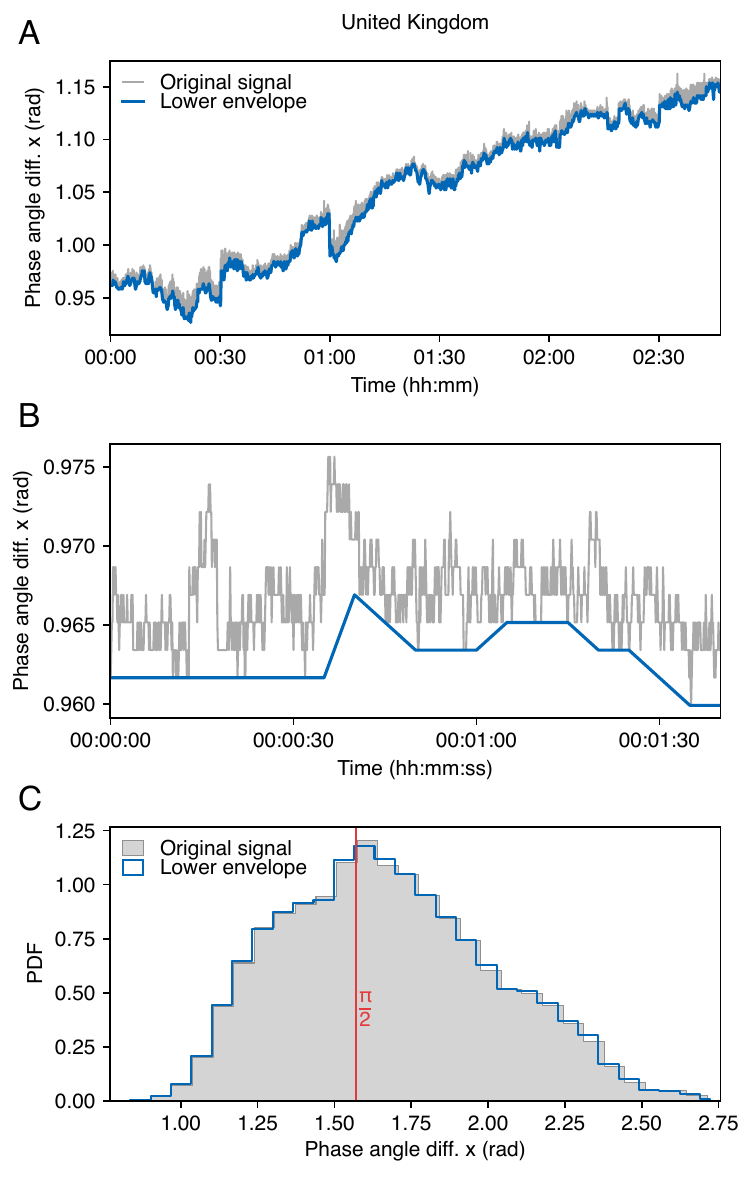}
\caption{Processing the phase-angle difference for the United Kingdom. See \Cref{apx-fig: phase processing sa} for a full caption. (C) Note that the distribution mode and a large portion of the distribution lie well above $\pi / 2$, which is the boundary of the stability region allowed by the tilted washboard potential (see \Cref{apx-fig: tilted washboard panel}B and \Cref{apx-sec: Analytical Derivations of Phase Angle Difference Distribution}). Consequently, the tilted washboard model is insufficient to describe or fit the data. In practice, MLE becomes unstable: as the distribution mode approaches $\pi / 2$, the tilted washboard potential $V(x)$ flattens (see \Cref{apx-fig: tilted washboard panel}A) and the data cannot be accurately captured by \Cref{apx-eq: stationary distribution phase difference}.}
\label{apx-fig: phase processing uk}

\end{figure*}

\subsection{Data Validation} We verify that the shape of the distributions in Figs. 2{B, E} (main text) is consistent over time by splitting the frequency data into six non-overlapping monthly intervals. We show the distributions for each interval in \Cref{apx-fig: frequency validation uk} for the United Kingdom and \Cref{apx-fig: frequency validation sa} for South Africa. The histogram profiles are largely consistent, with deviations occurring only in the relative height of the peaks formed at the boundaries between control regions.

\begin{figure*}[htpb]
\centering
\includegraphics[width=1.0\textwidth]{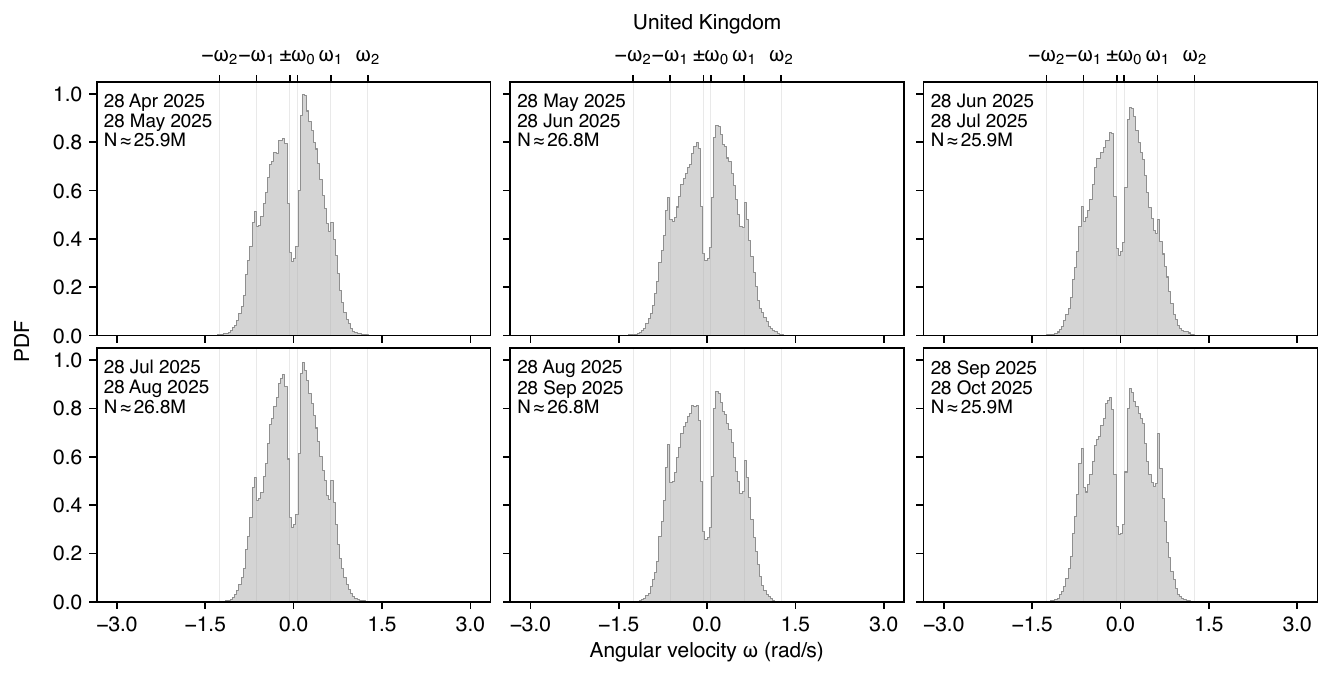}
\caption{Monthly frequency distributions for the United Kingdom. Each panel spans one month of measurements, and the windows are non-overlapping. Here, $N$ denotes the number of data points used for each histogram. The vertical lines denoting control boundaries are as per \Cref{apx-eq: deadband width uk}.}
\label{apx-fig: frequency validation uk}
\end{figure*}

\begin{figure*}[htpb]
\centering
\includegraphics[width=1.0\textwidth]{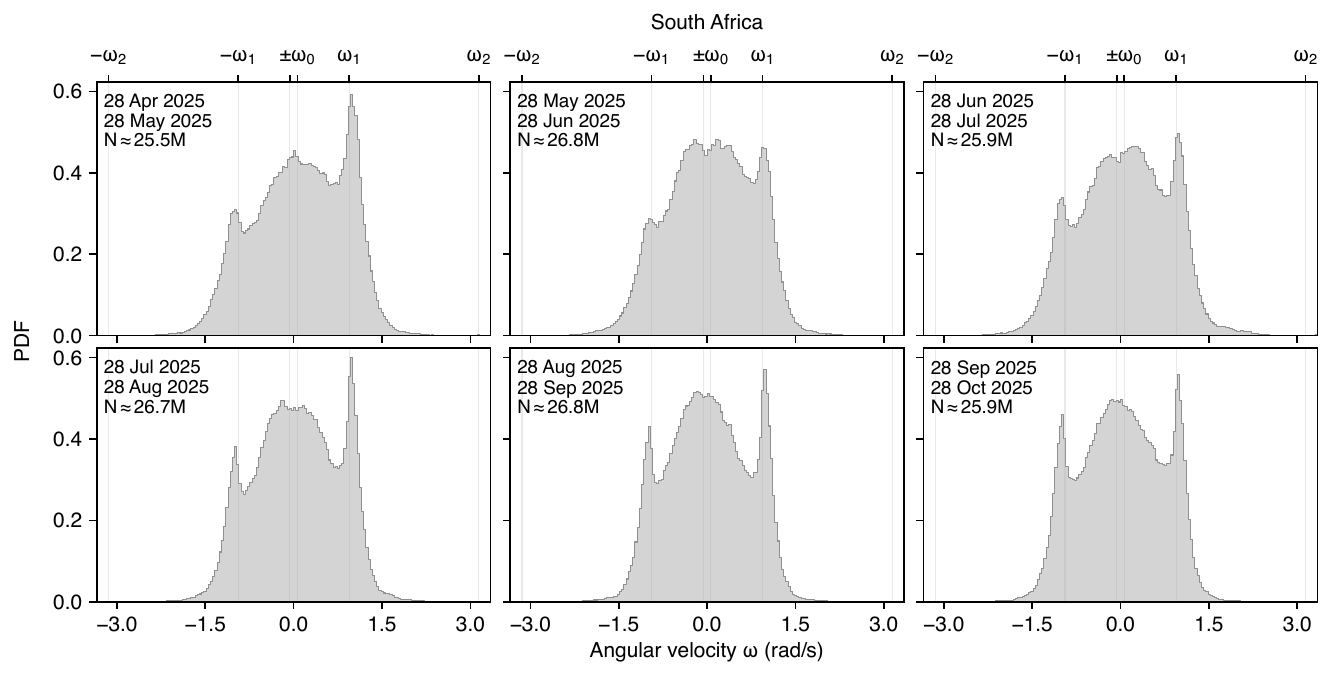}
\caption{Monthly frequency distributions for South Africa. Each panel spans one month of measurements, and the windows are non-overlapping. Here, $N$ denotes the number of data points used for each histogram. The vertical lines denoting control boundaries are as per \Cref{apx-eq: deadband width sa}.}
\label{apx-fig: frequency validation sa}
\end{figure*}

\section{Frequency Control}

Control mechanisms enable modern power grids to stabilize their operation following disturbances. Multiple control levels typically operate on distinct time scales, with different objectives and magnitudes \cite[for an extensive overview]{machowski2020power,kundur2007power}.

A power grid consists of synchronous generators: large rotating machines that produce electrical power. Following a disturbance, such as a generator disconnection, the remaining generators supply additional power to compensate for the perturbation. This leads to a temporary reduction in their angular velocity and, consequently, in their kinetic energy. Control mechanisms then restore regular grid operation as follows.
\begin{enumerate}
    \item Primary control starts acting within seconds. Contracted generators measure the frequency deviation from the nominal value and inject power proportional to the observed deviation, damping fast frequency fluctuations.
    \item Secondary control acts on a slower time scale, typically minutes. It gradually restores the generator rotation to its reference frequency, correcting any steady-state deviation left by primary control.
    \item On even longer time scales of minutes to hours, tertiary control may intervene. This, for instance, adjusts each generator's target power output and ensures efficient electricity supply while maintaining adequate reserves.
\end{enumerate}

We focus on primary control only, which provides the fastest and strongest response to stabilize the grid. We model it as a (nonlinear) continuous piecewise function \cite{kraljic2023towards,vorobev2019deadbands,mele2016impact}.

\subsection{Control in the United Kingdom}

In the United Kingdom, primary control is exerted through two mechanisms: ``Dynamic Regulation'' and ``Dynamic Moderation'' \cite{nesonote,ofgemnote}. The former linearly damps deviations by providing a linear power response to frequency deviations up to $\qty{0.2}{Hz}$. The latter gives an additional linear response for deviations between $\qty{0.1}{Hz}$ and $\qty{0.2}{Hz}$. Control is activated to a finite precision of $\qty{0.015}{Hz}$ \cite{rebours2007survey}, creating a deadband with no damping around the reference frequency. This scenario is accurately described by the nonlinear function $H(\omega)$ drawn in Fig. 2F (main text) and defined in Eq. 8 (main text). Here, the numerical values of the boundaries between control regions are
\begin{align}
    \label{apx-eq: deadband width uk}
    \omega_0 &= 2\pi \cdot \qty{0.015}{rad/s} \\
    \text{United Kingdom:} \quad \omega_1 &= 2\pi \cdot \qty{0.1}{rad/s} \nonumber \\
    \omega_2 &= 2\pi \cdot \qty{0.2}{rad/s} \nonumber \,.
\end{align}

In the United Kingdom, a post-fault control mechanism, called ``Dynamic Control'', is also activated when deviations exceed $\qty{0.2}{Hz}$. We can safely ignore it in our models, as we collect data during routine grid operations that rarely exceed this value. Note, indeed, that the tails of the frequency distribution rarely extend beyond $\qty{0.2}{Hz}$, as shown in Fig.~2G (main text). These large deviations are consistent with the description above, which specifies the nominal regulation bands of primary control. Control continues to operate beyond these bands, albeit possibly nonlinearly or partially.

\subsection{Control in South Africa}

Control in South Africa is less strict than in the United Kingdom. The first layer of primary control is provided by generators contracted with ESKOM, the country's main electricity utility. These generators provide an ``Instantaneous Reserve'' \cite{nersacode}, acting within $\qty{10}{s}$ to counter deviations exceeding $\qty{0.15}{Hz}$. Unlike in the United Kingdom, these generators are not tasked with actively damping all deviations within a range, specifically between the deadband and $\qty{0.15}{Hz}$. Instead, they are bound to respond only when the frequency deviation exceeds $\qty{0.15}{Hz}$, leaving the region between the deadband and $\qty{0.15}{Hz}$ more loosely regulated than its corresponding region in the United Kingdom. The second layer of control is provided by additional non-contracted generators, which respond to deviations between $\qty{0.15}{Hz}$ and $\qty{0.5}{Hz}$.

This staggered control makes frequency damping irregular in practice: for deviations below $\qty{0.15}{Hz}$, different generators may respond at different thresholds. Still, for simplicity, we model control as piecewise linear with three distinct regions. Linear control in the $\num{0.015}$-$\qty{0.15}{Hz}$ region can be interpreted as a net effect of the heterogeneous generator response. Accordingly, we set the control boundaries in South Africa as
\begin{align}
    \label{apx-eq: deadband width sa}
    \omega_0 &= 2\pi \cdot \qty{0.015}{rad/s}  \\
    \text{South Africa:} \quad \omega_1 &= 2\pi \cdot \qty{0.15}{rad/s} \nonumber \\
    \omega_2 &= 2\pi \cdot \qty{0.5}{rad/s} \nonumber\,.
\end{align}

This simplified control model facilitates numerical tractability while still capturing the essential features of the distributions shown in Figs.~2{B, E} (main text). In the South African frequency distribution, the deadband dip appears less pronounced than in the United Kingdom. This difference may be attributed to the staggered activation of generators. The discontinuity between control regimes at $\pm \omega_2$ in South Africa manifests as a secondary peak in the tails of the frequency histogram in Fig.~2{G} (main text), similar to the peaks observed at $\pm \omega_1$. However, too few data preclude a reliable MLE fit.

\section{Power Markets in the United Kingdom and South Africa}

Power grids worldwide operate within various energy markets, each with its own rules and trading mechanisms.

In the United Kingdom, energy trading occurs through Nominated Electricity Market Operators (NEMOs), which are regulated by the Office of Gas and Electricity Markets (Ofgem). NEMOs include the European Power Exchange (EPEX SPOT) SE and Nord Pool AS, operating across different geographical areas of Europe. Similar to most electricity markets worldwide, these markets allow for both day-ahead and intraday trading. Intraday transactions occur in fixed intervals every $\qty{15}{min}$ and $\qty{30}{min}$ \cite{epexnotes}. Large transactions every $\qty{30}{min}$ are visible in the frequency time series Fig. 2{A} (main text).

South Africa is the leading country of the Southern African Power Pool (SAPP), a regional cooperation that connects national grids in Southern Africa to facilitate cross-border electricity trade. SAPP also allows both day-ahead and intraday trading, with the intraday market operating on $\qty{1}{h}$ intervals \cite{sappmarket}. These transactions can be seen in Fig. 2{D} (main text).

To better visualize the contribution of market transactions to the frequency time series, we compute the Rate Of Change Of Frequency (ROCOF) for the signal aggregated over $\qty{24}{h}$. This is defined as the finite-difference derivative $\text{ROCOF}(t) = {\Delta \omega}/{\Delta t_{\mathrm{PMU}}} := {(\omega(t+\Delta t_{\mathrm{PMU}})-\omega(t))}/{\Delta t_{\mathrm{PMU}}}$. We display it over time windows centered around market transaction events in both countries in \Crefrange{apx-fig: rocof uk}{apx-fig: rocof sa}. Measuring its value exactly at transaction times, taking its absolute value, and averaging, we obtain
\begin{align}
    \label{apx-eq: rocof uk}
    \text{United Kingdom:} \quad |\text{ROCOF}(t=\text{transaction})| &= \num{0.03 \pm 0.02}\,\unit{{rad/s}^2} \\
    \label{apx-eq: rocof sa}
    \text{South Africa:} \quad |\text{ROCOF}(t=\text{transaction})| &= \num{0.007 \pm  0.004}\,\unit{{rad/s}^2} \,.
\end{align}
The United Kingdom values are consistent with similar estimates reported in the literature \cite{gorjao2020data}. The relatively high standard deviations in \Crefrange{apx-eq: rocof uk}{apx-eq: rocof sa} arise from the strong variability in the magnitude of power transactions throughout the day, as shown in \Crefrange{apx-fig: rocof uk}{apx-fig: rocof sa} and Figs. 2{A, D} (main text). For example, in the United Kingdom, between 2:30 and 5:30 p.m., market transactions exhibit smaller ROCOF spikes in \Cref{apx-fig: rocof uk} because the influence of heterogeneous consumer loads dominates the frequency profile (Fig. 2A (main text)). In South Africa, the frequency time series is less regular than in the United Kingdom, which explains the lower magnitude of the ROCOF estimate in \Cref{apx-eq: rocof sa}.

\begin{figure*}[htpb]
\centering
\includegraphics[width=1.0\textwidth]{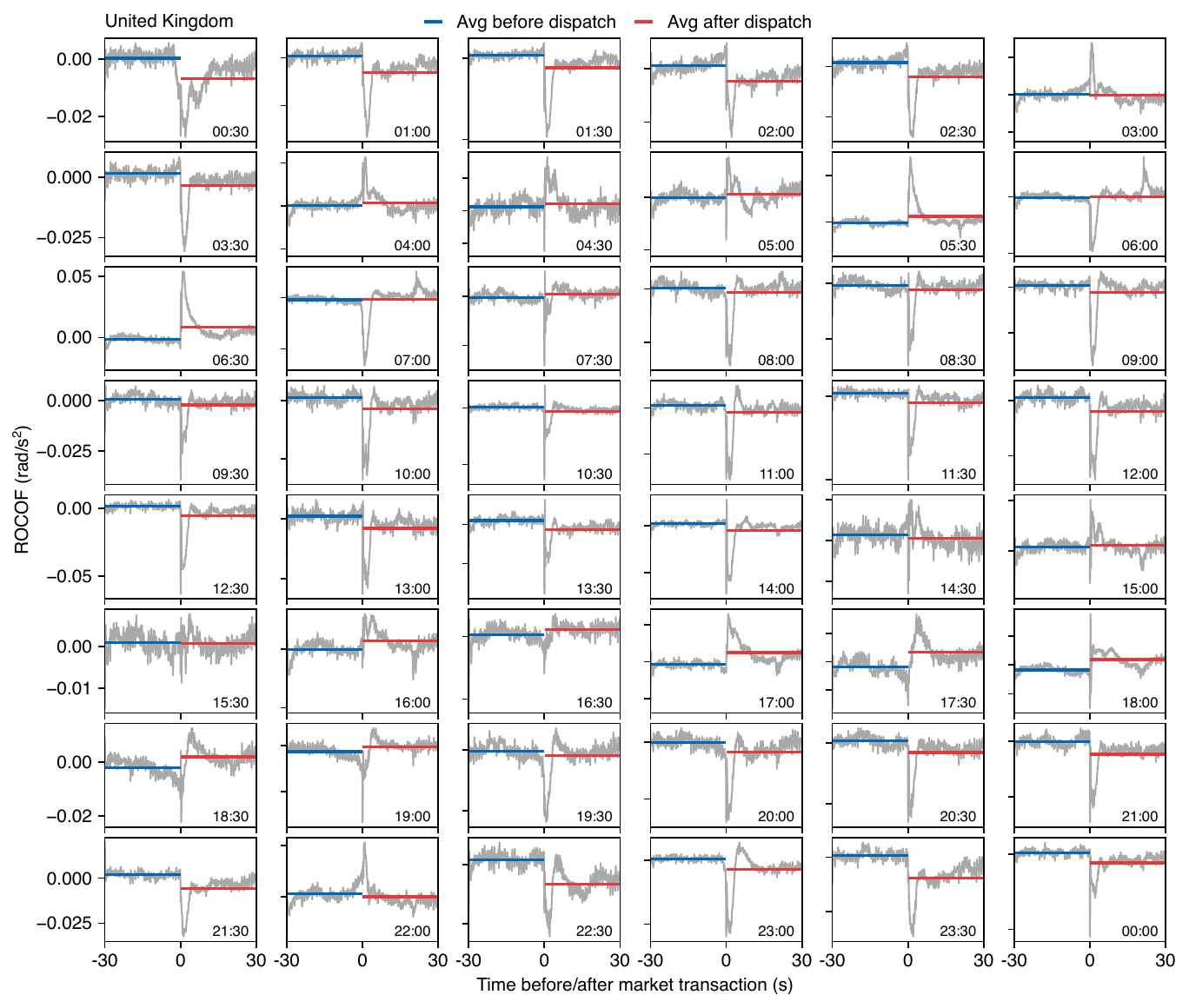}
\caption{United Kingdom ROCOF around market transactions. The ROCOF is computed using finite-difference derivatives. We show data for 48 windows of $\qty{1}{min}$, spanning $\qty{30}{s}$ before each energy market transaction, recorded every $\qty{30}{min}$ daily. The red and blue lines show the average ROCOF for the $\qty{30}{s}$ before and after the transaction.}
\label{apx-fig: rocof uk}
\end{figure*}

\begin{figure*}[htpb]
\centering
\includegraphics[width=1.0\textwidth]{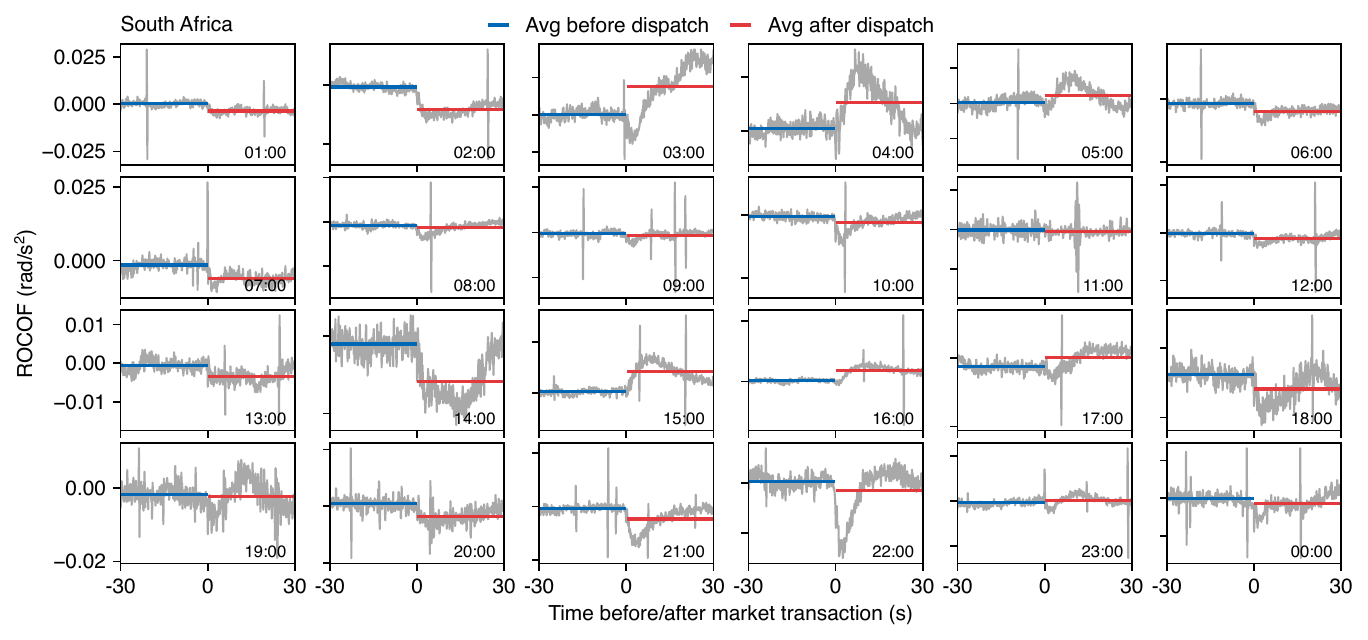}
\caption{South Africa ROCOF around market transactions. The ROCOF is computed using finite-difference derivatives. We show data for 48 windows of $\qty{1}{min}$, spanning $\qty{30}{s}$ before each energy market transaction, recorded every $\qty{1}{h}$ daily. The red and blue lines show the average ROCOF for the $\qty{30}{s}$ before and after the transaction.}
\label{apx-fig: rocof sa}
\end{figure*}

\section{Derivation of the Aggregated Swing Equations}

We derive the aggregated swing equation in Eq. 1 (main text) from Eqs. 5-6 (main text). The swing equations read, for each generator $i = 1,\dots,N$,
\begin{alignat}{2}
    \label{apx-eq: swing equation 1} 
    \frac{\mathrm{d} \theta_i}{\mathrm{d} t} &= \omega_i \\
    \label{apx-eq: swing equation 2}
    M_i \frac{\mathrm{d}  \omega_i}{\mathrm{d} t} &= H_i(\omega_i) + P_i(t )+\sum_{j = 1}^{N} K_{ij} \sin (\theta_j -  \theta_i) +\epsilon_i \xi_i \,.
\end{alignat}
We focus on \Cref{apx-eq: swing equation 2}. We define the bulk angular velocity as
\begin{equation*}
    \omega(t) := \frac{1}{M}\sum_{i=1}^N M_i \omega_i(t) \qquad M:= \sum_{i=1}^N M_i \,.
\end{equation*}
During regular operation, $\omega_i(t) = \omega(t)$ for all $i$. Therefore, we aim to derive a single aggregate equation from \Cref{apx-eq: swing equation 2}, controlling $\omega(t)$ instead of $N$ separate equations for each $\omega_i(t)$. The bulk angular velocity can be recovered on the left-hand side of \Cref{apx-eq: swing equation 2} by summing over $i$ and dividing by $M$:
\begin{equation}
    \label{apx-eq: swing equation step 1}
    \frac{\mathrm{d} \omega}{\mathrm{d} t} = \frac{1}{M} \sum_{i=1}^N H_i(\omega_i) + \frac{1}{M} \sum_{i=1}^N P_i(t) +  \frac{1}{M} \sum_{i=1}^N\sum_{j = 1}^{N} K_{ij} \sin (\theta_j -  \theta_i) +\frac{1}{M} \sum_{i=1}^N \epsilon_i\xi_i  \,.
\end{equation}
To simplify \Cref{apx-eq: swing equation step 1}, we make two standard constructive assumptions \cite{schaefer2018non,schafer2018dynamically}.
First, we take a uniform damping-to-inertia ratio across all control regions for all generators. Let \begin{align}
    \label{apx-eq: control region 0}
    \Omega_{0} &:= \{ \omega \, : \, |\omega| \leq \omega_{0} \} \\
    \label{apx-eq: control region 1}
    \Omega_1 &:= \{ \omega \, : \, \omega_{0} < |\omega| \leq\omega_{1} \} \\
    \label{apx-eq: control region 2}
    \Omega_2 &:= \{ \omega \, : \, |\omega| > \omega_{1} \} \,
\end{align}
denote the deadband, the inner control region, and the outer control region of Fig. 2F (main text), which we assume are the same size for all generators $i$.
Taking a constant  damping-to-inertia ratio amounts to having
\begin{align*}
    H_i(\omega_i) &:= - \sign{|\omega_i|} \,h_i(|\omega_i|)\\
    h_i(x) &:= \begin{cases}
    0 \quad &x \in \Omega_0\\
    D_{1,i} \, (x - \omega_0) \quad &x \in \Omega_1 \\
     D_{2,i} \, (x - \omega_1) + D_{1,i} \, (\omega_1 - \omega_0) \quad &x \in \Omega_2 \,.
    \end{cases}
\end{align*}
satisfy for all $i$
\begin{align}
    \label{apx-sex: uniform damp inertia ratio 1}
    \gamma_1 &:= \frac{D_{1,i}}{M_i} \\
    \label{apx-sex: uniform damp inertia ratio 2}
    \gamma_2 &:= \frac{D_{2,i}}{M_i} \,,
\end{align}
where $D_{1,i}$ and $D_{2,i}$ are the damping coefficients of the two control regions specific to each generator. \Crefrange{apx-sex: uniform damp inertia ratio 1}{apx-sex: uniform damp inertia ratio 2} imply
\begin{align}
    \frac{1}{M} \sum_{i=1}^N H_i(\omega_i) &= \frac{1}{M} \sum_{i=1}^N M_i H(\omega_i) \nonumber  \\
    \label{apx-eq: assumption 1 step}
    &= H \left( \frac{1}{M} \sum_{i=1}^N M_i \omega_i \right) \\
    \label{apx-eq: assumption 1}
    &= H (\omega) \,.
\end{align}
Here, $H(\omega) = -\sign{\omega} h (|\omega|)$ and $h (|\omega|) $ is defined as in Eq. 8 (main text). $H(\omega)$ has coefficients independent of $i$ and is piecewise linear, hence the step in \Cref{apx-eq: assumption 1 step} is valid.
Second, we assume a symmetric coupling satisfying for all $i,j$:
\begin{equation}
    \label{apx-eq: symmetric coupling}
    K_{ij} = K_{ji} \,.
\end{equation}

Putting \Crefrange{apx-eq: assumption 1}{apx-eq: symmetric coupling} together, and substituting $P(t)$ as defined in Eq. 7 (main text) gives Eq. 1 (main text). Indeed, \Cref{apx-eq: swing equation 2} becomes
\begin{align}
    \label{apx-eq: swing equation to match}
    \frac{\mathrm{d} \omega}{\mathrm{d} t} &= H(\omega) + P(t) +\frac{1}{M}\sum_{i=1}^N \epsilon_i \xi_i \,,
\end{align}
where the sine contributions in \Cref{apx-eq: swing equation to match} vanish because of \Cref{apx-eq: symmetric coupling}.

To conclude, we are only left to evaluate the noise. Assuming that the Gaussian random variables $\xi_i \sim \mathcal{N}(0,1)$ are i.i.d. allows us to sum their contribution to define
\begin{align}
 \epsilon &= \frac{1}{M}\sqrt{\sum_{i=1}^N \epsilon_i^2 } \nonumber \\
 \label{apx-eq: gaussian noise}
 \xi &\sim \mathcal{N}(0,1) \,.
\end{align}
This last step completes the derivation.

\section{Modeling Noise for Power Grid Dynamics}

We assume the noise $\xi$ in Eq. 1 (main text) to be Gaussian white noise, as specified by \Cref{apx-eq: gaussian noise}. This choice is convenient, as it allows us to carry forward analytical derivations for both the frequency distribution and the autocorrelation. Gaussian white noise alone, however, cannot reproduce heavy-tailed empirical distributions. To account for these, we introduce a slowly varying driving term, $P(t)$, and employ superstatistics.

Alternatives adopted in the literature include replacing Gaussian noise with non-Gaussian (Lévy) noise to capture heavy-tailed frequency fluctuations, or using superstatistics with Gaussian white noise but assuming the control coefficient not constant \cite{schaefer2018non,gorjao2021spatio,anvari2016short,kashima2015modeling,anvari2020stochastic,schafer2018isolating}. These approaches lead to Lévy-stable \cite{samoradnitsky2017stable} and $q$-Gaussian distributions \cite{tsallis1988possible} for $p(\omega)$, respectively. While such distributions successfully capture heavy tails, they have been studied without accounting for the nonlinearity of control to maintain analytical tractability. Therefore, they do not adequately describe multimodal frequency distributions.

Works incorporating nonlinear control have thus far relied exclusively on numerical simulations \cite{oberhofer2023non,vorobev2019deadbands}, precluding direct data fitting. Interestingly, yet only with simulations, fractional Lévy noise has been employed to capture heavy tails, multimodality, and slow autocorrelation decays at long time lags \cite[as well as \Cref{apx-sec: Frequency Autocorrelation}, \Cref{apx-fig: autocorrelation}B]{kraljic2023towards}, all of which are features observed in high-resolution frequency measures.

\section{Analytical Derivations of the Frequency Distribution}

We derive Eq. 10 (main text). We begin with the Fokker--Planck equation
\begin{equation*}
    \frac{\partial f}{\partial t} = -\frac{\partial}{\partial \omega} \left[ (H(\omega)+P ) f \right] + \frac{\epsilon^2}{2} \frac{\partial^2 f}{\partial \omega^2} \,,
\end{equation*}
which follows from the aggregated swing equation in Eq. 1 (main text) and is valid for short time intervals during which the slow power can be treated as constant, i.e., $P(t)=P$. We impose ``stationarity'', ${\partial f}/{\partial t} = 0$, which is also meaningful only over short time intervals. This yields
\begin{equation}
    \label{apx-eq: fp stationary to integrate}
    \frac{\partial}{\partial \omega} \left[ (H(\omega)+P ) f \right] = \frac{\epsilon^2}{2} \frac{\partial ^2 f}{ \partial \omega^2} \,.
\end{equation}
Integrating \Cref{apx-eq: fp stationary to integrate} twice, we obtain its general solution, namely
\begin{align}
    \label{apx-eq: quasi static distr general solution}
    f(\omega \, | \, P) &=\frac{1}{Z}\exp  (-\Phi(\omega \, | P)) \\
    \label{apx-eq: quasi static distr general solution integral}
    \Phi(\omega \, | P) &:= -\frac{2}{\epsilon^2} \int ( H(\omega)+P )\, \mathrm{d} \omega \,.
\end{align}
We set to zero the integration constant arising from the first integration of \Cref{apx-eq: fp stationary to integrate} by assuming ${\partial f}/{\partial \omega} = 0$ as $\omega \to \pm \infty$.

Before beginning the calculations, we illustrate how the distribution and the potential in \Crefrange{apx-eq: quasi static distr general solution}{apx-eq: quasi static distr general solution integral} change with their parameters. We show in \Cref{apx-fig: control potential drawing}B the potential $\Phi(\omega \, | \, P)$ that has a single well whose minimum yields the most probable frequency. If $P = 0$, then $\Phi(\omega \, | \, P)$ is symmetric the minimum is at $\omega = 0$. Physically, $\Phi(\omega \, | \, P)$ can be interpreted as a potential energy arising from the control $H(\omega)$ in \Cref{apx-fig: control potential drawing}A and the mechanical power $P$. The noise amplitude $\epsilon$ controls the width of the potential, as shown in \Cref{apx-fig: control potential drawing}C, and thereby the spread of the distribution $f(\omega \, | \, P)$.

With superstatistics, the multimodality observed in frequency histograms can be understood as a superposition of the single-well states described by $f(\omega \, | \, P)$. Such an explanation has been noted before in the literature \cite{wen2023non}, but has not been formalized before our work. We draw $f(\omega \, | \, P)$ in \Cref{apx-fig: control potential drawing}D. The shape of $f(\omega \, | \, P)$ is an exponential distribution modulated by $P$ in $\Omega_0$ (uniform when $P = 0$), joined to two Gaussian branches in $\Omega_1$ and $\Omega_2$. The effect of piecewise linear control is to damp the Gaussian branches more strongly in the in $\Omega_2$ than in $\Omega_1$.

To get closed-form expressions, we integrate \Cref{apx-eq: quasi static distr general solution integral} separately over the three control regions $\Omega_0$, $\Omega_1$, and $\Omega_2$ in \Crefrange{apx-eq: control region 0}{apx-eq: control region 2}. Then, we impose continuity between each expression of $f(\omega \, | \, P)$ at $\pm \omega_0$ and $\pm \omega_1$. Finally, we compute the normalization $Z$ in the deadband and observe that the normalization in the linearly damped regions $\Omega_1$, $\Omega_2$ does not admit an elementary closed-form solution. We write all calculations in detail, the final results are summarized in \Crefrange{apx-eq: c1 final result}{apx-eq: c2 final result} for the continuity constants, and \Cref{apx-eq: zeta 0 final sol} for the normalization.

\begin{figure*}[htpb]
\centering
\includegraphics[width=0.7\textwidth]{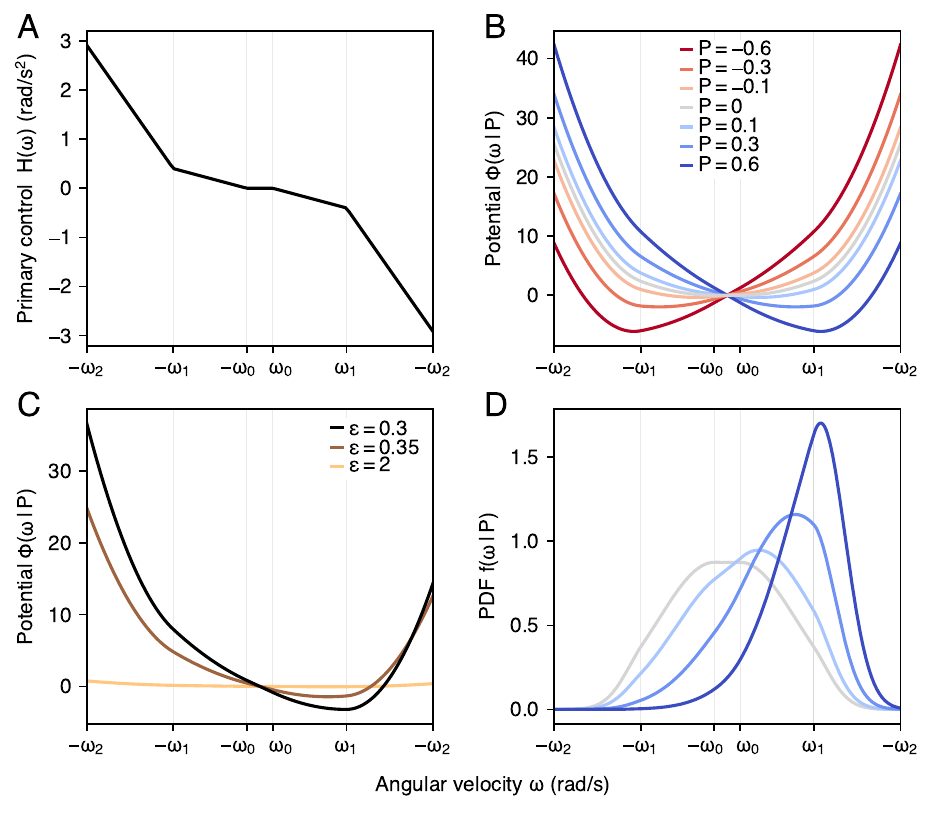}
\caption{Nonlinear control, potential, and PDF. For all panels, the axes are arbitrarily scaled. The damping coefficients are conventionally set to $\gamma_1 = 0.75$ and $\gamma_2 = 0.4$. Control regions are separated by $\pm \omega_0$, $\pm \omega_1$, and $\pm \omega_2$ as in \Cref{apx-eq: deadband width uk}. (A) Nonlinear control function $H(\omega)$ as in Fig. 2F (main text). (B) Potential $\Phi(\omega \, | \, P)$ obtained by integrating $H(\omega)$ as per \Cref{apx-eq: quasi static distr general solution integral}. We plot the potential as a function of $P$ while conventionally fixing $\epsilon = 0.3$. When integrating the control function, we choose arbitrary integration constants to ensure the continuity of $\Phi(\omega \, | \, P)$. (C) Potential $\Phi(\omega \, | \, P)$ as $\epsilon$ varies. We focus on the single value $P = 0.3$, as the shrinking effect of $\epsilon$ is qualitatively identical for all $P$. (D) PDF $f(\omega \, | \, P)$ for the non-negative values of $P$ in panel B. The color scheme matches that of panel B. Here, the integration constants are as in \Crefrange{apx-eq: c1 final result}{apx-eq: c2 final result}, namely, they are set to ensure the continuity of $f(\omega \, | \, P)$. The PDF is normalized numerically.}
\label{apx-fig: control potential drawing}
\end{figure*}

\subsection{Fokker--Planck Integrals} 

In the deadband $\Omega_0$, there is no damping. Therefore \Crefrange{apx-eq: quasi static distr general solution}{apx-eq: quasi static distr general solution integral} give
\begin{equation}
    \label{apx: integral f result 1}
    f_0(\omega \, | \, P) \propto \exp \left( \frac{2 P}{\epsilon^2} \omega \right) \,,
\end{equation}
where we introduced the convention $f_0(\omega \, | P) = f(\omega \, | \, P)$ for $\omega \in \Omega_0$ (and will do analogously for the control region $\Omega_1$, $\Omega_2$).

In $\Omega_1$, the control function takes the form $H(\omega) = -\gamma_1 \sign{\omega} (|\omega| - \omega_{0})$, i.e., it is separately affine on the two branches with $\omega > 0$ and $\omega < 0$. As a result, $\Phi(\omega \, | \, P)$ is quadratic and \Cref{apx-eq: quasi static distr general solution} becomes Gaussian. Formally,
\begin{align}
    f_1(\omega \, | \, P) &\propto \exp \left[ \frac{2}{\epsilon^2} \int ( -\gamma_1 \, \sign{\omega} (|\omega| - \omega_{0})+P )\, \mathrm{d}  \omega \right] \nonumber \\
    &= \exp \left[ \frac{2}{\epsilon^2} \int ( -\gamma_1 (\omega - \sign{\omega} \omega_{0})+P )\, \mathrm{d}  \omega \right] \nonumber  \\
    &= \exp \left\{ \frac{2}{\epsilon^2} \left[ -\gamma_1 \left( \frac{\omega^2}{2} - |\omega| \, \omega_{0} \right) + P \omega   \right] \right\} \nonumber  \\
    \label{apx-eq: completing square 1}
    &\propto \exp \left\{ \frac{2}{\epsilon^2} \left[ -\frac{\gamma_1}{2} \left( \omega - \sign\omega \omega_{0} \right)^2 + P \omega   \right] \right\}\\
    \label{apx-eq: completing square 2}
    &\propto \exp \left[ - \frac{\gamma_1}{\epsilon^2} \left( \omega - \sign\omega \omega_{0} - \frac{P}{\gamma_1} \right)^2  \right] \,.
\end{align}
In \Cref{apx-eq: completing square 1}, we completed the squares in the innermost parentheses. Similarly, in \Cref{apx-eq: completing square 2}, we completed the squares in the squared brackets. Defining $\Lambda_1 = \Lambda_1( \sign{\omega}, P)$ as in Eq. 11 (main text) gives
\begin{equation}
    \label{apx: integral f result 2}
    f_1(\omega \, | \, P) \propto \exp \left[ -\frac{\gamma_1}{\epsilon^2} \left( \omega - {\Lambda_1} \right)^2 \right] \,.
\end{equation}

In $\Omega_2$, control is again separately affine for $\omega > 0$, $\omega < 0$ and of the form $H(\omega) = - \sign{\omega} \left[ \gamma_2  (|\omega| - \omega_{1}) + \gamma_1 (\omega_{1} - \omega_{0}) \right]$. Once again \Cref{apx-eq: quasi static distr general solution}, becomes a Gaussian distribution. In detail,
\begin{align}
    f_2(\omega \, | \, P) &\propto \exp \left( \frac{2}{\epsilon^2} \int \{ - \sign{\omega} \left[ \gamma_2  (|\omega| - \omega_{1}) + \gamma_1 (\omega_{1} - \omega_{0}) \right] +P \} \, \mathrm{d}  \omega \right) \nonumber \\
    \label{apx-eq: term omega introduced}
    &= \exp \left( \frac{2}{\epsilon^2} \int \{ - \left[ \gamma_2  (\omega - \sign{\omega} \omega_{1}) + \gamma_1 \, \sign{\omega} \Delta \right] +P \}\, \mathrm{d}  \omega \right) \\
    &= \exp \left( \frac{2}{\epsilon^2} \left\{ - \left[ \gamma_2  \left( \frac{\omega^2}{2} - |\omega| \omega_{1} \right) + \gamma_1 |\omega| \Delta \right] +P \omega \right\} \right) \nonumber \\
    \label{apx-eq: derivation quasi stat}
    &= \exp \left\{ \frac{2}{\epsilon^2} \left[ - \frac{\gamma_2}{2} \omega^2 + (\gamma_2 \, \sign\omega \omega_{1} - \gamma_1 \, \sign\omega \Delta + P) \omega \right] \right\} \,.
\end{align}
In \Cref{apx-eq: term omega introduced} we introduced $\Delta:=\omega_{1}-\omega_{0}$. To further unclutter notation, we defined $\Lambda_2 = \Lambda_1( \sign{\omega}, P)$ as in Eq. 12 (main text), which is constant over the two branches $\omega > 0$ and $\omega < 0$, and therefore can be treated as constant when completing the squares. Substituting it into \Cref{apx-eq: derivation quasi stat} gives
\begin{equation*}
    f_2(\omega \, | \, P) \propto \exp \left[ \frac{2}{\epsilon^2} \left( - \frac{\gamma_2}{2} \omega^2 + \gamma_2\Lambda_2 \omega \right) \right] \,.
\end{equation*}
We complete the squares again to get
\begin{equation}
    \label{apx: integral f result 3}
    f_2(\omega \, | \, P) \propto \exp \left[ -\frac{\gamma_2}{\epsilon^2} \left( \omega - {\Lambda_2} \right)^2 \right] \,.
\end{equation}

\Cref{apx: integral f result 1}, \Cref{apx: integral f result 2}, and \Cref{apx: integral f result 3} correspond the right-hand side of Eq. 10 (main text) without the continuity factors and normalization constant. Using the notation of \Crefrange{apx-eq: quasi static distr general solution}{apx-eq: quasi static distr general solution integral}, we may write the potential
\begin{align}
    \label{apx-eq: potential close-form expression}
    \Phi(\omega \, | P) = \begin{cases}
        \displaystyle-\frac{2 P}{\epsilon^2} \omega &\omega \in \Omega_0 \\[8pt]
        \displaystyle\frac{\gamma_1}{\epsilon^2}(\omega - \Lambda_1)^2 &\omega \in \Omega_1 \\[8pt]
        \displaystyle\frac{\gamma_2}{\epsilon^2}(\omega - \Lambda_2)^2 &\omega \in \Omega_2 \,.
    \end{cases}
\end{align}

\subsection{Continuity Factors}

We impose the continuity of $f(\omega \, | \, P)$ at $\pm \omega_{0}$ and $\pm \omega_{1}$ for all values of $P$. Formally, using the inequality conventions of \Crefrange{apx-eq: control region 0}{apx-eq: control region 2}, we impose
\begin{alignat}{2}
    \label{apx-eq: limit continuity quasi static 1}
     f_0( \omega_{0} \, | \, P) &= \lim_{\omega \to  \omega_{0}^{+}} f_1(\omega \, | \, P) \qquad & f_0( -\omega_{0} \, | \, P) &= \lim_{\omega \to - \omega_{0}^{-}} f_1(\omega \, | \, P) \\
     \label{apx-eq: limit continuity quasi static 2}
     f_1( \omega_{1} \, | \, P) &= \lim_{\omega \to  \omega_{1}^{+}} f_2(\omega \, | \, P) \qquad & f_1( -\omega_{1} \, | \, P) &= \lim_{\omega \to - \omega_{1}^{-}} f_2(\omega \, | \, P) \,.
\end{alignat}
\Crefrange{apx-eq: limit continuity quasi static 1}{apx-eq: limit continuity quasi static 2} return two multiplicative constants $C_1 = C_1( \sign{\omega}, P)$, $C_2 = C_2( \sign{\omega}, P)$ to scale $f_1( \omega \, | \, P)$, $f_2( \omega \, | \, P)$, respectively, and obtain a continuous distribution $f( \omega \, | \, P)$ over all $\omega$.

Continuity at $\pm \omega_{0}$, i.e., \Cref{apx-eq: limit continuity quasi static 1}, gives
\begin{equation*}
    \exp \left(\pm \frac{2P}{\epsilon^2}\,\omega_{0} \right) = C_1(\pm 1, P) \exp \left[ - \frac{\gamma_1}{\epsilon^2} \left( \pm \omega_{0} \mp \omega_{0} - \frac{P}{\gamma_1} \right)^2  \right] \,.
\end{equation*}
Rearranging terms, we obtain
\begin{align*}
    C_1(\pm 1, P) &= {\exp \left(\pm \frac{2P}{\epsilon^2}\,\omega_{0} \right)} \bigg/{\exp \left( -\frac{P^2}{\gamma_1 \epsilon^2}\right)} \\
    &= {\exp \left(\pm \frac{2P}{\epsilon^2}\,\omega_{0} + \frac{P^2}{\gamma_1 \epsilon^2}\right)} \,.
\end{align*}

Continuity at $\pm \omega_{1}$, i.e, \Cref{apx-eq: limit continuity quasi static 2}, yields an analogous derivation. This time $C_1$ and $C_2$ must obey
\begin{equation*}
     C_1(\pm 1, P) \exp \left( - \frac{\gamma_1}{\epsilon^2} \left( \pm \omega_{1} - \Lambda_1 \right)^2  \right) = C_2(\pm 1, P) \exp \left( - \frac{\gamma_2}{\epsilon^2} \left( \pm \omega_{1} - \Lambda_2 \right)^2  \right)
\end{equation*}
where, we employed $\Lambda_1$ and $\Lambda_2$ to ease notation. Rearranging terms and using $\Delta := \omega_{1} - \omega_{0}$, we obtain
\begin{align*}
    C_2(\pm 1, P) &= C_1(\pm 1, P) \, \exp \left( - \frac{\gamma_1}{\epsilon^2} \left( \pm \omega_{1} - \Lambda_1 \right)^2  \right) \bigg/ \exp \left( - \frac{\gamma_2}{\epsilon^2} \left( \pm \omega_{1} - \Lambda_2 \right)^2  \right) \\
    &= C_1(\pm 1, P) \, \exp \left[-\frac{\gamma_1}{\epsilon^2}\!\left(\pm\Delta - \frac{P}{\gamma_1}\right)^{\!2}\right] \, \bigg/ \exp \left[ - \frac{\gamma_2}{\epsilon^2} \left( \pm\frac{\gamma_1}{\gamma_2}\Delta - \frac{P}{\gamma_2} \right)^2  \right] \\
    &= C_1(\pm 1, P)\, \exp \left[
    -\frac{\gamma_1}{\epsilon^2} \left(\pm \Delta - \frac{P}{\gamma_1}\right)^{2}
    +\frac{\gamma_2}{\epsilon^2}\left(\pm \frac{\gamma_1}{\gamma_2}\Delta - \frac{P}{\gamma_2}\right)^{2}
    \right] \,.
\end{align*}

In summary, $f(\omega \, | \, P)$ is continuous for
\begin{align}
    \label{apx-eq: c1 final result}
    C_1\left( {\sign{\omega}} , P \right) &= \exp\left(\sign{\omega} \frac{2 P\, \omega_{0}}{\epsilon^{2}}+\frac{P^{2}}{\gamma_{1}\epsilon^{2}}\right) \\
    \label{apx-eq: c2 final result}
    C_{2}\left( {\sign{\omega}} , P \right) &=C_{1}\left( {\sign{\omega}} , P \right)\,\exp\left[-\frac{\gamma_{1}}{\epsilon^{2}}\left( \sign{\omega}(\omega_{1} - \omega_{0})-\frac{P}{\gamma_{1}}\right)^{2}+\frac{\gamma_{2}}{\epsilon^{2}}\left(\frac{\gamma_{1}}{\gamma_{2}}\sign{\omega} (\omega_{1} - \omega_{0}) -\frac{P}{\gamma_{2}}\right)^{2}\right] \,.
\end{align}

\subsection{Normalization Constant}

To conclude our derivation of $f(\omega \, | \, P)$, we are left with the calculation of $Z = Z(P)$. Again, we divide the discussion into three parts for $\Omega_0$, $\Omega_1$, and $\Omega_2$, and express $Z$ as the sum $Z = Z_0 + Z_1 + Z_2$.

In the deadband, we separate the two cases where $P \neq 0$ and  $P = 0$. In the former, we evaluate $Z_{0}$ as follows
\begin{align*}
    Z_{0} &= \int_{-\omega_{0}}^{\omega_{0}} \exp \left( \frac{2 P }{\epsilon^2} \omega\right) \, \mathrm{d} \omega \\
    &= \frac{\epsilon^2}{2P} \left[ \exp \left( \frac{2 P }{\epsilon^2} \omega_{0} \right) - \exp \left( -\frac{2 P }{\epsilon^2} \omega_{0} \right) \right] \\
    &= \frac{\epsilon^2}{P} \sinh\left( \frac{2 P }{\epsilon^2} \omega_{0} \right) \,.
\end{align*}
In the latter, instead, $f(\omega \, | \, P)$ is constant, hence
\begin{equation*}
    Z_{0} = 2 \omega_{0} \,.
\end{equation*}
Combining the two cases, we obtain
\begin{align}
    \label{apx-eq: zeta 0 final sol}
    Z_{0} = \begin{cases}
        \displaystyle\frac{\epsilon^2}{P} \sinh\left( \frac{2 P }{\epsilon^2} \omega_{0} \right) &\quad P \neq 0\\
        \displaystyle 2\omega_{0} &\quad P = 0 \,.
    \end{cases}
\end{align}

In $\Omega_1$ and $\Omega_2$, the integration of $f(\omega \, | \, P)$ yields the Gaussian integrals
\begin{align}
    \label{apx-eq: normalization 1}
    Z_{1} &= \int_{\Omega_1} C_1(\sign{\omega}, P) \exp \left[ - \frac{\gamma_1}{\epsilon^2} \left( \omega - \Lambda_1(\sign{\omega},P) \right)^2  \right] \, \mathrm{d} \omega \\
    \label{apx-eq: normalization 2}
    Z_{2} &= \int_{\Omega_{2}} C_2(\sign{\omega} , P) \exp \left[ - \frac{\gamma_2}{\epsilon^2} \left( \omega - \Lambda_2(\sign{\omega},P) \right)^2  \right] \, \mathrm{d} \omega \,.
\end{align}
over bounded domains. One can carry calculations further to obtain a normalization factor $Z$ that depends on the error function
\begin{equation*}
    \erf{x} := {\frac {2}{\sqrt {\pi }}}\int _{0}^{x}e^{-t^{2}}\, \mathrm{d}  t \,.
\end{equation*}
However, for our discussion, it is sufficient to note that $Z$ cannot be expressed in terms of elementary functions due to the dependence on $Z_1$ and $Z_2$. To proceed with further analytical results, we resort to approximations.

\section{Frequency Distribution: Deadband, Peaks at Control Boundaries, Heavy Tails}

\subsection{Deadband}

For large values of $P$, the distribution $f(\omega \, | \, P)$ is concentrated away from the deadband $\Omega_0$. To get an intuition of how large $P$ should be we assume $P > 0$ (the case $P < 0$ is analogous) and impose
\begin{equation}
    \label{apx-eq: power bound}
    \Lambda_1(1, P) \geq \omega_0 + \sigma_1 \,,
\end{equation}
where $\sigma_1 = \epsilon / \sqrt{2 \gamma_1}$ is the standard deviation of $f_1(\omega \, | \, P)$. In other words, we search for the values of $P$ such that the Gaussian $f_1(\omega \, | \, P)$ is centered at least one standard deviation beyond the control boundary $\omega_0$. Substituting the full expression of $\Lambda_1$ into \Cref{apx-eq: power bound} yields
\begin{equation}
    \label{apx-eq: power bound final}
    P \geq P_c := \epsilon \sqrt{\frac{\gamma_1}{2}} \, .
\end{equation}
indicating that any power fluctuations larger than $P_c$ might suffice to suppress the deadband contribution $f_0(\omega \, | \, P)$ of $f(\omega \, | \, P)$. Consequently, to model the deadband, we choose $P$ to be arbitrarily small, for example $|P| \leq \delta/2$ with $\delta/2 \ll P_c$, and assume $Z$ to be constant.

We get a numerical estimate of $P_c$ in the right-hand side of \Cref{apx-eq: power bound final} for the United Kingdom, where we perform MLE. The parameter $\epsilon$ is obtained via KR (\Cref{apx-eq: epsilon uk kr}), and $\gamma_1$ is computed by fitting the frequency autocorrelation (\Cref{apx-eq: autocorr sa fit}, we use the single-exponential fit as it gives an effectively identical parameter to the double-exponential one but with lower standard deviation). This gives $P_c = (\num{ 7 \pm 2 }) \cdot 10^{-5}\,\unit{{rad/s}^2}$.

Taking the normalization $Z$ to be constant allows us to carry out analytical computations for $p(\omega)$ in the deadband. In particular, for $\omega \in \Omega_0$, we can write
\begin{equation*}
p(\omega) \propto M_{\varphi} \left(\frac{2\omega}{\epsilon^{2}}\right),
\end{equation*}
where $M_{\varphi}$ is the moment-generating function
\begin{equation*}
M_{\varphi}(s) := \mathbb{E} [\exp(s P)] \propto\int \exp(s P) \varphi(P) \, \mathrm{d} P \,.
\end{equation*}
Expanding $M_{\varphi}$ in a Taylor series around $\omega=0$ gives
\begin{align}
M_{\varphi} \left(\frac{2\omega}{\epsilon^{2}} \right)
&= M_{\varphi}(0)
+ \frac{\mathrm{d} M_{\varphi}}{\mathrm{d} \omega}\bigg|_{\omega=0}\,\omega
+ \frac{1}{2} \frac{\mathrm{d}^{2}M_{\varphi}}{\mathrm{d} \omega^{2}}\bigg|_{\omega=0}\,\omega^{2}
+ O(\omega^{3}) \nonumber \\
\label{apx-eq: taylor expansion mgf}
&= 1 + \frac{2}{\epsilon^{2}}\mathbb{E}[P] \,\omega
+ \frac{2}{\epsilon^{4}}\mathbb{E}[P^{2}]\,\omega^{2} + O(\omega^{3}) \,.
\end{align}
\Cref{apx-eq: taylor expansion mgf} shows that the first and second derivatives of $p(\omega)$ at $\omega=0$ are
\begin{align*}
\frac{\mathrm{d} p}{\mathrm{d} \omega} \bigg|_{\omega=0}
&\propto \mathbb{E}[P] \\
\frac{\mathrm{d} ^{2}p}{\mathrm{d} \omega^{2}}\bigg|_{\omega=0}
&\propto \mathbb{E}[P^{2}] \,.
\end{align*}
The first condition, $\mathbb{E}[P] = 0$, is expected to hold because of energy conservation imposed by Eq. 7 (main text) and assuming ergodicity. As a result, the first derivative of $p(\omega)$ vanishes at $\omega = 0$, and the second derivative is strictly positive for any non-degenerate distribution with $\mathbb{E}[P^2] > 0$. This establishes that $\omega = 0$ is a local minimum of $p(\omega)$ for arbitrarily small, but non-zero, values of $P$.

To better quantify the impact of the power-induced dip at $\omega = 0$, we assume that for $|P| \leq \delta/2$
\begin{equation*}
\varphi(P) = c \,.
\end{equation*}
Since we approximated the normalization $Z$ with a constant, we can integrate Eq. 2 (main text) for $\omega \in \Omega_0$. This yields
\begin{align}
    p(\omega) &\propto \int_{-\delta/2}^{\delta/2} c \exp \left( \frac{2 P }{\epsilon^2} \omega\right) \, \mathrm{d} P \nonumber \\
    \label{apx-eq: stationary ansatz deadband}
    &= \frac{c \epsilon^2}{\omega} \sinh \left( \frac{\delta \omega}{\epsilon^2} \right) \,.
\end{align}
\Cref{apx-eq: stationary ansatz deadband} is the ansatz we fit with MLE on the data from the United Kingdom, as discussed in detail in \Cref{apx-sec: mle}.

\subsection{Peaks at Control Boundaries} Since the frequency control changes its magnitude between $\Omega_1$ and $\Omega_2$, the frequency distributions in Figs. 2{B, E} (main text) have two peaks at $\pm \omega_1$. Large frequency fluctuations that populate the histograms around the peaks can happen when, in Eq. 2 (main text), the slow power is large enough to drift the system's average beyond $\pm \omega_1$. Similarly to \Cref{apx-eq: power bound}, considering $P > 0$, this happens if
\begin{equation*}
        \Lambda_1(1, P) \geq \omega_1 \,,
\end{equation*}
which is to say
\begin{equation*}
        P \geq P_c:=\gamma_1(\omega_1 - \omega_0) \,,
\end{equation*}
where we reused $P_c$ for simplicity.

Without a closed-form expression for $\varphi(P)$, we cannot directly fit the peaks at $\pm \omega_1$. Roughly speaking, these peaks are expected to appear when the marginalization in Eq. 2 (main text) receives a significant contribution from values of $P$ near the boundaries $\pm P_c$. This requires that $\varphi(P)$ has sufficient probability mass in a neighborhood of $\pm P_c$ of width $\sigma_2$ and does not decay too rapidly in this region, so that the Gaussian contributions $f_2(\omega \, | \, P)$ centered beyond $\pm \omega_1$ contribute visibly to $p(\omega)$ after marginalization.

\subsection{Heavy Tails}

We observe that a $q$-Gaussian fit fits the frequency tails more accurately than a Gaussian one. For details, see \Cref{apx-sec: mle}. In \Cref{apx-fig: qq plot tails}, we plot a quantile-quantile (Q-Q) plot, intended as a companion figure to Fig. 2G (main text).

Examining the plot, we observe that the $q$-Gaussian data points fall along the main diagonal, indicating that the $q$-Gaussian distribution closely matches the empirical distribution across the entire range of quantiles. In contrast, the Gaussian fit deviates from the diagonal in the tails, around $\omega_2$, falling below the empirical quantiles. This indicates that the Gaussian fit underestimates the tails of the frequency distribution.

%%%%%%%%%%%%%%%%%%%%%%%%%%%%%%%%%%%%%%%%%%
\begin{figure*}[htpb]
\centering
\includegraphics[width=0.32\textwidth]{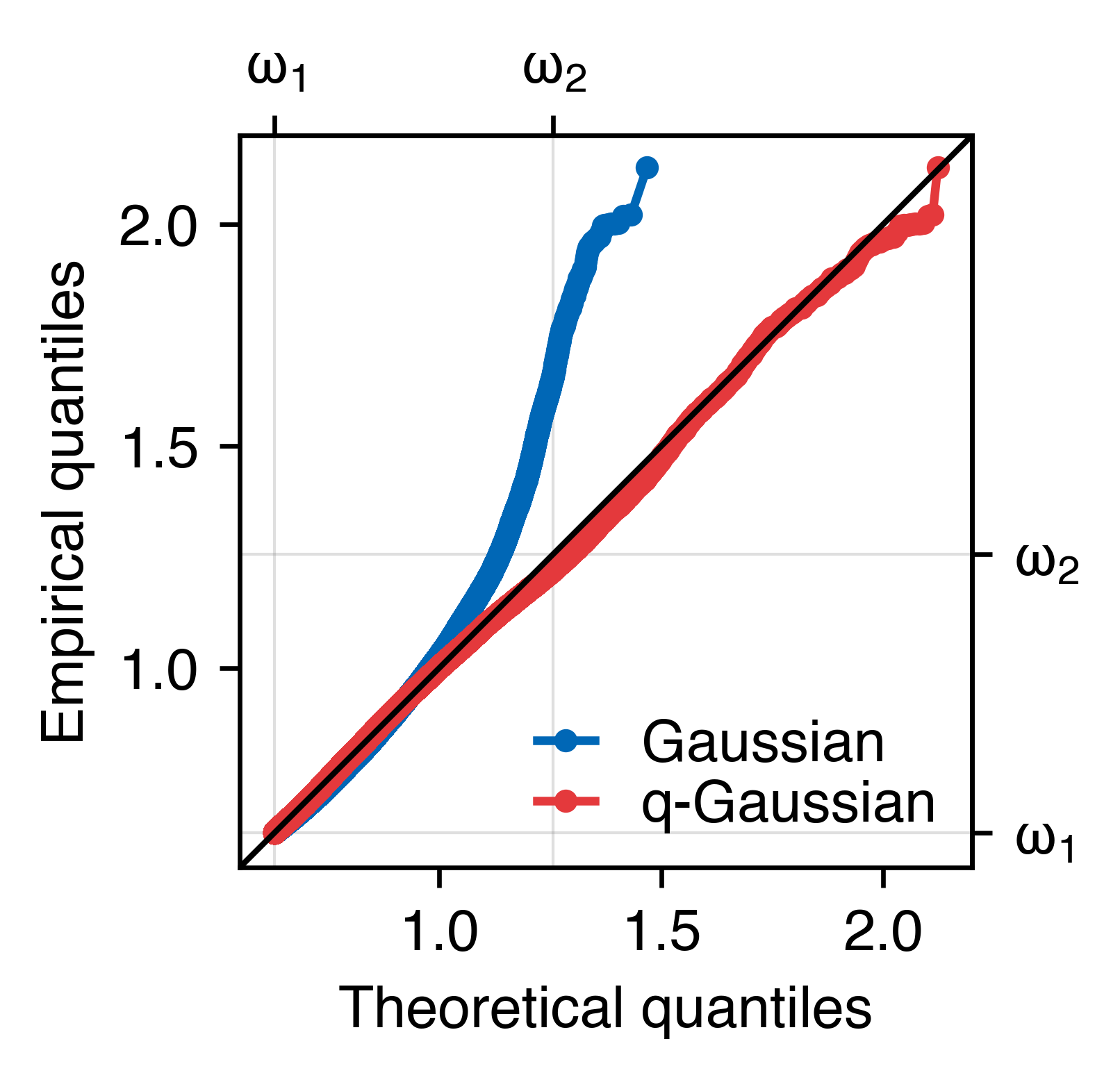}
\caption{Q-Q plot for the frequency tails in the United Kingdom.  The $x$-axis shows the theoretical quantiles, calculated as the quantiles of the fitted distributions (a Gaussian and a $q$-Gaussian). The $y$-axis shows the empirical quantiles taken from the empirical data.}
\label{apx-fig: qq plot tails}
\end{figure*}
%%%%%%%%%%%%%%%%%%%%%%%%%%%%%%%%%%%%%%%%%%

Although recovering a closed-form expression for a $q$-Gaussian distribution from our statistical model is difficult, we can carry forward an analytical argument for the tails. Heavy tails arise from integrating $f(\omega \, | \, P)$ against a heavy-tailed distribution $\varphi(P)$. We demonstrate this by employing saddlepoint approximation \cite{butler2007saddlepoint}. The goal of this derivation is to demonstrate that heavy tails in $p(\omega)$ are inherited from the market power distribution.

To simplify calculations, we consider linear control across all frequencies with a single damping parameter $\gamma$. Such an assumption is sensible as we are only interested in the tail of $p(\omega)$. Therefore, we can neglect boundary effects caused by the separation between the deadzone and the two control regions. For uniform control, our model becomes analytically integrable since the distribution $f(\omega \, | \, P)$ reduces to the Gaussian
\begin{equation}
    \label{apx-eq: gaussian for tails}
    f(\omega \, | \, P) \propto  \exp \left[ -\frac{\gamma}{\epsilon^2} \left( \omega - \frac{P}{\gamma}\right)^2\right] \,
\end{equation}
with a constant normalization factor.
Consequently, $p(\omega)$ in Eq. 2 (main text) can be written as a convolution between \Cref{apx-eq: gaussian for tails} and $\psi(y) := \gamma \, \varphi(\gamma y)$, which is a rescaling of $\varphi(P)$ obtained with the change of variable $y = P / \gamma$. Formally,
\begin{equation*}
    p(\omega) \propto \int \exp \left( -\frac{\gamma}{\epsilon^2} (\omega - y)^2 \right) \, \psi(y) \, \mathrm{d} y \,.
\end{equation*}
We assume a heavy-tailed ansatz of the form $\psi(y) \sim |y|^{-\beta}$ with $\beta > 0$. This allows us to rewrite the convolution in as
\begin{align}
    \label{apx-eq: saddle point integral}
    p(\omega) &\propto \int |y|^{-\beta} \exp \left(-\frac{\lambda}{2}\Psi(y, \omega) \right) \, \mathrm{d} y \\
    \Psi(y, \omega) &:= (\omega-y)^2 \nonumber \,,
\end{align}
where $\lambda = {1}/{\sigma^2} = {2\gamma}/{\epsilon^2}$ denotes the precision of the Gaussian in the convolution.

To apply saddlepoint approximation, we must first verify that $\Psi(y,\omega)$ is a convex function. This is straightforward: setting $d\Psi/dy = 0$ yields the minimizer $y_{\text{min}} = \omega$, which is unique since $d^2\Psi/dy^2 = 2 > 0$. We also restrict the discussion to $\omega \neq 0$, which is not a limiting assumption when analyzing the tails of $p(\omega)$. In practice, we focus on the regions where $|\omega| > \omega_{1}$.

Then, we solve \Cref{apx-eq: saddle point integral} with the change of variable $z = \sqrt{\lambda}(y-\omega)$. This yields 
\begin{align}
    p(\omega)  &\propto \int \frac{1}{\sqrt{\lambda}} \left|\omega + \frac{z}{\sqrt{\lambda}}\right|^{-\beta} \exp \left(- \frac{z^2}{2} \right) \, \mathrm{d} z \nonumber \\
    \label{apx-eq: laplace approx to expand}
    &= \frac{|\omega|^{-\beta}}{\sqrt{\lambda}}\int \left|1 + \frac{z}{\sqrt{\lambda} \omega}\right|^{-\beta} \exp \left(- \frac{z^2}{2} \right)  \, \mathrm{d} z \,.
\end{align}
We can perform a Taylor expansion to the integrand in \Cref{apx-eq: laplace approx to expand} as long as $r(z) := {|z|} / {\sqrt{\lambda}\,|\omega|}  \ll 1$. Since the Gaussian factor localizes $z$ to values $|z| = O(1)$, we have $r(z) \sim 1 /\sqrt{\lambda} |\omega| \ll 1$. This condition is equivalent to
\begin{equation*}
    \lambda \gg \frac{1}{|\omega|^2} \quad \iff \quad \omega \gg \frac{\epsilon}{\sqrt{2\gamma}} \,.
\end{equation*}
Notice that as $|\omega|$ increases, constraint on the shape of the Gaussian becomes less stringent in the tails: the Gaussian does not need to be as sharply peaked to keep the integral localized around $\omega$. Performing a Taylor expansion, we get, up to the second order
\begin{align}
    \label{apx-eq: taylor expansion to solve}
    p(\omega)  &\propto \frac{|\omega|^{-\beta}}{\sqrt{\lambda}}\int \left[ 1 - \frac{\beta z}{\sqrt{\lambda} \omega} + \frac{\beta(\beta +1)}{2} \frac{z^2}{\lambda \omega^2} + O\left( \frac{1}{(\sqrt{\lambda} \omega)^3} \right)\right] \exp \left(- \frac{z^2}{2} \right)  \, \mathrm{d} z \,.
\end{align}
Odd terms in \Cref{apx-eq: taylor expansion to solve} vanish when integrating and, finally, we obtain the approximation of Eq. 14 (main text).

Our derivation is consistent with power-law tails of $q$-Gaussian distributions discussed in the literature of superstatistics on power grids \cite{schaefer2018non}. Matching parameters, typically a $q$-Gaussian has power-law tails $p(\omega) \sim |\omega|^{-2/(q-1)}$ \cite{touchette2005asymptotics}. The tail exponent $\beta$ of $\varphi(P)$ and the parameter $q$ are therefore related by $q = 1 + {2}/{\beta}$. The limit case of $p(\omega)$ Gaussian is obtained as $q \to 1$, or equivalently, $\beta \to + \infty$.

\section{Frequency Autocorrelation}
\label{apx-sec: Frequency Autocorrelation}

We initially derive a closed-form expression for the stationary frequency autocorrelation at short time lags. Specifically, we focus on $\Delta t \leq \qty{20}{min}$ of the stochastic frequency trajectory, where $\Delta t = t - t_0$ denotes the time lag from an initial time $t_0$ (conventionally $t_0 = 0$). In this regime, the superstatistical assumption treats $P(t)$ as approximately constant, i.e., we might consider $P(t) = P$ over intervals during which the system relaxes. While $\qty{20}{min}$ is not substantially smaller than the characteristic timescale of power variations, e.g., market intervals of $\qty{30}{min}$, the time range choice appears to be empirically justified: the autocorrelation exhibits a stationary exponential decay over this range. This observation might suggest that variations in $P(t)$ can still be regarded as small over these time scales. Under this approximation, the evolution of $\omega(t)$ is governed by effectively stationary coefficients, allowing a closed-form analytical expression for the autocorrelation to be derived.

For longer lags $\Delta t$, the stochastic frequency trajectories correlate with the slow power $P(t)$, violating the assumption $P(t)=P$. Beyond this point, our derivation no longer applies. Here, the autocorrelation profile changes due to the influence of $P(t)$ on $\omega(t)$. We observe a power-law decay of the correlation at long lags and discuss its relation with fractional noise \cite{mandelbrot1968fractional}.

Empirically, to compute and plot the autocorrelation in Fig. 2{H} (main text), we aggregate frequency time series over 4-day batches and analyze data with lags $\Delta t$ up to $\qty{2}{h}$ ($\Delta t \leq \qty{20}{min}$ are short lags, and $\qty{20}{min} \leq \Delta t \leq \qty{2}{h}$ are long lags). This procedure prevents spurious finite-size effects and ensures the robustness of our analysis. We show the full autocorrelation profile in \Cref{apx-fig: autocorrelation}A.

\subsection{Analytical Derivation of the Autocorrelation at Short Lags}

First, we tackle the calculation of the conditional expectation 
\begin{align}
\label{apx-eq: conditional expectation omega}
m(\Delta t \, | \, \omega(0),P) &:= \mathbb{E} \left[ \omega(\Delta t) \, | \, \omega(0), P \right] \\
&= \int_{-\infty}^{+\infty} \omega(\Delta t) \, p(\omega(\Delta t)\, | \, \omega(0), P) \, \mathrm{d} \omega(\Delta t) \nonumber \,,
\end{align}
where $p(\omega(\Delta t) \, | \, \omega(0), P)$ is the probability of observing $\omega(\Delta t)$ given the initial condition $\omega(0)$ and the slow power dispatch $P$. We denote as $\omega(\Delta t)$ the dependence of the frequency on $\Delta t$, although one should keep in mind that its trajectories are stochastic as per Eq. 1 (main text).
Calculating \Cref{apx-eq: conditional expectation omega}
has one main challenge. Boundary crossings between control regions may occur; for example, $\omega(0)$ might lie in $\Omega_1$ while $\omega(\Delta t)$ lies in $\Omega_2$. Therefore, one should take into account all initial conditions $\omega(0)$ giving rise to all potential crossings between control regions. To alleviate this burden, we approximate the conditional expectation as 
\begin{equation}
    m(\Delta t \, | \, \omega(0), P)  \approx \widehat{m}(\Delta t \, | \, \omega(0), P)\,,
\end{equation}
where $\widehat{m}(\Delta t; \omega(0), P)$ is computed separately for $\Omega_1$ and $\Omega_2$ by neglecting boundary crossings. In principle, this approximation is expected to break down for short lags $\Delta t$ when the initial condition $\omega(0)$ lies close to the boundaries $\pm \omega_{1}$. Still, we find that it works well in practice when fitting empirical data.
Additionally, we neglect the deadband, since most of the recorded frequency data lie outside it. Formally, we consider the control function $H(\omega)$ to be approximated by two affine functions in $\Omega_1$ and $\Omega_2$, and set $\omega_0 = 0$ in \Crefrange{apx-eq: control region 0}{apx-eq: control region 1}.

Second, using $\widehat{m}(\Delta t; \omega(0), P)$, we approximate the second moment of $\omega(\Delta t)$ using the law of total expectation:
\begin{align}
\label{apx-eq: double expectaction}
\mathbb{E}\left[\omega(0) \omega(\Delta t) \, | \, P\right] 
&= \mathbb{E} \left[ \mathbb{E} \left[\omega(0) \omega(\Delta t) \, | \, \omega(0), P \right] \, | \, P \right] \\
&\approx \mathbb{E} \left[ \omega(0)  \widehat{m}(\Delta t \, | \, \omega(0), P) \, | \, P \right] \nonumber  \\
\label{apx-eq: integral second moment}
&= \int_{-\infty}^{+\infty} \omega(0) \, \widehat{m}(\Delta t \, | \, \omega(0), P) f (\omega(0) \, | \, P) \, \mathrm{d}  \omega(0) \,.
\end{align}
In \Cref{apx-eq: double expectaction}, the inner expectation is taken with respect to the transition probability $p(\omega(\Delta t) \, | \, \omega(0), P)$ while the outer one is with respect to $f (\omega(0) \, | \, P)$. Since $\widehat{m}(\Delta t; \omega(0), P)$ has separate expressions for the control regimes, the integral in \Cref{apx-eq: integral second moment} can be split into a sum over $\Omega_1$ and $\Omega_2$.

We calculate $\widehat{m}(\Delta t \, | \, \omega(0), P)$. To simplify the notation, we introduce two indices: $r = 1,2$ to label the control regions, and $\sigma = \sign{\omega(0)}$ to indicate the sign of $\omega(0)$ and, in turn, the branch (positive or negative) of $\Omega_1$ and $\Omega_2$. Using these indices, we define
\begin{alignat}{2}
\label{apx-eq: division of omega conditional exp}
\Omega_{r}(\sigma) &:= \begin{cases}
     \Omega_{{1}}(1) \quad& r=1, \sigma = 1\\
     \Omega_{{1}}(-1) \quad& r=1, \sigma = -1\\
     \Omega_{{2}}(1) \quad& r=2, \sigma = 1\\
     \Omega_{{2}}(-1) \quad& r=2, \sigma = -1\,.
\end{cases}
\end{alignat}
With \Cref{apx-eq: division of omega conditional exp}, we introduce
\begin{equation}
    \label{apx-eq: conditional expectation control branch}
    \widehat{m}_{r\sigma}(\Delta t \, | \, \omega(0), P) := \mathbb{E} \left[ \omega(\Delta t) \, | \, \omega(0), P, \, \omega(s) \in \Omega_{r}(\sigma) \, \forall s \in [0,\Delta t] \right] 
\end{equation}
that separates $\widehat{m}(\Delta t \, | \, \omega(0), P)$ over the four signed control branches in \Cref{apx-eq: division of omega conditional exp} by taking trajectories $\omega(s)$ that for $s \in [0,\Delta t]$ remain in the control branch where they are initially located.

The rewriting in \Cref{apx-eq: conditional expectation control branch} allows us to carry out calculations independently for each control branch as Eq. 2 (main text) reduces to an Ornstein-Uhlenbeck process. The difference between $\Omega_1$ and $\Omega_2$ lies only in the damping coefficients $\gamma_1$, $\gamma_2$ and in the contribution of $\pm \omega_0$, $\pm \omega_1$. Conveniently, we rewrite Eq. 2 (main text) in its Itô differential form
\begin{equation}
    \label{apx-eq: ito diff equation}
    \mathrm{d} \omega(s) = - \gamma_r (\omega(s) -  \Lambda_r (\sigma, P) ) \mathrm{d} s + \epsilon \, \mathrm{d} W(s) \,,
\end{equation}
where $s \in [0,\Delta t]$, $W(s)$ is a Wiener process whose distributional derivative corresponds to the Gaussian white noise $\xi$, and $\gamma_r$ and $\Lambda_r$ denote either $\gamma_1, \Lambda_1$ or $\gamma_2, \Lambda_2$, depending on the control region. $\Lambda_1$ and $\Lambda_2$ are defined in Eq. 11-12 (main text). Integrating \Cref{apx-eq: ito diff equation} between $0$ and $\Delta t$, and $\omega(0)$ and $\omega(\Delta t)$, yields
\begin{equation*}
    \omega(\Delta t) = \exp(-\gamma_r \Delta t) \, \omega(0) +\gamma_r \Lambda_r (\sigma, P) \int_0^{\Delta t} \exp(-\gamma_r (\Delta t-s)) \,  \, \mathrm{d} s + \epsilon \int_0^{\Delta t} \exp({-\gamma_r (\Delta t-s)}) \, \mathrm{d} W(s) \,
\end{equation*}
from which the approximation of the conditional expectation follows:
\begin{align}
    \label{apx-eq: evolution conditional mean step}
    \widehat{m}_{r\sigma}(\Delta t \, | \, \omega(0), P) &= \exp(-\gamma_r \Delta t) \, \omega(0) + \gamma_r \Lambda_r (\sigma, P) \int_0^{\Delta t}  \exp(-\gamma_r (\Delta t-s)) \, \mathrm{d} s \\
    \label{apx-eq: evolution conditional mean}
    &= \Lambda_r (\sigma, P)+ (\omega(0)-\Lambda_r(\sigma, P))\exp(-\gamma_r \Delta t) \,.
\end{align}
In \Cref{apx-eq: evolution conditional mean step}, the Itô integral of the noise vanishes because the Wiener process has a zero conditional mean. \Cref{apx-eq: evolution conditional mean} is a standard result for an Ornstein-Uhlenbeck process where the conditional mean relaxes exponentially towards its mean $\Lambda_r$ with characteristic decay rate $\gamma_r$.

We calculate the second moment. This amounts to evaluating the integral in \Cref{apx-eq: integral second moment} separately for $\sigma = \pm 1$, $r=1,2$.

We substitute \Cref{apx-eq: evolution conditional mean} into \Cref{apx-eq: integral second moment}, and using the quasi-static distribution in Eq. 10 (main text), we get
\begin{align}
    \label{apx-eq: second moment integral}
    \mathbb{E} \left[\omega(0) \omega(\Delta t) \, | \, P \right] &\approx \frac{1}{Z} \sum_{\sigma = \pm 1} \int_{\Omega_{r}(\sigma)} C_r \, \omega(0) [ \Lambda_r + (\omega(0) - \Lambda_r(\sigma,P)) \exp(-\gamma_r \Delta t) ] \exp \left[ - \frac{\lambda_r}{2} \left( \omega(0) - \Lambda_r(\sigma,P) \right)^2  \right] \,\mathrm{d} \omega(0) \,.
\end{align}
Here, we unclutter notation with the precision $\lambda_r := 2{\gamma_r}/{\epsilon^2}$. Also, recall that $Z = Z(P)$, $C_r = C_r(\sigma,P)$, with $r=1,2$. The sum over $\sigma = \pm 1$ merges contributions coming from the two signed branches of $\Omega_1$ and $\Omega_2$ to get only a dependence on $r$ on the left-hand side of the equation. Once again, the approximation arises from using $\widehat{m}(\Delta t \, | \, \omega(0), P)$ instead of ${m}(\Delta t \, | \, \omega(0), P)$.

Note that the dependence on $\Delta t$ in \Cref{apx-eq: second moment integral} comes only from specific terms of the integrand. All other terms yield constants that are subsequently normalized in the autocorrelation. Therefore, we simplify \Cref{apx-eq: second moment integral} as
\begin{align}
    \label{apx-eq: second moment integral 2}
    \mathbb{E} \left[\omega(0) \omega(\Delta t) \, | \, P \right] &\approx\frac{1}{Z} \exp(-\gamma_r \Delta t)  \sum_{\sigma = \pm 1}\int_{\Omega_{r}(\sigma)}  C_r \,  \omega(0)  (\omega(0) - \Lambda_r(\sigma,P)) \exp \left[ - \frac{\lambda_r}{2} \left( \omega(0) - \Lambda_r(\sigma, P) \right)^2  \right] \, \mathrm{d} \omega(0)  + \text{const.}
\end{align}
The integral in \Cref{apx-eq: second moment integral 2} involves calculating the moments of a truncated Gaussian distribution which could, in principle, be expressed exactly. Performing these computations is not strictly necessary, as the profile of the autocorrelation can still be found while keeping the integrals unsolved.

Although a closed-form expression for $\varphi(P)$ is not available, since we have neglected boundary crossings, the terms over $\Omega_1$ and $\Omega_2$ can be integrated separately. This gives rise to two distinct exponentially decaying contributions. In particular, we observe that \Cref{apx-eq: second moment integral 2} can be conveniently rewritten by multiplying and dividing it by the region-specific normalizations $Z_{r}(P)$ of \Crefrange{apx-eq: normalization 1}{apx-eq: normalization 2}. This gives
\begin{align}
        \mathbb{E} \left[\omega(0) \omega(\Delta t) \, | \, P \right] &\approx\frac{1}{Z} \exp(-\gamma_r \Delta t)  \int_{\Omega_{r}} \omega(0)  [\omega(0) - \Lambda_r(\sign{\omega(0)},P)] \, f_{r}(\omega(0) \, | \, P) \, \mathrm{d} \omega(0)  + \text{const.} \nonumber \\
       \label{apx-eq: expected second moment} 
        &= \exp(-\gamma_r {\Delta t})  \, w_{r}(P) \, \mathbb{E} \big[\omega(0)[\omega(0) - \Lambda_r(\sign{\omega(0)}, P)] \, | \, P\big]  + \text{const.}
\end{align}
Here, $w_{r}(P) := {Z_r}(P)/{Z(P)}$, and the conditional expectations are computed with respect to the truncated Gaussian distributions in Eq. 10 (main text). These distributions are defined over both signed branches of $\Omega_1$ and $\Omega_2$, and therefore marginalize $\sigma = \pm 1$. The factor $w_{r}(P)$ represents the probability that the power grid resides in the $\Omega_1$ or $\Omega_2$ region for a given $P$.

Integrating \Cref{apx-eq: expected second moment} against $\varphi(P)$ gives
\begin{align}
    \label{apx-eq: second moment final apx}
    \mathbb{E} [\omega(0) \omega(\Delta t) ] &\approx \sum_{r=1,2} I_r \exp(-\gamma_r\Delta t) +\text{const.} \\
    I_{r} &:= \frac{1}{Z_{\varphi}} \int w_{r}(P) \, \mathbb{E} \big[\omega(0)[\omega(0) - \Lambda_r(\sign{\omega(0)}, P)] \, | \, P\big] \, \varphi(P) \, \mathrm{d} P \,, \nonumber
\end{align}
where $Z_{\varphi}$ is the normalization constant of $\varphi(P)$. \Cref{apx-eq: second moment final apx} shows that the autocorrelation is well described by the weighted sum of two exponentials at short lags. Indeed,
\begin{equation*}
C(\Delta t) := \frac{\text{Cov}[\omega(0), \omega(\Delta t)]}{\text{Var}[\omega(0)]} \,
\end{equation*}
which, using \Cref{apx-eq: second moment final apx}, can be fitted by
\begin{equation}
\label{apx-eq: biexp decay}
    C(\Delta t) = A_1 \exp(-\gamma_1 \Delta t) + A_2 \exp(-\gamma_2 \Delta t) \,
\end{equation}
with the normalization condition $C(0)=1$ requiring $A_1 + A_2 = 1$, i.e., $A_1 = A$ and $A_2 = 1 - A$.

\subsection{Empirical Fit}

We fit \Cref{apx-eq: biexp decay} with \texttt{curve\_fit()} from SciPy. For comparison, we also fit our data with a single-exponential decay $C(\Delta t) = \exp(-\gamma \Delta t)$, which one obtains by assuming linear control with a single damping coefficient $\gamma$.

Results for the double-exponential decay fit are as follows:
\begin{alignat}{2}
    %%%%%%
    \nonumber
    \text{United Kingdom:} \qquad\qquad  \gamma_1 &= \dots
    \qquad\qquad\qquad\qquad\qquad\quad\quad\;\; \gamma_2 &&= (\num{1.45 \pm 0.06} ) \cdot 10^{-4}\,\unit{{ rad/s}^2} \\
    %%%%%%
    \nonumber
    \quad A_1 &= \num{2} \cdot 10^{-16} \approx 0  \qquad\qquad\qquad\qquad A_2 &&\approx 1 \\
    %%%%%%
    \nonumber
    \text{SSR} &= \num{0.0790814} \\
    %%%%%%
    \label{apx-eq: autocorr sa fit}
    \text{South Africa:} \qquad\qquad  \gamma_1 &= (\num{8.6 \pm 0.2}) \cdot 10^{-4} \, \unit{ rad/s}^2 \qquad\quad\;\; \gamma_2 &&= ( \num{1.15 \pm 0.02} ) \cdot 10^{-4}\unit{ rad/s}^2 \\
    %%%%%%
    \nonumber
    \quad A_1 &= \num{0.446 \pm 0.007} \qquad\qquad\qquad\qquad A_2 &&= \num{0.554 \pm 0.007} \\ 
    %%%%%%
    \nonumber
    \text{SSR} & = \num{0.0227342} \,.
\end{alignat}
Results for the single-exponential decay fit are as follows:
\begin{alignat}{2}
    \label{apx-eq: autocorr uk fit}
    \text{United Kingdom:} \quad \gamma &= (\num{1.452 \pm 0.005} ) \cdot 10^{-4}\,\unit{{ rad/s}^2} \quad \text{SSR} &&= \num{0.0790815}\\
    %%%%%%
    \nonumber
    \text{South Africa:} \quad \gamma &= (\num{2.28 \pm 0.04} ) \cdot 10^{-4} \,\unit{{ rad/s}^2} \quad \quad \text{SSR} &&= \num{1.1892102} \,.
\end{alignat}

We find that a single-exponential decay accurately fits the frequency autocorrelation in the United Kingdom. In fact, the single coefficient $\gamma$ in \Cref{apx-eq: autocorr uk fit} is effectively identical to $\gamma_2$ found by fitting the double-decay ansatz. Numerically, in the double-decay fit, all of the weight is given to one parameter, $\gamma_2$, rendering $\gamma_1$ idle. The Sum of Squared Residuals (SSR) of the two fits is also identical. In South Africa, the double-decay ansatz fits the empirical autocorrelation much better than the single-decay ansatz.

\subsection{Autocorrelation at Long Lags}

For long lags $\Delta t \geq \qty{20}{min}$, the autocorrelation exhibits a profile that deviates from the exponential behavior described above. We show its profile in \Cref{apx-fig: autocorrelation}B. First, we observe recurrent deviations arising from market transactions. Second, we observe a slow power-law decay. This finding aligns with previous observations \cite{kraljic2023towards}, where the power-law decay is explained through fractional noise \cite{mandelbrot1968fractional}. This type of noise causes the correlation to decay slowly with a power-law dependence on the Hurst exponent $\mathcal{H}$: for a standard Wiener process, $\mathcal{H} = 1/2$; if $\mathcal{H} > 1/2$, the increments are positively correlated; and if $\mathcal{H} < 1/2$, the increments are negatively correlated. In particular, with fractional noise, the correlation scales as $C(\Delta t) \sim {\Delta t}^{2\mathcal{H} - 2}$ \cite{kraljic2023towards}. Fitting the data in a log-log plot, we obtain
\begin{align*}
    \text{United Kingdom:} \quad \mathcal{H} &= \num{0.6975  \pm 0.0007} \\
    \text{South Africa:} \quad \mathcal{H} &= \num{0.6243 \pm 0.0005}
\end{align*}
which agree well with previous observations \cite{kraljic2023towards}.

\begin{figure*}[htpb]
\centering
\includegraphics[width=0.7\textwidth]{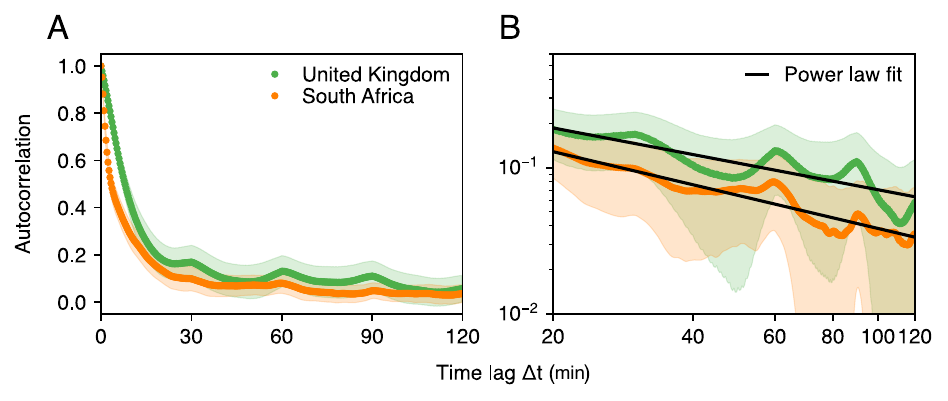}
\caption{Autocorrelation. (A) Autocorrelation at all time lags $\qty{0}{h} \leq \Delta t \leq \qty{2}{h}$. Lighter areas are standard deviations computed over the 4-day batches used to calculate the autocorrelation. (B) Log-log plot of the autocorrelation at long lags, i.e., $\qty{20}{min} \leq \Delta t \leq \qty{2}{h}$. We fit the log-log plot with a power-law ansatz $\sim \Delta t^{a}$, we use $a$ to compute the Hurst exponents as $\mathcal{H} = (a+2)/2$.}
\label{apx-fig: autocorrelation}
\end{figure*}

\section{Superstatics Validation}

\subsection{Deadband Escape Time}

Since there is no damping in the deadband, at a finer scale, the relaxation time formula $\Delta t = 1/\gamma$ breaks down. We obtain a finer estimate of the characteristic time spent by $\omega(t)$ in the deadband to validate superstatistics. To do so, we employ mean first-passage time theory \cite{redner2001guide}. Our goal is to calculate the mean first-passage exit time $\tau_{\text{exit}}$, which is the average time taken by a stochastic trajectory starting at $\omega(0)$ in the deadband, i.e., $\omega(0) \in \Omega_0$, to cross either $\pm \omega_{0}$.

The deadband is not damped, hence $H(\omega) = 0$ for $\omega \in \Omega_0$. Additionally, for short time scales, the slow power $P(t)$ can be kept constant. The Langevin equation in Eq. 1 (main text) then becomes, for $\omega \in \Omega_0$,
\begin{equation}
    \label{apx-eq: brownian motion}
    \frac{\mathrm{d} \omega}{\mathrm{d} t} = P + \epsilon \xi \,.
\end{equation}
In particular, we can take $P = 0$ as this yields pure Brownian motion in \Cref{apx-eq: brownian motion}, where there is no drift and the mean first-passage exit time is maximized. Using the backward Kolmogorov approach \cite{redner2001guide} on \Cref{apx-eq: brownian motion}, we can write an ordinary differential equation for the mean first-passage time $\tau_{\text{exit}}(\omega(0))$:
\begin{equation}
    \label{apx-eq: exit time ode}
    \frac{\epsilon^2}{2}\frac{\mathrm{d} ^2 \tau_{\text{exit}}}{\mathrm{d}  \omega(0)^2} = -1 \,,
\end{equation}
with boundary conditions being $\tau_{\text{exit}}(\pm \omega_{0}) = 0$. Integrating \Cref{apx-eq: exit time ode} twice gives
\begin{equation*}
    \tau_{\text{exit}}(\omega(0)) = \frac{\omega_0^2-\omega(0)^2}{\epsilon^2} \,.
\end{equation*}
We evaluate it at $\omega(0) = 0$, where it is maximized, as doing so gives us a worst-case estimate to validated superstatistics. In particular, we get
\begin{equation}
    \label{apx-eq: escape time final}
   \tau_{\text{exit}} = \frac{\omega_0^2}{\epsilon^2} \,.
\end{equation} 

Setting $\omega_0 = 2 \pi \cdot \qty{0.015}{rad/s}$ as per \Crefrange{apx-eq: deadband width uk}{apx-eq: deadband width sa} and using the values of $\epsilon$ extracted via KR in \Crefrange{apx-eq: epsilon uk kr}{apx-eq: epsilon sa kr}, we obtain (rounding at the first decimal)
\begin{align}
    \label{apx-eq: tay escape uk}
    \text{United Kingdom:} \quad \tau_{\text{exit}} &= \qty{2.3}{min} \\
    \label{apx-eq: tay escape sa}
    \text{South Africa:} \quad \tau_{\text{exit}} &= \qty{1.3}{min} \,.
\end{align}
These values are compatible with the control relaxation times $\tau$ reported in the main text.

\subsection{Kernel Regression for Grid Parameters}

To calculate the control relaxation rates, we estimate the parameters governing the frequency dynamics, $\gamma$ and $\epsilon$. As discussed in the main text, obtaining reliable estimates of these parameters with KR requires using continuous trajectories $\omega(t)$. This constraint makes it difficult to extract region-specific damping coefficients $\gamma_1$ and $\gamma_2$, since doing so would require selecting trajectories that do not cross control boundaries. Such trajectories do not contain sufficient data to estimate the damping coefficients robustly. Consequently, for validation purposes only, we simplify $H(\omega)$ by assuming linear damping with a single parameter, $\gamma$.

First, we detrend the frequency measurements using a Gaussian kernel with standard deviation $\sigma = 60$ s \cite{drewnick2025analyzing,gorjao2020data}, yielding $\omega_{\text{det}}(t) = \omega(t) - \omega_{\text{filter}}(t)$. We show a snapshot of the filtering process in \Cref{apx-fig: kramers-moyal-panel}A covering $\qty{15}{min}$ of data. For the filtered signal, the slow power drift can be neglected, and the dynamics is well described by
\begin{equation}
    \label{apx-eq: langevin kr}
    \frac{\mathrm{d} \omega_{\text{det}}}{\mathrm{d} t} = \gamma \omega_{\text{det}} + \epsilon  \xi \,.
\end{equation}
\Cref{apx-eq: langevin kr} yields the Fokker--Planck equation
\begin{equation*}
    \frac{\partial p}{\partial t} = -\gamma \frac{\partial}{\partial \omega_{\text{det}}}( \omega_{\text{det}} p) + \frac{\epsilon^2}{2} \frac{\partial^2 p}{\partial \omega_{\text{det}}^2} \,,
\end{equation*}
which can be written equivalently by making explicit its first two Kramers--Moyal coefficients \cite{risken1989fokker},  $\mathcal{D}_1 (\omega_{\text{det}})$ and $\mathcal{D}_2 (\omega_{\text{det}})$, as 
\begin{equation}
    \label{apx-eq: fp kramers}
    \frac{\partial p}{\partial t} = -\frac{\partial}{\partial \omega_{\text{det}}}( \mathcal{D}_1 (\omega_{\text{det}}) p) + \frac{\partial^2}{\partial \omega_{\text{det}}^2} (D_2 (\omega_{\text{det}}) p) \,.
\end{equation}
In \Cref{apx-eq: fp kramers}, $\mathcal{D}_1(\omega_{\text{det}}) = -\gamma \omega_{\text{det}}$ is the drift coefficient, while  $\mathcal{D}_2(\omega_{\text{det}}) = \epsilon^2 / 2$ is the diffusion coefficient.

We estimate the Kramers--Moyal coefficients using the \texttt{kramersmoyal} software package \cite{gorjao2019kramersmoyal}, which allows us to efficiently compute the estimators \cite{lamoroux2009kernel}
\begin{align}
    \label{apx-eq: km 1}
    {D}_1(\omega) &= \frac{1}{\Delta t_{\mathrm{PMU}}} \left\langle \omega_{\text{det}}(t + \Delta t_{\mathrm{PMU}}) - \omega_{\text{det}}(t) \, | \, \omega_{\text{det}}(t) = \omega \right\rangle \\
     \label{apx-eq: km 2}
    {D}_2(\omega) &= \frac{1}{2\Delta t_{\mathrm{PMU}}} \left\langle \left( \omega_{\text{det}}(t + \Delta t_{\mathrm{PMU}}) - \omega_{\text{det}}(t) \right)^2 \, | \, \omega_{\text{det}}(t) = \omega \right\rangle \,
\end{align}
for $\mathcal{D}_1(\omega)$ and $\mathcal{D}_2(\omega)$, respectively. Here, $
\left\langle \cdot \, | \, \omega_{\text{det}}(t) = \omega \right\rangle
$
denotes a kernel-weighted conditional average at $\omega_{\text{det}}(t) = \omega$. Following the literature \cite{gorjao2020data}, we choose a scaled Epanechnikov kernel $K_h(x) = 3 [1 - {(x/h)}^2 ]/4h$, $|x| < h$ \cite{gorjao2023stochastic} with $h = \qty{0.1}{rad/s}$.

As shown in \Cref{apx-fig: kramers-moyal-panel}C, the drift estimator ${D}_1(\omega)$ exhibits a negative slope in both South Africa and the United Kingdom. In the United Kingdom, the slope is larger, reflecting tighter grid control. To obtain reliable estimates of $\gamma$, we fit the data $\omega_{\text{det}}$ populating the central portion of the distribution $p(\omega_{\text{det}})$, where KR is stable. More specifically, we use linear least squares on those values of $D_1(\omega)$ populated by at $\num{50}$ thousand data points in the conditional averages of \Crefrange{apx-eq: km 1}{apx-eq: km 2}. These vales lie in the central darker region of the histogram in \Cref{apx-fig: kramers-moyal-panel}B. This procedure returns
\begin{align}
    \label{apx-eq: damping uk km}
    \text{United Kingdom:} \quad \gamma &= (\num{4.10 \pm 0.03} ) \cdot 10^{-3}\,\unit{ rad/s} \\
    \label{apx-eq: damping sa km}
    \text{South Africa:} \quad \gamma &= (\num{1.838 \pm 0.009}) \cdot 10^{-3}\unit{ rad/s} \,,
\end{align}
which we use to calculate the values of $\tau = 1 / \gamma$ reported in the manuscript.

We show the estimated $D_2(\omega)$ in \Cref{apx-fig: kramers-moyal-panel}D. Its value shows a slight inflection as $\omega_{\text{det}}$ departs from zero. Similar to previous findings \cite{gorjao2020data,drewnick2025analyzing}, we take $\epsilon$ extracted to be
\begin{equation}
    \label{apx-eq: epsilon from diff}
    \epsilon = \sqrt{2 D_2(\omega_{\det} = 0)} \,.
\end{equation}
This is close to the minimum of $D_2(\omega)$ and, consequently, it yields a good worst-case estimate of the escape time $\tau_{\text{exit}}$ in \Cref{apx-eq: escape time final}. Empirically, we obtain (rounding at the first significant decimal)
\begin{align}
    \label{apx-eq: epsilon uk kr}
    \text{United Kingdom:} \quad \epsilon &= \qty{0.008}{{ rad}\text{/s}^{3/2}} \\
    \label{apx-eq: epsilon sa kr}
    \text{South Africa:} \quad \epsilon &= \qty{0.01}{{ rad}\text{/s}^{3/2}} \,.
\end{align}

All our estimates are consistent with those found for several grids worldwide  \cite{gorjao2020data,drewnick2025analyzing,schaefer2018non,kruse2023physics,oberhofer2023non}. Additionally, \Crefrange{apx-eq: damping uk km}{apx-eq: damping sa km} are compatible with the damping coefficient found by fitting the frequencies autocorrelation in \Cref{apx-sec: Frequency Autocorrelation}.

\begin{figure*}[htpb]
\centering
\includegraphics[width=0.7\textwidth]{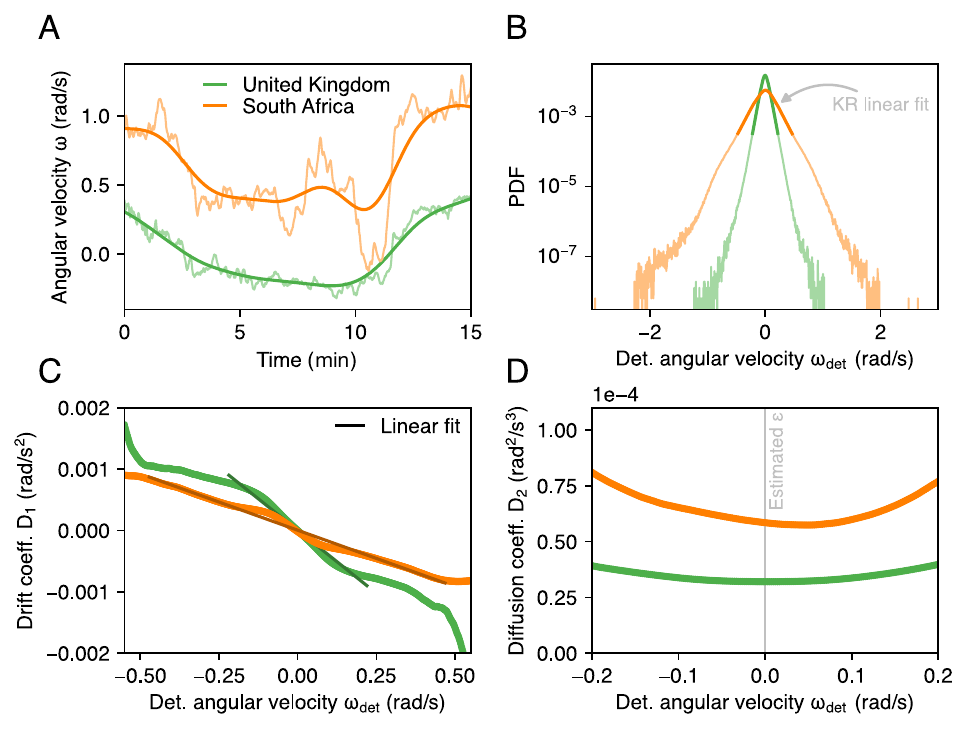}
\caption{KR to estimate control coefficients and noise amplitude. (A) Gaussian filtering of the frequency time series. The thicker and smoother line is $\omega_{\text{filter}}(t)$ obtained by applying a Gaussian smoothing on $\omega(t)$, which is drawn with a lighter and thinner line. (B) PDF of $\omega_{\text{det}}(t)$. The central regions, denoted by a thicker line, are where the histogram bins are populated by more than 50k data points. We fit the drift in panel C over such data. (C) Drift coefficient of \Cref{apx-eq: km 1} against $\omega_{\text{det}}$. (D) Diffusion coefficient of \Cref{apx-eq: km 2} against $\omega_{\text{det}}$. The value of $D_2$ at $\omega_{\text{det}} = 0$ is used to calculate $\epsilon$ as per \Cref{apx-eq: epsilon from diff}.}
\label{apx-fig: kramers-moyal-panel}
\end{figure*}

\section{Analytical Derivations of Phase Angle Difference Distribution}
\label{apx-sec: Analytical Derivations of Phase Angle Difference Distribution}

\subsection{Stochastic Damped-Driven Harmonic Oscillator}

We derive Eq. 3-4 (main text) from the swing equations in \Crefrange{apx-eq: swing equation 1}{apx-eq: swing equation 2}. We assume a reduced grid with $N = 2$ nodes and consider the power outputs $P_1$ and $P_2$ to be constant in time. Additionally, we make two constructive assumptions: $\gamma = D_1 / M_1 = D_2 / M_2$ and $\kappa = K_{12} / M_1 = K_{21} / M_2$ \cite{schafer2017escape,manik2014supply}. This allows us to rewrite the swing equations as second-order stochastic differential equations:
\begin{align}
    \label{apx-eq: swing equations 2 nodes 1}
    \frac{\mathrm{d} ^2 \theta_1}{\mathrm{d} t^2} &= -\gamma \frac{\mathrm{d} \theta_1}{dt} + \frac{P_{1}}{{M_1}} + \kappa \sin (\theta_2 -  \theta_1) + \frac{\epsilon_1}{M_1} \xi_1\\
    \label{apx-eq: swing equations 2 nodes 2}
    \frac{\mathrm{d} ^2 \theta_2}{\mathrm{d} t^2} &= -\gamma \frac{\mathrm{d} \theta_2}{\mathrm{d} t} + \frac{P_{2}}{{M_2}} + \kappa \sin (\theta_1 -  \theta_2) +\frac{\epsilon_2}{M_2} \xi_2 \,.
\end{align}
Denoting the difference of the two phase angles as $x := \theta_1 - \theta_2 \in (-2 \pi, 2 \pi)$, we subtract \Cref{apx-eq: swing equations 2 nodes 1} from \Cref{apx-eq: swing equations 2 nodes 2} and readily obtain Eqs. 3-4 (main text).

We are only left with characterizing the noise. If $\xi_1$ and $\xi_2$ are i.i.d. Gaussian random variables with amplitudes $\epsilon_1$ and $\epsilon_2$, then $\xi$ in Eq. 3 (main text) is also Gaussian with amplitude
\begin{align*}
    \epsilon = \sqrt{\frac{\epsilon_1^2}{M_1} + \frac{\epsilon_2^2}{M_2}}
\end{align*}

Eqs. 3-4 (main text) describes the motion of a stochastic damped-driven harmonic oscillator in a tilted washboard potential and serves as the pivotal equation of our reduced model for phase-angle differences. The tilted washboard potential
\begin{equation*}
        V(x) := - \delta x - 2 \kappa \cos x 
\end{equation*}
governs the system's stability, with the stationary states of the phase angle difference $x$ distributed around its minima. We illustrate $V(x)$ for different parameter choices in \Cref{apx-fig: tilted washboard panel}A.

The potential's tilt depends on the mechanical power difference $\delta$ between the two generators. Because of energy conservation, $P_1 + P_2 = 0$ holds. Hence, $\delta = 0$ if and only if $P_1 = P_2 = 0$. Any other value of $P_1$, $P_2$ produces a tilt in the potential.

We also note that $V(x + 2\pi) = V(x) - 2\pi \delta$. In other words, the potential is periodic in $2\pi$, up to a constant shift set by the linear tilt $\delta$. Consequently, configurations of $x(t)$ that differ by integer multiples of $2\pi$ are equivalent, since their evolution is governed by $dV/dx$. Physically, only phase differences enter the exchanged-power (sine) terms of \Crefrange{apx-eq: swing equations 2 nodes 1}{apx-eq: swing equations 2 nodes 2}, and therefore $2\pi$ translations in $x(t)$ do not affect the power exchange. From a practical standpoint, we can restrict our analysis to $x \in [0,2\pi)$.

\subsection{Equilibria}

We characterize the stationary points of $V(x)$ focusing on Eq. 3 (main text) as a dynamical system \cite{manik2014supply}, i.e., setting the noise to zero. For convenience, we define the relative power difference
\begin{equation*}
    \label{eq: relative power exchange}
    \rho := {\delta}/{2 \kappa} \,.
\end{equation*}
Using $\rho$, we classify equilibria as follows.
\begin{enumerate}
    \item If $\rho > 0$ and $|\rho| < 1$, there is a positive power difference between generator $1$ and $2$. The power grid admits stable working points that are the minima of $V(x)$:
    \begin{alignat*}{2}
     x_{\text{min},1} &= \arcsin \rho \qquad\qquad
    x_{\text{min},2} &&= x_{\text{min},1} - 2 \pi \\
    x_{\text{max},1} &= - x_{\text{min},1} + \pi \qquad
    x_{\text{max},2} &&= - x_{\text{min},1} - \pi  \,.
    \end{alignat*}
    \item If $\rho < 0$ and $|\rho| < 1$, the setup is analogous to (1), but with a negative power difference:
    \begin{alignat*}{2}
     x_{\text{min},1} &= \arcsin \rho \qquad\qquad
    x_{\text{min},2} &&= x_{\text{min},1} + 2 \pi  \\
    x_{\text{max},1} &= - x_{\text{min},1} - \pi \qquad
    x_{\text{max},2} &&= - x_{\text{min},1} + \pi  \,.
    \end{alignat*}
    \item If $\rho = 0$, then $\delta = 0$. Physically, this means that both generators balance their own power demand. If this happens, the transmission line between $1$ and $2$ becomes idle. In fact, the stable points of the grid are
    \begin{align*}
        x_{\text{min},1} &= x_{\text{min},2} = 0, 2 \pi\\
        x_{\text{max},1} &= x_{\text{max},2} = \pm \pi \,,
    \end{align*}
    which make the sine term in Eq. 3 (main text) vanish. 
    \item If $\rho = \pm 1$, the power exchanged between $1$ and $2$ is the maximum allowed by the transmission line. Maxima and minima collapse into the saddle points:
    \begin{align*}
            x_{\text{saddle},1} &= \pm \frac{\pi}{2} \\ x_{\text{saddle},2} &= \mp \frac{3\pi}{2} \,.
    \end{align*}
    \item $|\rho| > 1$ would imply a power exchange larger than what is physically allowed by the transmission line. In this case, there is no stable fixed point for the grid to operate, but only runaway solutions.
\end{enumerate}

We show all stability regions where the equilibria in (1)-(5) lie in \Cref{apx-fig: tilted washboard panel}B.

We restrict our discussion to case (1), which is valid during regular operations and physically equivalent to (2).

\begin{figure*}[htpb]
\centering
\includegraphics[width=0.7\textwidth]{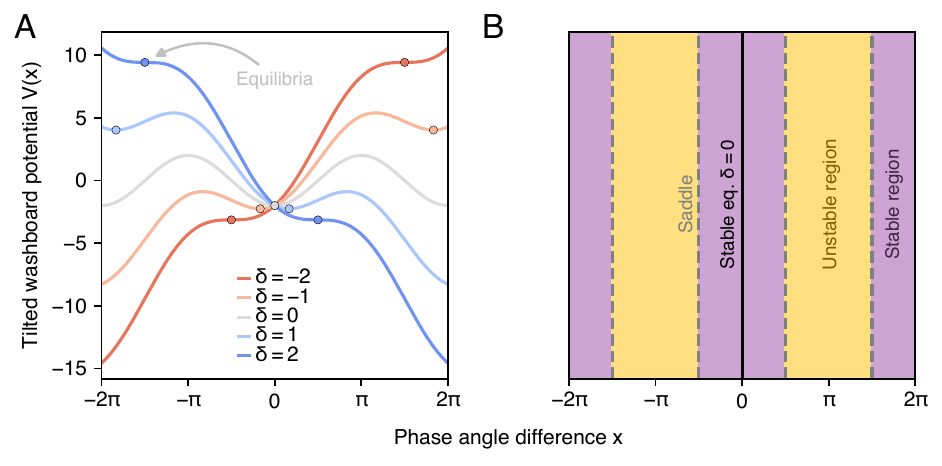}
\caption{Tilted washboard potential. The $ y$-axis is arbitrarily scaled, and we conventionally set $\kappa = 1$. (A) We display $V(x)$ for different values of $\delta$. The circles represent the equilibria of $V(x)$. Notice how, for $\delta \neq 0$, two equilibria arise, one being the translation by $2\pi$ of the other. (B) We plot the stability regions of $V(x)$. Unstable equilibria lie in the purple stable region $\{|x|\leq\pi /2\} \cup \{ |x| \geq 3\pi/2 \}$. Conversely, in the unstable yellow region, $V(x)$ admits only runaway solutions and no equilibria. Saddles between the two regions are denoted by gray dashed lines. When $\delta = 0$, the potential is symmetric, and the only equilibrium within $(-2\pi, 2 \pi)$ is indicated by the black line at $x = 0$.}
\label{apx-fig: tilted washboard panel}
\end{figure*}

\subsection{Solving the Smoluchowski Equation}

We derive the closed-form equilibrium distribution $p(x)$ fitted in the main text. To do so, we assume that the grid operates in the overdamped limit where
\begin{align}
    \label{apx-eq: overdamped limit}
    \gamma \gg \left|\frac{\mathrm{d}^2V}{\mathrm{d} x^2} \bigg|_{x={x_{\text{min}}}}\right| = 2 \kappa \sqrt{1-\rho^2} \,.
\end{align}
In words, the damping coefficient $\gamma$ is large compared to the characteristic frequency of the undamped system. Physically, this implies that after a perturbation, phase differences readjust slowly and monotonically, without oscillating. In the stable operating regime where $|\rho| < 1$, \Cref{apx-eq: overdamped limit} shows that if $|\rho| \to 1$, the transmission line is almost at capacity and cannot change power flow much in response to phase deviations. As a consequence, a smaller damping is sufficient to control phase deviations. Conversely, when $|\rho|$ is small, the transmission line is underloaded and reacts strongly to deviations. To suppress perturbations in this case, $\gamma$ must be larger.

In the high-damping limit, we can neglect the second derivative in Eq. 3 (main text) \cite[Chapter 11.2]{risken1989fokker} and write
\begin{align}
    \label{apx-eq: swing equation difference high friction velocity}
    v &= \frac{\mathrm{d} x}{\mathrm{d} t} \\
    \label{apx-eq: swing equation difference high friction}
    \gamma \frac{\mathrm{d}  x}{\mathrm{d} t} &= -\frac{\mathrm{d} V}{\mathrm{d} x} + \epsilon \xi\,.
\end{align}

The system in \Crefrange{apx-eq: swing equation difference high friction velocity}{apx-eq: swing equation difference high friction} includes the tilt $\delta$. Therefore, the system generally settles into a steady state with a non-zero probability current. In formula, let $W(x,v,t)$ be the probability density of the phase space described by $x$ and $v$. We can define the marginal probability distribution and the probability current
\begin{align*}
p(x,t) &:= \int W(x,v,t) \, \mathrm{d} v \\
J(x,t) &:= \int v \, W(x,v,t) \, \mathrm{d} v \,.
\end{align*}
These two variables satisfy the continuity equation
\begin{equation}
    \label{apx-eq: continuity equation equation}
    \frac{\partial p}{\partial t} + \frac{\partial J}{\partial x} = 0 \,,
\end{equation}
which we couple with the Smoluchowski equation given by \Cref{apx-eq: swing equation difference high friction}: \cite{risken1989fokker}
\begin{equation}
    \label{apx-eq: smoluchowski equation}
    \frac{\partial p}{\partial t}  = \frac{1}{\gamma} \frac{\partial}{\partial x} \left( \frac{\mathrm{d} V}{\mathrm{d} x} + \frac{2}{\lambda} \frac{\partial p}{\partial x} \right) \,,
\end{equation}
where $\lambda := 2\gamma / \epsilon^2$.

A general solution of \Crefrange{apx-eq: continuity equation equation}{apx-eq: smoluchowski equation} is found by imposing the stationarity condition $\partial p / \partial t = 0$ in \Cref{apx-eq: continuity equation equation} to get $J(x) \equiv J_{\text{eq}}$. By combining this condition with \Cref{apx-eq: smoluchowski equation}, we get
\begin{equation*}
    \gamma J_{\text{eq}} = \frac{\mathrm{d} V}{\mathrm{d} x} p - \frac{2}{\lambda} \frac{\partial p }{\partial x} \,.
 \end{equation*}
 Integrating in $x$, we get the general stationary probability $p(x)$ for a non-zero probability current
 \begin{equation}
    \label{apx-eq: solution with current}
     p(x) = \exp{ \left(- \frac{\lambda}{2} V(x)
    \right)} \left[ Z - \frac{\lambda \gamma}{2} J_{\text{eq}}  \int_0^x \exp{ \left( \frac{\lambda}{2} V(x')
    \right)} \, \mathrm{d} x' \right] \,.
 \end{equation}
It is sufficient to require that $p(x)$ is bounded to ensure that it is $2\pi$-periodic \cite{risken1989fokker}. Using periodicity, one can determine the normalization constant $Z$ by imposing normalization, namely,
\begin{equation*}
    \int_0^{2 \pi} p(x) \,\mathrm{d} x = 1 \,,
\end{equation*}
which gives
\begin{equation*}
    Z = \frac{
        1 + \displaystyle\frac{\lambda \gamma}{2} J_{\rm eq} \displaystyle \int_0^{2\pi} 
        \exp\left( - \frac{\lambda}{2} V(x) \right) 
        \left[ \int_0^x \exp \left( \frac{\lambda}{2} V(x') \right) \, \mathrm{d} x' \right] \mathrm{d} x
    }{
        \displaystyle \int_0^{2\pi} \exp \left( - \frac{\lambda}{2} V(x) \right) \, \mathrm{d} x
    } \,.
\end{equation*}

These equations are hard to fit numerically. However, experimental data in SA reveal that phase differences $x$ are tightly localized near a potential minimum well described by $J(x) \equiv 0$. This allows us to set the probability current to zero and fit our data with a simpler distribution. If there is no probability current, the stationary distribution of \Cref{apx-eq: solution with current} reduces to
\begin{equation}
    \label{apx-eq: stationary distribution phase difference}
    p(x) =  \frac{1}{Z}\exp{ \left(- \frac{\lambda}{2} V(x)
    \right)}  \,,
\end{equation}
where $Z$ is the normalization constant 
\begin{equation*}
    Z = \int_0^{2\pi} \exp{ \left( - \frac{\lambda}{2} V(x)
    \right)} \, \mathrm{d} x \,.
\end{equation*}
This simplification greatly reduces the computational burden of running MLE to fit the data.

\section{Maximum-Likelihood Estimation}
\label{apx-sec: mle}

We use MLE to fit the deadband and the tails of the frequency distribution in the United Kingdom, as well as the phase-difference distribution in South Africa. That is, we optimize the fit parameters, subject to constraints specific to the variable under study, by minimizing the negative log-likelihood of the measured data.

The deadband is fitted with \Cref{apx-eq: stationary ansatz deadband} using a single parameter $a = \epsilon^2 / \delta$, which is bounded loosely as $a \in [10^{-8}, 10^{4}]$ to ensure positivity. After fitting, the distribution is normalized numerically in $\Omega_0$.

The tails are fitted with a Gaussian and a $q$-Gaussian ansatz. Specifically, for the Gaussian, we take a distribution centered at $\mu = \omega_1$ and fit the aggregated tails beyond $\omega_1$ with
\begin{align*}
    p(\omega) = 2\times\begin{cases}
        0 \quad &\omega < \omega_1 \\
        \displaystyle\frac{1}{\sqrt{2 \pi \sigma^2}} \exp \left[ -\frac{1}{2} \left( \frac{\omega - \omega_1}{\sigma} \right)^2 \right] \quad & \omega \geq \omega_1 \,.
    \end{cases}
\end{align*}
We optimize $\sigma$ that is loosely bounded as $\sigma \in [10^{-6}, 10^{4}]$. For the $q$-Gaussian, we take a distribution centered at $\mu = \omega_1$ and fit the aggregated tails beyond $\omega_1$ with
\begin{align*}
    p(\omega) = 2 \times \begin{cases}
        0 \quad & \omega < \omega_1 \\
        \displaystyle \frac{\sqrt{\beta}}{C_q} \left[ 1 + (1-q) \beta (\omega - \omega_1)^2 \right]_+^{{1}/({1-q})} \quad & \omega \geq \omega_1 \,,
    \end{cases}
\end{align*}
where, $[\cdot]_+$ is the positive part function, $q$ is the $q$-Gaussian nonextensivity parameter, $\beta$ is its inverse temperature, and $C_q$ its normalization constant.
We optimize $q$ and $\beta$ that are loosely bounded as $q \in [10^{-6}, 2.9999]$ and $\beta \in [10^{-6}, 10^{2}]$.

The tilted washboard distribution is fitted with two parameters, $a = {\delta \gamma}/{\epsilon^2}$ and $b= {2 \kappa \gamma}/{\epsilon^2}$, which are again loosely bounded as $a \in [10^{-6}, 10^{4}]$ and $b \in [10^{-6}, 10^{4}]$ to ensure their positivity.

MLE over $\num{1}$ million data points for each ansatz gives the parameters
\begin{alignat*}{2}
    \text{Deadband:} \quad a &= \num{0.0518 \pm 0.0007}\\
    \text{Tails Gaussian:} \quad \sigma &= \num{0.1671 \pm 0.0001}  \qquad\qquad\qquad\qquad\qquad\qquad\quad\, \text{neg-log-lik} &&= -\num{1061318.9897214} \\
    \text{Tails $q$-Gaussian:} \quad q &= \num{1.261 \pm 0.002} \qquad\qquad \beta = \num{29.10 \pm 0.09} \qquad\quad \text{neg-log-lik} &&= -\num{1080778.6347023}\\
    \text{Phase difference:} \quad a &= \num{420.8  \pm 0.6} \\
    \quad b &= \num{435.1  \pm 0.6} \,.
\end{alignat*}

We obtain parameters' standard deviations by inverting the Hessian of the log-likelihood evaluated at the optimum. We add iterative Tikhonov regularization when the Hessian is ill-conditioned. That is, we invert $H_{\text{reg}} = H + \lambda \mathbb{I}$ where $H$ is the log-likelihood Hessian, $\mathbb{I}$ the identity matrix, and $\lambda$ a regularization parameter that gets increased iteratively. To ensure convergence to the global minimum, we repeat each optimization with 50 random parameter initializations and select the fit with the lowest negative log-likelihood. We also accept only fits whose parameter variances remain below a threshold, ensuring that the regularization has not overly smoothed the log-likelihood. In practice, such precautions are not needed for the fits above, but make the routine more stable for potential future utilizations.

For the tails fit, we also report the negative log-likelihood values (truncated to arbitrary precision). Note how the negative log-likelihood is lower for the $q$-Gaussian fit, confirming its higher accuracy in modeling heavy tails, as also corroborated by the Q-Q plot of \Cref{apx-fig: qq plot tails}.

\end{document}